\documentclass[a4paper,10pt,]{article}
\usepackage{imakeidx}
\makeindex[columns=2, title=Alphabetical Index, intoc]
\usepackage{jheppub} 
\usepackage{lineno}
\usepackage{amsmath}
\usepackage{dsfont} 
\usepackage{nicefrac}
\usepackage[dvipsnames]{xcolor}
\usepackage{braket}
\usepackage{enumitem}
\usepackage{mathtools}
\usepackage{cleveref}
\usepackage{bbold}
\usepackage{physics}
\usepackage{graphicx}

\usepackage{multirow}
\usepackage{caption}
\usepackage{makecell}
\usepackage{subcaption}

\usepackage{tikz-feynman}
\usepackage{mathdots}
\usepackage{textgreek}

\usepackage{rotating}
\usepackage{cancel}
\usepackage{slashed}

\usepackage{longtable} 

\makeatletter
\newcommand\ps@indexpagestyle{
  \renewcommand\@oddfoot{\hfill-- \thepage\ --\hfill}
  \renewcommand\@oddhead{}
}
\makeatother
\indexsetup{firstpagestyle=indexpagestyle}

\def\SLAC{SLAC National Accelerator Laboratory, Stanford University, Stanford, CA 94039, USA}

\def\KITA{Institute for Theoretical Particle Physics, KIT, Wolfgang-Gaede-Straße 1, 76131, Karlsruhe, Germany}

\def\TIF{Tif Lab, Dipartimento di Fisica, Università di Milano and
INFN, Sezione di Milano, Via Celoria 16, I-20133 Milano, Italy}
\def\MP{Max-Planck-Institut für Physik, Boltzmannstrasse 8, 85748 Garching, Germany}

\preprint{
\begin{flushright} 
MPP-2025-168,
P3H-25-065,
SLAC-PUB-250309,
TIF-UNIMI-2025-19,
TTP25-030
\end{flushright}
}

\title{\boldmath Integrated subtraction terms and finite remainders for arbitrary processes with massless partons at colliders in the nested soft-collinear subtraction scheme}

\author[a]{Federica Devoto,}
\author[b]{Kirill Melnikov,}
\author[c]{Raoul R{\"o}ntsch,}
\author[d]{Chiara Signorile-Signorile,}
\author[b]{Davide Maria Tagliabue,}
\author[b]{Matteo Tresoldi}

\emailAdd{federica@slac.stanford.edu}
\emailAdd{kirill.melnikov@kit.edu}
\emailAdd{raoul.rontsch@unimi.it}
\emailAdd{chiara.signorile-signorile@cern.ch}
\emailAdd{davide.tagliabue@kit.edu}
\emailAdd{matteo.tresoldi@partner.kit.edu}

\affiliation[a]{\SLAC}
\affiliation[b]{\KITA}
\affiliation[c]{\TIF}
\affiliation[d]{\MP}

\abstract{We present integrated subtraction terms and finite remainders for \emph{arbitrary} processes with massless partons at hadron and lepton colliders in the context of the nested soft-collinear subtraction scheme.
These results provide the very last ingredients needed to make this scheme a fully local, analytic and process-independent framework for treating infrared singularities at next-to-next-to-leading order in perturbative QCD.
The explicit infrared finiteness of all required contributions, as well as their process-independence, puts these results on par with subtraction schemes developed for next-to-leading order computations and opens up a clear path towards the  automation of next-to-next-to-leading order computations in QCD. }


\usepackage{bbding}
\usepackage{mathtools}
\usepackage{booktabs}
\usepackage{lscape}
\usepackage{url}
\usepackage{bbold}
\usepackage{csquotes}

\usepackage{xspace}
\usepackage{makecell}
\usepackage{xifthen}

\usepackage{pgfplots}
\usepgfplotslibrary{colorbrewer}
\pgfplotsset{compat=newest}
\pgfplotsset{colormap/violet}

\usepackage{dowith}

\newcommand{\UpperCase}[1]{
  \expandafter\newcommand\csname bb#1\endcsname{{\mathbb{#1}}}
  \expandafter\newcommand\csname cal#1\endcsname{{\mathcal{#1}}}   
  \expandafter\newcommand\csname rm#1\endcsname{{\mathrm{#1}}}
  \expandafter\newcommand\csname bf#1\endcsname{{\mathbf{#1}}}
  \expandafter\newcommand\csname bold#1\endcsname{{\boldsymbol{#1}}}
  \expandafter\newcommand\csname hat#1\endcsname{\hat{#1}}
  \expandafter\newcommand\csname tilde#1\endcsname{\widetilde{#1}}
  \expandafter\newcommand\csname bar#1\endcsname{\overline{#1}}
  \expandafter\newcommand\csname frak#1\endcsname{\mathfrak{#1}}
  }
\DoWith\UpperCase ABCDEFGHILJKMNOPQRSTUVWXYZ\StopDoing

\newcommand{\LowerCase}[1]{
  \expandafter\newcommand\csname rm#1\endcsname{{\mathrm{#1}}} 
  \expandafter\newcommand\csname bf#1\endcsname{{\mathbf{#1}}} 
  \expandafter\newcommand\csname bold#1\endcsname{{\boldsymbol{#1}}}
  \expandafter\newcommand\csname hat#1\endcsname{\hat{#1}}
  \expandafter\newcommand\csname tilde#1\endcsname{\tilde{#1}}
  \expandafter\newcommand\csname bar#1\endcsname{\bar{#1}}
  \expandafter\newcommand\csname frak#1\endcsname{\mathfrak{#1}}
  \expandafter\newcommand\csname vec#1\endcsname{\vec{#1}}
  }
\DoWith\LowerCase abcdefghijklmnopqrstuvwxyz\StopDoing

\newcommand{\HP}{\calH}
\newcommand{\HPf}{{\calH_{\rmf}}}

\newcommand{\Born}{\calB}

\newcommand{\inotj}{(ij)}

\newcommand{\Q}[1]{Q_{#1}}

\newcommand{\Qab}{\Q{\inF\inS}}

\newcommand{\setg}[1]{\{g\}_{#1}}

\newcommand{\setq}[1]{\{q\}_{#1}}
\newcommand{\setqb}[1]{\{\qb\}_{#1}}

\newcommand{\setf}[3]{\calB_{#1,#2}^{#3}}

\newcommand{\Ng}{{N_g}}
\newcommand{\Nq}{{N_q}}
\newcommand{\Nqb}{{N_\qb}}
\newcommand{\barotimes}{\bar{\otimes}}
\newcommand{\colorprod}{\cdot}
\newcommand{\conv}{\otimes}

\newcommand{\eq}{Eq.~}
\newcommand{\eqdef}{\overset{\mathrm{def}}{=}}

\newcommand{\mydots}{...}

\newcommand{\oS}{\barS}
\newcommand{\oC}{\barC}

\newcommand{\qb}{{\bar{q}}}
\newcommand{\qp}{{q'}}
\newcommand{\qbp}{{\qb'}}

\newcommand{\np}{n'}

\newcommand{\iden}{\mathbb{1}}

\newcommand{\T}{\boldsymbol{T}}

\newcommand{\THmn}{\Theta_{\Fp \Sp}}   
\newcommand{\THnm}{\Theta_{\Sp \Fp}}

\newcommand{\fin}{{\mathrm{fin}}}

\newcommand{\gen}{{\mathrm{gen}}}

\newcommand{\lhs}{left-hand side}

\newcommand{\LV}{{\rm LV}}

\newcommand{\pdf}{{\mathrm{pdf}}}

\newcommand{\rhs}{right-hand side}
\newcommand{\R}{{\mathrm{R}}}
\newcommand{\RR}{{\mathrm{RR}}}
\newcommand{\RV}{{\mathrm{RV}}}

\newcommand{\Sec}{Section~}

\newcommand{\V}{\mathrm{V}}
\newcommand{\VV}{\mathrm{VV}}

\newcommand{\TR}{T_\rmR}
\newcommand{\nf}{{n_\rmf}}

\newcommand{\Ca}{C_\rmA}
\newcommand{\Cf}{C_\rmF}

\newcommand{\LO}{\mathrm{LO}}
\newcommand{\NLO}{\mathrm{NLO}}
\newcommand{\NNLO}{\mathrm {NNLO}}
\newcommand{\ONLO}{\mathcal{O}_\text{NLO}}

\newcommand{\colsing}{X}

\newcommand{\Emax}{E_{\rm max}}
\newcommand{\ep}{\epsilon}

\newcommand{\muF}{\mu_{\rmF}}
\newcommand{\muR}{\mu_{\rmR}}

\newcommand{\as}{\alpha_{\rms}}

\newcommand{\asbr}{[\alpha_{\rms}]}

\newcommand{\gsb}{g_{\rms, \rmb}}

\newcommand{\dE}{\rmd E}
\newcommand{\dt}{\rmd t}

\newcommand{\dx}{\rmd x}
\newcommand{\dz}{\rmd z}

\newcommand{\dxi}{\rmd \xi}

\newcommand{\dsigma}{\rmd\sigma}
\newcommand{\dsigmahat}{\rmd \hat{\sigma}}

\newcommand{\inF}{a} 
\newcommand{\inS}{b}
\newcommand{\Fp}{\mathfrak{m}}
\newcommand{\Sp}{\mathfrak{n}}
\newcommand{\barFp}{\bar{\Fp}}
\newcommand{\barSp}{\bar{\Sp}}
\newcommand{\fgflin}{\rho}  
\newcommand{\sgflin}{\tau}  
\newcommand{\tgflin}{\upsilon}  

\newcommand{\lint}{\big\langle}
\newcommand{\rint}{\big\rangle}
\newcommand{\llint}{\Big\langle}
\newcommand{\rrint}{\Big\rangle}
\newcommand{\Lint}{\bigg\langle}
\newcommand{\Rint}{\bigg\rangle}

\newcommand{\qarrowqb}{q \leftrightarrow \qb}

\newcommand{\PAP}{\hatP^{(0)}}
\newcommand{\PAPone}{\hatP^{(1)}}
\newcommand{\CalPgen}{\calP^{\gen}}

\newcommand{\CalPTC}{\calP^{\rm TC}}
\newcommand{\calPW}{\calP^{\calW}}

\newcommand{\ISR}[1]{\text{ISR}_{#1}}
\newcommand{\FSR}[1]{\text{FSR}_{#1}}
\newcommand{\FSRtilde}[1]{\widetilde{\text{FSR}}_{#1}}

\newcommand{\PNLO}{\calP^{\mathrm{NLO}}}
\newcommand{\PNNLO}{\calP^{\mathrm{NNLO}}}

\newcommand{\PgenoxPgen}[2]{\big[\CalPgen_{#1} \, \bar{\otimes} \, \CalPgen_{#2}\big]}

\newcommand{\GammaTC}{\Gamma^{\rm TC}}
\newcommand{\GFSR}[3]{G_{#1}\big|{\substack{ {#2} \\ {#3} }}}

\newcommand{\partFuncNLOfp}[1]{\omega^{\Fp #1 }}

\newcommand{\partFuncACsp}[1]{\omega^{\Fp #1, \Sp #1}_{#1 \parallel \Sp}}
\newcommand{\partFuncBD}[1]{\omega^{\Fp #1, \Sp #1}_{\Fp \parallel \Sp}}

\newcommand{\Wacfin}[1]{\calW_{#1}^{#1 \parallel \Sp, \fin}}

\newcommand{\Wbdfin}[1]{\calW_{#1}^{\Fp \parallel \Sp, \fin}}
\newcommand{\Wr}[1]{\calW_\rmr^{(#1)}}

\newcommand{\IVirt}{I_{\rm V}}

\newcommand{\ISoft}{I_{\rm S}}

\newcommand{\IColl}{I_{\rm C}}

\newcommand{\ITot}{I_{\rm T}}

\newcommand{\Iccfin}{I_{\rm cc}^{\fin}}

\newcommand{\Itrifin}{I_{\mathrm{tri}}^\fin}
\newcommand{\Iuncfin}{I_{\rm unc}^\fin}

\newcommand{\FLM}{F_{\mathrm{LM}}}
\newcommand{\FLMlo}[2]{\FLM^{#1}[#2]}
\newcommand{\FLMun}[3]{\FLM^{#1}[#2|#3]}
\newcommand{\FLMmunu}{F_{\mathrm{LM},\mu\nu}}
\newcommand{\FLMlomunu}[2]{\FLMmunu^{#1}[#2]}

\newcommand{\bbFLM}{\bbF_{\rm LM}}
\newcommand{\bbFLMlo}[2]{\bbFLM^{#1}[#2]}
\newcommand{\bbFLMlonf}[3]{\bbF_{\rm LM,#1}^{#2}[#3]}

\newcommand{\FLVfin}{F_{\LV,\fin}}
\newcommand{\FLVfinlo}[2]{F_{\LV,\fin}^{#1}[#2]}

\newcommand{\FLVfinsq}{F_{\LV^2,\fin}}
\newcommand{\FLVfinsqlo}[2]{F_{\LV^2,\fin}^{#1}[#2]}

\newcommand{\FVVfin}{F_{\VV,\fin}}
\newcommand{\FVVfinlo}[2]{F_{\VV,\fin}^{#1}[#2]}

\newcommand{\FRVfinlo}[2]{F_{\RV,\fin}^{#1}[#2]}

\newcommand{\DS}{{\rm DS}}
\newcommand{\noDS}{\cancel{{\rm DS}}}

\newcommand{\SigmaDCdc}{\Sigma_{\rm DC}^{\rm dc}}
\newcommand{\SigmaDCtc}{\Sigma_{\rm DC}^{\rm tc}}

\newcommand{\db}{{\rm db}}
\newcommand{\inFsb}{{{\rm sb},\inF}}
\newcommand{\inSsb}{{{\rm sb},\inS}}
\newcommand{\el}{{\rm el}}

\newcommand{\du}{{\rm DU}}
\newcommand{\fr}{{\rm FR}}
\newcommand{\su}{{\rm SU}}

\newcommand{\finalresult}{\texttt{FinalResult.m}\xspace}
\newcommand{\readme}{\texttt{README.txt}\xspace}
\newcommand{\triplecollinearsplittings}{\texttt{TripleCollinearSplittings.m}\xspace}

\newcommand{\xa}{\mathfrak{m}}
\newcommand{\yb}{\mathfrak{n}}

\newcommand{\myfloor}[1]{{\left\lfloor {#1} \right\rfloor}}

\newcommand{\fl}[1]{f_{#1}}

\usepackage[normalem]{ulem}

\usepackage{dashrule}

\allowdisplaybreaks   
\begin{document}

\newpage 

\maketitle
\flushbottom





\newcommand{\SoftLimit}[2]{
    \scalebox{#1}{
    \begin{tikzpicture}[baseline=(ic.base)]
        \begin{feynman}
            \vertex (ic);
            \vertex [blob, right=0.3cm of ic] (c) {};
            \vertex [above=0.4cm of ic] (i1) [label={[left=0.0cm] {$\inF$}}];
            \vertex [below=0.4cm of ic] (i2) [label={[left=0.0cm] {$\inS$}}];
            
            \vertex [right=1.3cm of c] (fc);
            \vertex [above=0.8cm of fc] (fA);
            \vertex [below=0.8cm of fc] (fB);

            \vertex at ($(fc) + (-0.1cm, 0.5cm)$) (dotA1) {$\vdots$};
            \vertex at ($(fc) + (-0.1cm, -0.3cm)$) (dotA2) {$\vdots$};
            
            \diagram{
                (i1) -- [plain] (c) -- [plain] (i2);
                (c) -- [plain] (fA);
                (c) -- [plain] (fB);
                (fc) -- [red, #2, edge label={$\Fp$}] (c);
            };
        \end{feynman} 
    \end{tikzpicture}
    }
}

\newcommand{\SquaredMatrixElementGen}[2]{
    \scalebox{#1}{
    \begin{tikzpicture}[baseline=(c.base)]
        \begin{feynman}
            \vertex (ic);
            \vertex [blob, right=0.3cm of ic] (c) {};
            \vertex [above=0.4cm of ic] (i1) [label={[left=0.0cm] {$#2$}}];
            \vertex [below=0.4cm of ic] (i2) [label={[left=0.0cm] {$\inS$}}];
            \vertex [right=1.cm of c] (fc);
            \vertex [above=0.6cm of fc] (fA);
            \vertex [below=0.6cm of fc] (fB);

            \vertex at ($(fc) + (-0.1cm, 0.31cm)$) (dotA1) {$\vdots$};
            \vertex at ($(fc) + (-0.1cm, -0.11cm)$) (dotA2) {$\vdots$};
            
            \diagram{
                (i1) -- [plain] (c) -- [plain] (i2);
                (c) -- [plain] (fA);
                (c) -- [plain] (fB);
            };
        \end{feynman} 
    \end{tikzpicture}
    }
}


\newcommand{\CollLimitFSRgen}[1]{
    \scalebox{#1}{
    \begin{tikzpicture}[baseline=(c.base)]
        \begin{feynman}
            \vertex (ic);
            \vertex [blob, right=0.3cm of ic] (c) {};
            \vertex [above=0.4cm of ic] (i1) [label={[left=0.0cm] {$\inF$}}];
            \vertex [below=0.4cm of ic] (i2) [label={[left=0.0cm] {$\inS$}}];
            \vertex [right=1.5cm of c] (fc);
            \vertex [above=0.2cm of fc] (fA);
            \vertex [above right=1.cm and 1.2cm of c] (fAA);
            \vertex [below=0.2cm of fc] (fB);
            \vertex [below right=1.cm and 1.2cm of c] (fBB);

            \vertex at ($(c) + (1.3cm, 0.65cm)$) (dotA1) [rotate=-30] {$\ddots$};
            \vertex at ($(c) + (1.2cm, -0.5cm)$) (dotA2) [rotate=30] {$\iddots$};
            
            \diagram{
                (i1) -- [plain] (c) -- [plain] (i2);
                (c) -- [red, plain, edge label={$\Fp$}, sloped] (fA);
                (c) -- [plain] (fAA);
                (c) -- [plain, edge label'={$i$}, sloped] (fB);
                (c) -- [plain] (fBB);
            };
        \end{feynman} 
    \end{tikzpicture} 
    }
}

\newcommand{\CollLimitFSRmain}[1]{
    \scalebox{#1}{
    \begin{tikzpicture}[baseline=(c.base)]
    \begin{feynman}
        \vertex (ic);
        \vertex [blob, right=0.3cm of ic] (c) {};
        \vertex [above=0.4cm of ic] (i1) [label={[left=0.0cm] {$\inF$}}];
        \vertex [below=0.4cm of ic] (i2) [label={[left=0.0cm] {$\inS$}}];
        \vertex [above right=1.1cm and 1.2cm of c] (fAA);
        \vertex [below right=1.1cm and 1.2cm of c] (fBB);
        \vertex [right=1.3cm of c] (f1);
        \vertex [right=0.9cm of f1] (f2);
        \vertex [above=0.7cm of f2] (f3);
        \vertex at ($(c) + (1.35cm, 0.85cm)$) (dotA1) [rotate=-10] {$\ddots$};
        \vertex at ($(c) + (1.3cm, -0.6cm)$) (dotA2) [rotate=10] {$\iddots$};
               
        \diagram{
            (i1) -- [plain] (c) -- [plain] (i2);
            (c) -- [plain, edge label'=$~~{[i\Fp]}$] (f1) -- [plain, edge label'=$~~~~i$] (f2);
            (f3) -- [red, plain, edge label={$~~~~~\Fp$}, sloped] (f1);
            (c) -- [plain] (fAA);
            (c) -- [plain] (fBB);
        };
    \end{feynman} 
    \end{tikzpicture} 
    }
}

\newcommand{\SquaredMatrixElementFSR}[1]{
    \scalebox{#1}{
    \begin{tikzpicture}[baseline=(c.base)]
        \begin{feynman}
            \vertex (ic);
            \vertex [blob, right=0.3cm of ic] (c) {};
            \vertex [above=0.4cm of ic] (i1) [label={[left=0.0cm] {$\inF$}}];
            \vertex [below=0.4cm of ic] (i2) [label={[left=0.0cm] {$\inS$}}];
            
            \vertex [right=1.5cm of c] (fc) {${[i\Fp]}$};
            \vertex [above=0.9cm of fc] (fA);
            \vertex [below=0.9cm of fc] (fB);

            \vertex at ($(fc) + (0cm, 0.6cm)$) (dotA1) [rotate=-30] {$\ddots$};
            \vertex at ($(fc) + (-0.1cm, -0.4cm)$) (dotA2) [rotate=30] {$\iddots$};
            
            \diagram{
                (i1) -- [plain] (c) -- [plain] (i2);
                (c) -- [plain] (fA);
                (c) -- [plain] (fB);
                (fc) -- [plain] (c);
            };
        \end{feynman}
    \end{tikzpicture}
    }
}


\newcommand{\CollLimitISR}[1]{
    \scalebox{#1}{
    \begin{tikzpicture}[baseline=(c.base)]
    \begin{feynman}
        \vertex (ic);
        \vertex [blob, below right=0.4cm and 0.8cm of ic] (c) {};
        \vertex [above left=0.3cm and 0.47cm of ic] (i1) [label={[left=0.0cm] {$\inF$}}];
        \vertex [above right=0.5cm and 1.3cm of ic] (im);
        \vertex [below left=1.6cm and 0.47cm of ic] (i2) [label={[left=0.0cm] {$\inS$}}];
        \vertex [right=0.8cm of c] (fc);
        \vertex [right=1.cm of c] (fc);
        \vertex [above=0.6cm of fc] (fA);
        \vertex [below=0.6cm of fc] (fB);

        \vertex at ($(fc) + (-0.1cm, 0.31cm)$) (dotA1) {$\vdots$};
        \vertex at ($(fc) + (-0.1cm, -0.11cm)$) (dotA2) {$\vdots$};
        \diagram{
            (i1) -- [plain] (ic) -- [plain, edge label'=${[\inF\barFp]}$, sloped] (c);
            (im) -- [red, plain, edge label={$\Fp$}, sloped] (ic);
            (i2) -- [plain] (c);
            (c) -- [plain] (fA);
            (c) -- [plain] (fB);
        };
    \end{feynman} 
    \end{tikzpicture}
    }
}


\newcommand{\VirtualOneL}[1]{
    \scalebox{#1}{
    \begin{tikzpicture}[baseline=(ic.base)]
        \begin{feynman}
            \vertex (ic);
            \vertex [blob, fill=none, right=0.3cm of ic] (c) {1-L};
            \vertex [above=0.4cm of ic] (i1) [label={[left=0.0cm] {$\inF$}}];
            \vertex [below=0.4cm of ic] (i2) [label={[left=0.0cm] {$\inS$}}];
            \vertex [right=1.cm of c] (fc);
            \vertex [above=0.6cm of fc] (fA);
            \vertex [below=0.6cm of fc] (fB);

            \vertex at ($(fc) + (-0.1cm, 0.31cm)$) (dotA1) {$\vdots$};
            \vertex at ($(fc) + (-0.1cm, -0.11cm)$) (dotA2) {$\vdots$};
            
            \diagram{
                (i1) -- [plain] (c) -- [plain] (i2);
                (c) -- [plain] (fA);
                (c) -- [plain] (fB);
            };
        \end{feynman} 
    \end{tikzpicture}
    }
}

\newcommand{\SquaredMatrixElementGenRevers}[1]{
    \scalebox{#1}{
    \begin{tikzpicture}[baseline=(c.base)]
        \begin{feynman}
            \vertex (ic);
            \vertex [blob, right=0.63cm of ic] (c) {};
            \vertex [above=0.6cm of ic] (i1);
            \vertex [below=0.6cm of ic] (i2);
            \vertex [right=0.65cm of c] (fc);
            \vertex [above=0.4cm of fc] (fA) [label={[right=0.0cm] {$\inF$}}];
            \vertex [below=0.4cm of fc] (fB) [label={[right=0.0cm] {$\inS$}}];

            \vertex at ($(ic) + (0.1cm, 0.31cm)$) (dotA1) {$\vdots$};
            \vertex at ($(ic) + (0.1cm, -0.11cm)$) (dotA2) {$\vdots$};
            
            \diagram{
                (i1) -- [plain] (c) -- [plain] (i2);
                (c) -- [plain] (fA);
                (c) -- [plain] (fB);
            };
        \end{feynman} 
    \end{tikzpicture}
    }
}


\newcommand{\LOvsNLOISR}[6]{
    \scalebox{1.0}{
    \begin{tikzpicture}[baseline=(c.base)]
        \begin{feynman}
            \vertex (ic);
            \vertex [blob, right=0.4cm of ic] (c) {};
            \vertex [above=1.5cm of ic] (ic2);
            \vertex [left=1cm of ic2] (i1) [label={[left=0.0cm] {#1}}];
            \vertex [left=1cm of ic] (i2) [label={[left=0.0cm] {$\inS$}}];
            \vertex [above right=0.3cm and 1cm of ic2] (f1) [label={[right=0.0cm] {#2}}];
            \vertex [right=1.cm of c] (fc);
            \vertex [above=0.6cm of fc] (fA);
            \vertex [below=0.6cm of fc] (fB);

            \vertex at ($(fc) + (-0.1cm, 0.31cm)$) (dotA1) {$\vdots$};
            \vertex at ($(fc) + (-0.1cm, -0.11cm)$) (dotA2) {$\vdots$};
            
            \diagram{
                (i1) -- [red, #4] (ic2);
                (i2) -- [plain] (c);
                (c) -- [red, #5, edge label={\color{black} ~~#3}, sloped] (ic2) -- [red, #6] (f1);
                (c) -- [plain] (fA);
                (c) -- [plain] (fB);
            };

            \draw [decorate,decoration={brace,amplitude=5pt}]
            ($(fA.north west) + (1.5mm,0)$) -- ($(fB.south west) + (1.5mm,0)$)
            node[midway,xshift=.7cm] {$\setf{N}{n'}{}$};

        \end{feynman} 
    \end{tikzpicture}
    }
}


\newcommand{\LOvsNLOFSR}[6]{
    \scalebox{1.0}{
    \begin{tikzpicture}[baseline=(c.base)]
        \begin{feynman}
            \vertex (ic);
            \vertex [blob, right=0.5cm of ic] (c) {};
            \vertex [above=0.4cm of ic] (i1) [label={[left=0.0cm] {$\inF$}}];
            \vertex [below=0.4cm of ic] (i2) [label={[left=0.0cm] {$\inS$}}];
            \vertex [right=1.cm of c] (fc);
            \vertex [above=0.6cm of fc] (fA);
            \vertex [below=0.6cm of fc] (fB);

            \vertex at ($(fc) + (-0.1cm, 0.31cm)$) (dotA1) {$\vdots$};
            \vertex at ($(fc) + (-0.1cm, -0.11cm)$) (dotA2) {$\vdots$};
            \vertex [above right=1.2cm and 0.7cm of fA] (fc2);
            \vertex [above right=0.45cm and 0.7cm of fc2] (f1) [label={[right=0.0cm] {#2}}];
            \vertex [below right=0.45cm and 0.7cm of fc2] (f2) [label={[right=0.0cm] {#3}}];
            \vertex [above=0.8cm of fA] (whitebullet);
            
            \diagram{
                (i1) -- [plain] (c) -- [plain] (i2);
                (c) -- [plain] (fA);
                (c) -- [plain] (fB);
                (c) -- [red, #4, edge label={\color{black} \!\!\!#1}, sloped] (fc2);
                (f2) -- [red, #6] (fc2) -- [red, #5] (f1);
            };

            \draw[fill=white,draw=none] ($(whitebullet)+(2.35mm,-1.7mm)$) circle (3pt);
            \draw[fill=white,draw=none] ($(whitebullet)+(1.9mm,-0.9mm)$) circle (2pt);
            \draw[fill=white,draw=none] ($(whitebullet)+(2.4mm,-0.9mm)$) circle (1.7pt);

            \draw [decorate,decoration={brace,amplitude=5pt}]
            ($(whitebullet.north west) + (1.5mm,0)$) -- ($(fB.south west) + (1.5mm,0)$)
            node[midway,xshift=.7cm] {$\setf{N}{n'}{}$};
            
        \end{feynman} 
    \end{tikzpicture}
    }
}
\newcommand{\figureRules}[2]{
    \begin{figure}[#1]
        \centering
        \underline{\text{SINGLE-SOFT LIMIT}}  \vspace{-2mm}                
        \begin{gather*}
            \left\langle S_\Fp \Delta^{(\Fp)} \left| \hspace{-1.4mm} \SoftLimit{0.8}{gluon} \hspace{-0.6mm} \right|^2 \right\rangle
            = \asbr \left\langle \ISoft(\ep) \colorprod \left| \hspace{-1.4mm} \SquaredMatrixElementGen{0.8}{\inF} \hspace{-0.6mm} \right|^2 \right\rangle
        \end{gather*} 
        \\ \vspace{3mm}
        \underline{\text{FINAL-STATE COLLINEAR LIMIT}}  \vspace{-2.5mm}                
        \begin{gather*}
            \left\langle \oS_\Fp C_{i\Fp} \Delta^{(\Fp)} \left| \hspace{-1.4mm} \CollLimitFSRgen{0.8} \hspace{-2.5mm} \right|^2 \right\rangle
            \sim \left\langle \oS_\Fp C_{i\Fp} \Delta^{(\Fp)} \left| \hspace{-1.4mm} \CollLimitFSRmain{0.8} \right|^2 \right\rangle \\
            = \frac{\asbr}{\ep} \left\langle \Gamma_{[i\Fp],f_{[i\Fp]} \to f_i f_\Fp} \left| \hspace{-1.4mm} \SquaredMatrixElementFSR{0.8} \hspace{-2.1mm} \right|^2 \right\rangle 
        \end{gather*} 
        \\ \vspace{3mm}
        \underline{\text{INITIAL-STATE COLLINEAR LIMIT}} \vspace{-3mm}               
        \begin{gather*}
            \left\langle \oS_\Fp C_{\inF\Fp} \Delta^{(\Fp)} \left| \hspace{-1.4mm} \SoftLimit{0.8}{plain} \hspace{-0.6mm} \right|^2 \right\rangle
            \sim \left\langle \oS_\Fp C_{\inF\Fp} \Delta^{(\Fp)} \left| \hspace{-1.4mm}\CollLimitISR{0.8} \hspace{-0.6mm} \right|^2 \right\rangle = \\
            = \frac{\asbr}{\ep} \, \delta_{g \Fp} \left\langle \Gamma_{\inF,f_\inF}  \left| \hspace{-1.4mm} \SquaredMatrixElementGen{0.8}{\inF} \hspace{-0.6mm} \right|^2 \right\rangle
            + \frac{\asbr}{\ep} \left\langle \CalPgen_{[\inF\barFp]\inF} \conv \left| \hspace{-1.4mm} \SquaredMatrixElementGen{0.8}{z\cdot[\inF\barFp]} \hspace{-0.6mm} \right|^2 \right\rangle 
        \end{gather*}
        \\ \vspace{3mm}
        \underline{\text{VIRTUAL CORRECTIONS}} \vspace{-2mm}                     
        \begin{gather*}
            \left\langle 2\Re\left( \hspace{-0.5mm} \left. \hspace{-1.4mm} \VirtualOneL{0.8} \hspace{-0.6mm} \right| \hspace{-1.2mm} \SquaredMatrixElementGenRevers{0.8} \hspace{-1.5mm} \right) \right\rangle
            = \asbr \left\langle \IVirt(\ep) \colorprod \left| \hspace{-1.4mm} \SquaredMatrixElementGen{0.8}{\inF} \hspace{-0.6mm} \right|^2 \right\rangle
            + \lint\FLVfin^{\inF\inS}\rint
        \end{gather*}
        \caption{#2}
        \label{fig_main_collinear_factorization_FSR_ISR}
    \end{figure}
}


\newcommand{\figureNLOvsLO}[2]{
    \begin{figure}[#1]
    \centering
        \begin{subfigure}{0.30\textwidth}
            \centering
            \LOvsNLOISR{$\inF_g$}{$\Fp_{q_\fgflin}$}{$[\inF\Fp]_{\qb_\fgflin}$}{gluon}{plain}{plain}
            \caption{}
            \label{figureNLOvsLO_a}
        \end{subfigure}
        \begin{subfigure}{0.30\textwidth}
            \centering
            \LOvsNLOISR{$\inF_{q_\fgflin}$}{$\Fp_{q_\fgflin}$}{$[\inF\Fp]_g$}{plain}{gluon}{plain}
            \caption{}
            \label{figureNLOvsLO_b}
        \end{subfigure}
        \\ \vspace{5mm}
        \begin{subfigure}{0.30\textwidth}
            \centering
            \LOvsNLOFSR{$[i\Fp]_{q_\fgflin}$}{$\Fp_{q_\fgflin}$}{$i_g$}{plain}{plain}{gluon}
            \caption{}
            \label{figureNLOvsLO_c}
        \end{subfigure}
        \begin{subfigure}{0.30\textwidth}
            \centering
            \LOvsNLOFSR{$[i\Fp]_g$}{$\Fp_{q_\fgflin}$}{$i_{\qb_\fgflin}$}{gluon}{plain}{plain}
            \caption{}
            \label{figureNLOvsLO_d}
        \end{subfigure}
        
        \caption{#2}
        \label{fig_figureW}
\end{figure}
}

\newcommand{\tableFinalResult}[2]{
\begin{table}[#1] 
    \centering
    \caption{#2}
    
    \begin{tabular}{
    >{\centering\arraybackslash}m{35mm}  
    >{\centering\arraybackslash}m{44mm} 
    >{\centering\arraybackslash}m{44.5mm}
    }
        \hline
        \multicolumn{3}{c}{\textbf{Functions collected in the ancillary file \finalresult}} \\
        \hline\hline
            \emph{Function}  
            & \emph{Eq.~number} 
            &  \emph{Name in the ancillary file} 
            \\ \hline
        \multicolumn{3}{c}{\emph{Quantities in spin-correlated contributions}} \\ \hline
            $\gamma_g^\perp$    
            & \eq\eqref{eq_dsigmahat_SU_el_final}
            & \texttt{\textgamma gPerp}  
            \\
            $\gamma_g^{\perp,\rmr}$    
            & \eq\eqref{eq_dsigmahat_SU_el_final}
            & \texttt{\textgamma gPerpR}  
            \\
            $\delta^{(0)}$    
            & \eq\eqref{eq_dsigmahat_DU_el_final}
            & \texttt{\textdelta zero}  
            \\
            $\delta^{\perp,(0)}$    
            & \eq\eqref{eq_dsigmahat_DU_el_final}
            & \texttt{\textdelta Perpzero}  
            \\ \hline
        \multicolumn{3}{c}{\emph{Splitting functions}} \\ \hline
            $\PAP_{xy}(z)$           
            & Eqs (\ref{eq_dsigmahat_su_inFsb}, \ref{eq_dsigmahat_su_inSsb})
            & \texttt{PxyAP[z\_]}
            \\
            $\PNLO_{xy}(z,E)$           
            & Eqs (\ref{eq_dsigmahat_su_inFsb}, \ref{eq_dsigmahat_su_inSsb}, \ref{eq_final_formula_DU_DB} -- \ref{eq_dsigmahat_ab_du_inSsb_nf})
            & \texttt{PxyNLO[z\_,En\_]}
            \\
            $\calPW_{xx}(z,E)$           
            & Eqs (\ref{eq_dsigmahat_ab_du_inFsb_nf}, \ref{eq_dsigmahat_ab_du_inSsb_nf})
            & \texttt{PxxW[z\_,En\_]}
            \\
            $\PNNLO_{xy}(z,E)$           
            & Eqs (\ref{eq_dsigmahat_ab_du_inFsb_nf}, \ref{eq_dsigmahat_ab_du_inSsb_nf})
            & \texttt{PxyNNLO[z\_,En\_]}
            \\ \hline
        \multicolumn{3}{c}{\emph{Elastic functions}} \\ \hline
            $\gamma_{z,g \to gg}^{\calW}(E)$     
            & \eq\eqref{eq_dsigmahat_DU_el_final}
            & \texttt{\textgamma WgTOgg[En\_]}
            \\
            $\gamma_{z,q \to qg}^{\calW}(E)$     
            & \eq\eqref{eq_dsigmahat_DU_el_final}
            & \texttt{\textgamma WqTOqg[En\_]}
            \\
            $D_{T^2}$     
            & \eq\eqref{eq_Iuncfin_def}
            & \texttt{DT2}
            \\
            $D_g^{\rm ISR}(E)$     
            & \eq\eqref{eq_Iuncfin_def}
            & \texttt{DgISR[En\_]}
            \\
            $D_q^{\rm ISR}(E)$     
            & \eq\eqref{eq_Iuncfin_def}
            & \texttt{DqISR[En\_]}
            \\
            $D_g^{\rm FSR}(E)$    
            & \eq\eqref{eq_Iuncfin_def}
            & \texttt{DgFSR[En\_]}  
            \\
            $D_q^{\rm FSR}(E)$    
            & \eq\eqref{eq_Iuncfin_def}
            & \texttt{DqFSR[En\_]}
            \\ \hline
        \multicolumn{3}{c}{\emph{Double-soft finite remainders}} \\ \hline
            ${\rm DS}^{\rm fin}_{ij}$ 
            & \eq\eqref{eq_Issfin_def}
            & \texttt{DSfin[i\_,j\_]}
            \\
            \noalign{\vspace{1mm}} \hline
    \end{tabular}
    \label{table_final_result}
\end{table}
}


\newcommand{\tableTripleCollinear}[2]{
\begin{table}[#1]
    \centering
    \caption{#2} 
    \label{table_triplecollinearsplittings}
    
    \begin{tabular}{
    >{\centering\arraybackslash}m{40mm}  
    >{\centering\arraybackslash}m{40mm} 
    >{\centering\arraybackslash}m{40mm}
    }
    \hline
    \multicolumn{3}{c}{\textbf{\shortstack{\rule{0pt}{4mm}Integrated triple-collinear subtraction functions \\ collected in \triplecollinearsplittings}}} \\
    \hline\hline
        \emph{Function}  
        & \emph{Splitting} 
        &  \emph{Name in the ancillary file} 
        \\ \hline \hline
    \multicolumn{3}{c}{\emph{Initial-state splitting functions}} \\ \hline
        $\CalPTC_{gg}(z,E)$    
        & $g \to g g g^* + \sum_\sgflin (q_\sgflin \qb_\sgflin) g^*$
        & \texttt{PTCgg[z\_,En\_]}  
        \\
        $\CalPTC_{gq}(z,E)$    
        & $q \to q g g^*$
        & \texttt{PTCgq[z\_,En\_]}  
        \\
        $\CalPTC_{q q}(z,E)$     
        & $q \to g g q^* + \sum_\sgflin (q_\sgflin \qb_\sgflin) q^*$
        & \texttt{PTCqq[z\_,En\_]}  
        \\
        $\CalPTC_{q g}(z,E)$     
        & $g \to g \qb q^*$
        & \texttt{PTCqg[z\_,En\_]}  
        \\
        $\CalPTC_{q \qb}(z,E)$     
        & $q \to q q \qb^*$
        & \texttt{PTCqqb[z\_,En\_]}  
        \\
        $\CalPTC_{q \qp}(z,E)$    
        & $q \to q \qbp \qp^* ~ \text{ with } ~ q \neq \qp$
        & \texttt{PTCqqp[z\_,En\_]}  
        \\
        \hline
    \multicolumn{3}{c}{\emph{Final-state splitting functions}} \\ \hline
        $\GammaTC_{g}(E)$    
        & $g^* \to g g g + \sum_\sgflin g (q_\sgflin \qb_\sgflin)$
        & \texttt{\textGamma TCg[En\_]}  
        \\
        $\GammaTC_{q}(E)$    
        & $q^* \to qgg + \sum_\sgflin q (q_\sgflin \qb_\sgflin)$
        & \texttt{\textGamma TCq[En\_]}  
        \\
        $\GammaTC_{\qb}(E)$    
        & $\qb^* \to \qb gg + \sum_\sgflin \qb (q_\sgflin \qb_\sgflin)$
        & \texttt{\textGamma TCqb[En\_]}  
        \\ 
        \hline
\end{tabular}
\end{table}
}


\section{Introduction}

Over the last two decades, the evolving needs of the physics program at the Large Hadron Collider (LHC) have shaped theoretical studies aimed at improving the understanding of partonic scattering processes. 
Since the asymptotic freedom of quantum chromodynamics (QCD) allows the use of perturbation theory for the description of high-energy interactions, computing higher-order predictions for a wide range of partonic cross sections has become one of the major undertakings in contemporary theoretical particle physics.
Central to this goal is the efficient treatment of infrared singularities,  which arise separately in real-emission and virtual corrections and must cancel among themselves to give a finite physical result. This is challenging because the real and virtual corrections populate different phase spaces, and solving this problem relies on so-called subtraction schemes. At next-to-leading order (NLO) in perturbative QCD, such schemes were developed in a process-independent fashion nearly thirty years ago~\cite{Frixione:1995ms, Catani:1996vz, Nagy:2003qn, Bevilacqua:2013iha} and have been extensively used for theoretical predictions.  
However, a similar level of understanding at \emph{next-to-next-to-leading order} (NNLO) has not yet been achieved. 

Indeed, in spite of the fact that many different subtraction schemes for NNLO calculations are being developed~\cite{Frixione:2004is,Gehrmann-DeRidder:2005btv,Currie:2013vh,
Somogyi:2005xz,Somogyi:2006db,DelDuca:2016csb,DelDuca:2016ily,Czakon:2010td,
Czakon:2011ve,Czakon:2014oma,Anastasiou:2003gr,Caola:2017dug,Catani:2007vq,
Grazzini:2017mhc,Boughezal:2011jf,Gaunt:2015pea,Boughezal:2015dva,Sborlini:2016hat,
Herzog:2018ily,Magnea:2018hab,Magnea:2020trj,Chen:2022ktf,Bertolotti:2022aih,Magnea:2024jqg, Capatti:2019ypt,
Braun-White:2023sgd,Braun-White:2023zwd,Fox:2023bma,Gehrmann:2023dxm,Fox:2024bfp, DelDuca:2024ovc}, none is as advanced as the NLO methods.
What is missing is the explicit demonstration of the cancellation of infrared divergences and the derivation of finite remainders of the integrated subtraction terms for \emph{arbitrary collider processes}.   
However, it is important to stress that the absence of such general results has not hindered the impressive progress in NNLO QCD computations. 
In fact, such calculations for many very important and complex LHC processes have already been performed (see, e.g., Refs~\cite{Chen:2014gva,
Boughezal:2015dra,Caola:2015wna,Chen:2016zka,Campbell:2019gmd,Cacciari:2015jma,
Cruz-Martinez:2018rod,Gauld:2021ule,Catani:2022mfv,Chawdhry:2019bji,Chawdhry:2021hkp,
Czakon:2020coa,Gauld:2023zlv,Currie:2017eqf,Chen:2022tpk,Badger:2023mgf,
Czakon:2021mjy,Czakon:2015owf,Catani:2019hip,Buonocore:2023ljm,Brucherseifer:2014ama,
Berger:2016oht,Campbell:2020fhf,Bronnum-Hansen:2022tmr,Buonocore:2022pqq,Alvarez:2023fhi,Armadillo:2024ncf, Devoto:2024nhl, Buonocore:2025fqs, Bonino:2025qta} for a selection of phenomenological papers employing different theoretical methods), which implies that this issue is hardly a \emph{practical} limitation. Nevertheless, we believe that understanding the infrared structure of perturbative QCD at NNLO in full generality is an interesting \emph{theoretical} problem whose solution may also improve the efficiency of computations at this order and lead to their automation, as well as provide insight into the connection between fixed-order and all-order (i.e.\ resummed or parton shower) approaches.

In this paper, we solve this problem in the context of the nested soft-collinear (NSC) subtraction scheme \cite{Caola:2017dug} by deriving the finite remainders of the integrated NNLO subtraction terms for \emph{arbitrary} collider processes with massless partons.  
Along the way, we demonstrate \emph{analytically} the cancellation of all $1/\ep$ infrared poles for infrared-safe observables in a process-independent manner. 

Achieving this requires a good understanding of the many singular limits of various scattering amplitudes, as well as the interplay of these limits, which becomes rather intricate at NNLO. 
Furthermore, the integrals over the unresolved parts of phase space of universal quantities arising in these limits, such as eikonal and splitting functions, have to be calculated.
We have studied these issues in detail in Refs~\cite{Devoto:2023rpv,Devoto:2025kin}, focusing on  final states of increasing complexity and preparing a solid foundation for addressing the NNLO subtraction problem in \emph{full generality}.

Another outstanding obstacle that one has to face when crafting subtraction schemes at NNLO is the bookkeeping. This issue is somewhat unusual, as it originates from the need to keep track of the many partonic channels that contribute to an arbitrary process. The problem stems from the fact that the cancellation of collinear singularities involves all partonic channels at once, because collinear emissions by initial-state partons may change the initial state of a hard partonic process.
A different, but somewhat analogous problem also exists for the final-state collinear splittings, since particular combinations of various limits and various final states are needed to arrive at the physical splitting functions and collinear anomalous dimensions.  
Although some aspects of this problem have already been addressed in Ref.~\cite{Devoto:2025kin}, the fully general treatment that we present in this paper goes beyond these results. 

The remainder of the paper is organized as follows. 
In Section~\ref{sec:summary}, we set the stage by summarizing the results of the earlier papers on the NSC scheme \cite{Caola:2017dug,Caola:2019nzf,Caola:2019pfz,Asteriadis:2019dte,Devoto:2023rpv,Devoto:2025kin}. The goal of this section is to make the discussion in the following sections understandable without the need to consult earlier papers. 
In Sections \ref{NLO_any_process} and \ref{sec_NNLO_general}, we discuss the calculation of NLO and NNLO QCD corrections, respectively.  
In particular, Section~\ref{sec_NNLO_general} contains the final result for the finite remainders of the integrated subtraction terms for arbitrary process. 
We conclude in Section \ref{sec:concl}, where we also summarize how to use the final results scattered throughout Section~\ref{sec_NNLO_general}.  
Several appendices contain discussions of some aspects of the problem at a more technical level.  
\section{Summary of the nested soft-collinear subtraction scheme}
\label{sec:summary}
 
Before proceeding with the derivation of the integrated subtraction terms for arbitrary processes at colliders, both at NLO and at NNLO in QCD, we summarize the  aspects of earlier work on the NSC scheme~\cite{Caola:2017dug,Caola:2019nzf,Caola:2019pfz, Asteriadis:2019dte,Devoto:2023rpv,Devoto:2025kin} that provide the foundation for the following discussion. 
Our primary intention is to explain the basic approach of this method and introduce notation that will be used throughout the paper. With this out of the way, we will be able to focus on the problems of combinatorics and bookkeeping that will arise when discussing NLO and NNLO QCD corrections to general processes in Sections~\ref{NLO_any_process} and \ref{sec_NNLO_general}. 

Consider a process where $N$ jets and a color-singlet $X$ are produced in a hadronic collision.\footnote{Throughout the paper, we discuss hadronic collisions, but our results can
easily be modified to obtain formulas valid for leptonic or lepton-hadron collisions. We explain how to do so when we present our final result in Section~\ref{sec_NNLO_general}.} The cross section for this process is written as
\begin{equation}
\begin{split}
    \dsigma & = \sum_{\inF,\inS} \int_0^1 \dx_1 \dx_2 \, f_\inF(x_1,\muF)f_\inS(x_2,\muF)  \, \dsigmahat_{\inF \inS}(x_1,x_2,\muF,\muR; 
    \mathcal{O}) \\
    & = \sum_{\inF,\inS} \, (f_\inF \conv f_\inS) \conv \dsigmahat_{\inF \inS}(x_1,x_2,\muF,\muR; 
    \mathcal{O})   \,,
    \label{eq2.1}
\end{split}
\end{equation}
where $f_{\inF,\inS}$ are the parton distribution functions (pdfs), $\muR$ and $\muF$ are the renormalization and factorization scales, respectively, and $\mathcal{O}$ is an infrared-safe observable. 
The sum in Eq.~(\ref{eq2.1}) includes all initial-state partons $a$ and $b$ that contribute to the production of a particular final state. Throughout the paper, we set $\muF=\muR=\mu$.

It is conventional to expand partonic cross sections in series in the strong coupling $\as$, 
\begin{equation}
\label{eq:series_xsection}
    \dsigmahat_{\inF\inS} = \dsigmahat_{\inF\inS}^{\LO}+\dsigmahat_{\inF\inS}^{\NLO}+ \dsigmahat_{\inF\inS}^{\NNLO} + \mydots \,,
\end{equation}
where each subsequent term in the above equation is suppressed by an additional power of $\as$ with respect to $\dsigmahat_{\inF\inS}^{\LO}$.
The leading-order (LO) term $\dsigmahat_{\inF\inS}^{\LO}$ is defined as 
\begin{equation}
\begin{split}
    2 s_{\inF\inS} \, \dsigmahat^\LO_{\inF \inS} 
    = 
    \lint
    \FLM^{\inF\inS}[\,\mydots\,] \rint
    = \calN \int \rmd\Phi \, (2\pi)^4\delta^{(4)}(p_{\HP} + p_X - p_a-p_b) \, |\mathcal{M}_0(p_\inF,p_\inS;p_\HP ,p_X)|^2 \, \mathcal{O}(p_{\HP},p_X)  \, ,
    \label{eq_FLM_LO_defn_gen}
\end{split}
\end{equation}
where $\mathcal{N}$ is the appropriate symmetry factor, $p_\HP$ and $p_X$ denote the momenta of the outgoing partons and the color-singlet in the hard process, respectively, and ${\rm d} \Phi$ is the phase space for final-state particles. We do not display the arguments of the function $\FLM$ since a convenient way to introduce them will be discussed later.  
Further details about the function $\FLM$ can be found in Section~2 of Ref.~\cite{Devoto:2023rpv}. 
\\

\figureRules{t!}{
    Illustration of  soft and collinear factorizations for real emissions and  of virtual contributions. 
    The unresolved parton $\Fp$, shown in red, is either soft or collinear.  
    The first pane illustrates the single-soft limit $E_\Fp \to 0$, as defined in \eq\eqref{eq_soft_limit_review}.  
    The second pane shows the final-state collinear splitting $[i\Fp]^* \to i(z) + \Fp(1-z)$, with $z=1-E_\Fp/E_{[i\Fp]}$, described in \eq\eqref{eq_coll_limit_FS_review}.  
    The third pane depicts the initial-state collinear splitting $\inF \to [\inF\barFp]^* + \Fp$, where $z=1-E_\Fp/E_\inF$, from \eq\eqref{eq_coll_limit_IS_review}.  
    Finally, the fourth pane represents the virtual contributions described in point vii).
}

The well-known problem with constructing the perturbative expansion of the partonic cross section in Eq.~\eqref{eq:series_xsection} is that, at each perturbative order, one must combine contributions of partonic final states with different multiplicities to achieve results which are insensitive to long-distance physics. These long-distance effects manifest themselves as infrared divergences that appear in contributions with different numbers of emissions of off-shell (in virtual loops) and on-shell (i.e.\ real) partons. These divergences cancel when their combined effect on infrared-safe observables is evaluated.  The goal of many studies performed during the past thirty years aimed at developing subtraction schemes both at NLO and NNLO ~\cite{Frixione:1995ms, Catani:1996vz, Nagy:2003qn, Bevilacqua:2013iha,
Frixione:2004is,Gehrmann-DeRidder:2005btv,Currie:2013vh,
Somogyi:2005xz,Somogyi:2006db,DelDuca:2016csb,DelDuca:2016ily,Czakon:2010td,
Czakon:2011ve,Czakon:2014oma,Anastasiou:2003gr,Caola:2017dug,Catani:2007vq,
Grazzini:2017mhc,Boughezal:2011jf,Gaunt:2015pea,Boughezal:2015dva,Sborlini:2016hat,
Herzog:2018ily,Magnea:2018hab,Magnea:2020trj,Chen:2022ktf,Bertolotti:2022aih,Capatti:2019ypt,
Braun-White:2023sgd,Braun-White:2023zwd,Fox:2023bma,Gehrmann:2023dxm,Fox:2024bfp}
was to establish a general, process- and observable-independent procedure, where the cancellation of the divergences is achieved prior to nontrivial integrations over the phase space of hard partons. 

Restricting our discussion to NLO and NNLO in the perturbative expansion, it is fair to say that the origin of infrared divergences is well understood. 
They arise from three sources: i) from the integration over loop momenta in virtual corrections, where their general form is encapsulated by the well-known formulas due to Catani~\cite{Catani:1998bh}, ii) from the emission of low-energy (\emph{soft}) gluons, and iii) from the emission of (\emph{collinear}) partons at small angles relative to other incoming or outgoing partons.
The individual singular limits, both at NLO and at NNLO, have been known for more than twenty years~\cite{Catani:2000ef, Catani:1999ss}, yet the question of how to combine them into a working subtraction scheme at NNLO continues to be the subject of active research. 
This includes our work \cite{Devoto:2023rpv,Devoto:2025kin} on the development of the nested soft-collinear subtraction scheme, introduced in Ref.~\cite{Caola:2017dug}. 

Our goal in this section is to provide the reader with a minimal background to understand the following discussion of NLO and NNLO subtraction-based calculations in this scheme. To this end, we need to explain how we identify, manipulate, and isolate singular contributions that arise from emissions of soft and collinear partons. Below we summarize the important steps required to accomplish this in the NSC scheme.

\begin{enumerate}[label=\roman*)]

    \item As the name of the subtraction scheme suggests, the singular limits of the real-emission contributions are removed \emph{sequentially}, starting with the soft ones, and  
    continuing with subtracting the collinear singularities from the soft-regulated expressions.   

    \item  To isolate singularities, we define soft and collinear operators that act on functions $\FLM$. We denote them as $S_i$, $C_{ij}$, $S_{ij}$ and $C_{ij,k}$, where the indices identify partons that become soft (in case of $S_i$ and $S_{ij}$) or collinear to each other (in case of $C_{ij}$ and $C_{ij,k}$). When these operators act on the product of the matrix element squared, the observable, and the phase space, they pick up the leading asymptotic behavior of this product in the respective limit that is non-integrable in four dimensions.\footnote{We use dimensional regularization throughout this paper, working in $d = 4 -2 \ep$ dimensional space-time.}
    Hence, if any of these operators acts on a quantity that does not possess a non-integrable singularity, the result vanishes. 

    \item  When considering processes with a large number of final-state partons, one needs to account for the fact that \emph{all}  partons can contribute to singular limits of matrix elements.
    However,  since we work at a particular order in perturbation theory, we need to ensure that the number of hard partons does not drop below the number of jets in the LO processes. Hence, the observable ${\cal O}$ that is contained in $\FLM$ vanishes if more than one parton at NLO, and more than two  partons at NNLO, are ``lost'' to various infrared limits (i.e., become unresolved). 
    
    We need to find a way to divide the final-state partons into those that can become unresolved, causing  singularities, and those that remain resolved and define physical jets. 
    To accomplish this, we introduce \emph{damping factors}. They are constructed in such a way that they vanish if a resolved parton becomes soft or collinear to \emph{another} resolved parton, and thus the integrand (which includes a damping factor) is not singular. On the contrary, if a potentially unresolved parton becomes soft, or collinear to any of the resolved partons or to another unresolved parton, the integrand remains singular. 
    
    We will refer to the unresolved partons as $\xa$ and $\yb$, and to the damping factors as $\Delta^{(\xa)}$ if only one parton is potentially-unresolved, or $\Delta^{(\xa \yb)}$ if two are unresolved. 
    We can use the symmetry of the matrix elements with respect to different types of partons (gluons, quarks, antiquarks etc.), to  minimize  the number of unresolved partons that we have to consider. 
    We then write the cross section as a sum over the contributions with  different unresolved partons, i.e.    
     \begin{equation}
        \dsigma \sim \hspace{-4mm} \sum_{\Fp \in \{q,\qb,g\}} \hspace{-4mm}  \lint \Delta^{(\xa)}  \FLMlo{ab}{\HPf|\xa} \rint \,,
        \qquad \!\!{\rm or} \qquad
        \dsigma \sim \hspace{-5mm} \sum_{\Fp,\Sp \in \{q,\qb,g\}} \hspace{-5mm} \lint \Delta^{(\xa \yb)} \FLMlo{ab}{\HPf|\xa,\yb} \rint \,,
        \label{eq2.4}
    \end{equation}
    where $\HPf$ represents the list of final-state resolved partons. We note that the damping factors are explicitly constructed in Appendix~B of  Ref.~\cite{Devoto:2023rpv}.

    \item  At NLO, soft singularities in \eq\eqref{eq2.4} appear when $\xa = g$, while at NNLO, they arise when $\xa=g$ and/or $\yb=g$, as well as when $\xa = q, \yb = \bar q$.
    In the NNLO case, when both $\xa$ and $\yb$ become soft, it is important to order them in energy, as this makes the approach of the double-soft limit unambiguous. 
    Therefore, when double-soft singularities are present, we require the energy of the parton $\yb$ to be smaller than the energy of the parton $\xa$, $E_{\yb} < E_\xa$, and introduce double- and single-soft operators $S_{\xa \yb}$ and $S_\yb$.  
    The former extracts the soft limit $E_{\xa, \yb} \to 0$ with the ratio $E_\yb/E_\xa$ fixed, while the latter extracts  the soft limit $E_\yb \to 0$ at fixed $E_\xa$. 
    
    The action of the soft operators on the function $\FLM$ can be described by compact formulas. 
    The single-soft operator acting on $\Delta^{(\Fp)} \FLMlo{}{\Fp}$ returns the phase-space element of the gluon $\xa$ and an eikonal factor, multiplied with the $\FLM$ function that does not depend on  $\xa$ anymore. 
    Integrating over the soft gluon momentum with an upper cut-off on the gluon energy $E_{\rm max}$, we find 
    \begin{equation}
        \lint S_{\Fp} \Delta^{(\Fp)} \FLMlo{\inF\inS}{\Fp}\rint 
        =
        \delta_{\Fp g} \,
        \asbr \, \lint \ISoft(\ep,E_{\rm max} ) \cdot \FLM^{\inF\inS} \rint \, .
    \label{eq_soft_limit_review}
    \end{equation}
    Here, $\asbr$ is defined as
    \begin{equation}
        \asbr = \frac{\alpha_s(\mu)}{2
        \pi} \frac{e^{\ep\gamma_\rmE}}{\Gamma(1-\ep)} \,,
    \end{equation}
   where $\as(\mu)$ is the renormalized strong coupling constant and $\gamma_\rmE$ the Euler–Mascheroni constant,
    and $\ISoft$ is an operator in color space that contains sums over color  matrices $\T_i \cdot \T_j$. Its explicit expression can be found in Eq.~(A.38) of Ref.~\cite{Devoto:2025kin}. The relation in Eq.~\eqref{eq_soft_limit_review} is shown schematically in the first pane of Fig.~\ref{fig_main_collinear_factorization_FSR_ISR}; we will use it extensively throughout this paper. We note that \eq\eqref{eq_soft_limit_review} can also be used to describe single-soft limits in NNLO contributions with the unresolved parton $\Sp$. In this case, if the energy ordering $E_{\Sp} < E_{\Fp}$ is present, one needs to replace
    $E_{\rm max}$ in the expression for $\ISoft$ with $E_\xa$.
    
    We can write a similar formula for the double-soft limits that appear at NNLO. It reads 
    \begin{equation}
    \lint S_{\xa \yb} \THmn \Delta^{(\xa \yb)}
    \FLMlo{ab}{\xa \yb} \rint 
    \sim
    \asbr^2 \lint I_{\rm DS}(\ep, \Emax) \cdot \FLM^{\inF\inS} \rint \,,
    \end{equation}
    where $\THmn = \Theta(E_\Fp - E_\Sp)$, and the $I_{\rm DS}$-operator on the right-hand side can be extracted from Ref.~\cite{Caola:2018pxp} for both $gg$ and $q\bar q$ unresolved partons.

    \item Once the soft singularities are removed, one needs to extract the hard-collinear ones. 
    These arise when an unresolved parton $\Fp$ or a pair of unresolved partons $(\Fp,\Sp)$ becomes collinear to initial-state or hard final-state partons, or they become collinear to each other. 
    
    To ensure that we can focus on a minimal subset of collinear singularities at a time, we partition the phase space by means of angular functions.
    We refer to them as $\omega^{\Fp i}$ at NLO, and $\omega^{\xa i, \yb j}$ at NNLO. 
    Their properties and definitions are reported in Appendix~B of Ref.~\cite{Devoto:2023rpv}.
    Here we only mention that $C_{\Fp j} \, \omega^{\Fp i}=\delta^{ij}$, $\forall \, i,j \in \{a, b, \HPf\}$, where $C_{\Fp j}$ is the operator which extracts the leading singular behavior in the collinear limit $\Fp \parallel j$.
    In the case of the NNLO partitions with $i=j$, i.e. $\omega^{\Fp i, \Sp i}$, we need to further divide the angular phase space into sectors in order to fully isolate the collinear divergences. A parametrization of the angular phase space that achieves this sectoring is given in Refs~\cite{Czakon:2010td,Czakon:2011ve}.

    \item At NLO, the hard-collinear divergences are extracted by acting with  the operator $\oS_{\Fp} C_{i \Fp} \equiv (\iden - S_{\Fp})\, C_{i \Fp}$ on the product of the relevant function $\FLM$, the damping factor $\Delta^{(\xa)}$ and the partition functions. Depending on whether $i$ belongs to the final state, $i \in \HPf$, or to the initial state, i.e.\ $i\in\{a, b\}$, the hard-collinear limits evaluate to 
     \begin{align}   
        & \lint  \oS_{\Fp} C_{i \Fp} \Delta^{(\Fp)} 
        \omega^{\Fp i} \FLMun{\inF\inS}{... \,, i, ...}{\Fp} \rint
        = \frac{\asbr}{\ep}  \, 
        \lint \Gamma_{[i\Fp],\fl{[i\Fp]} \to \fl{i} \fl{\Fp}} \, \FLMlo{\inF\inS}{... \,, [i\Fp], ...} \rint  \, , 
        \label{eq_coll_limit_FS_review}
        \\
        & \lint \oS_{\Fp} C_{\inF \Fp} \Delta^{(\Fp)} 
        \omega^{\Fp \inF} \, \FLMlo{\inF\inS}{\Fp} \rint 
        = \frac{\asbr}{\ep} \, \delta_{g\Fp} \,  \lint \Gamma_{a,\fl{\inF}} \FLM^{\inF\inS} \rint 
        +\frac{\asbr}{\ep} \, 
        \lint \CalPgen_{\fl{[\inF \barFp]}\fl{\inF}} \conv \FLM^{[\inF \barFp] \inS} \rint  \, ,
    \label{eq_coll_limit_IS_review}
    \end{align}
    where $[i \Fp]$ and $[a \barFp]$ are the final- and initial-state clustered partons, respectively, and $\barFp$ is the anti-particle corresponding to $\Fp$ (i.e.\ $\bar q$ for $q$, $q$ for $\bar q$ and $g$ for $g$).  
    The explicit definition of the various functions appearing in Eqs (\ref{eq_coll_limit_FS_review}, \ref{eq_coll_limit_IS_review}) can be found in Ref.~\cite{Devoto:2025kin}.
    In particular, the generalized splitting functions $\CalPgen$ are given in Eq.~(A.18), the generalized collinear anomalous dimensions $\Gamma_{i,\fl{i}}$ are reported in Eq.~(A.17), the weighted anomalous dimensions $\Gamma_{[i\Fp],\fl{[i\Fp]} \to \fl{i} \fl{\Fp}}$ are given in Eq.~(A.19), and the convolution denoted by $\otimes$ is defined in Eq.~(3.13) of that reference, and reads
    \begin{equation}
    \begin{aligned}
        & \CalPgen_{\alpha\beta} \conv \FLM  
        = 
        \int_{0}^{1} \dz \, \CalPgen_{\alpha\beta}(z) \, \frac{\FLMlo{}{z \cdot p_\inF, p_\inS; \HPf}}{z}  \,, \\
        & \FLM \conv \CalPgen_{\alpha\beta}
        = 
        \int_{0}^{1} \dz \, \CalPgen_{\alpha\beta}(z) \, \frac{\FLMlo{}{p_\inF, z \cdot p_\inS; \HPf}}{z} \,,
    \end{aligned}
    \label{eq_convolution}
    \end{equation}
    for the left and right convolutions, respectively.
    We emphasize that the order of the partons in the weighted anomalous dimension $\Gamma_{i,f_1\to f_2 f_3}$ is important, with $f_2$ being the hard parton, and $f_3$ being the potentially-unresolved one. 
    The action of the hard-collinear operators on $\FLM$ in Eqs~(\ref{eq_coll_limit_FS_review}, \ref{eq_coll_limit_IS_review}) is illustrated by the second and third panes in Fig.~\ref{fig_main_collinear_factorization_FSR_ISR}. 

    It is clear from Eqs.~\eqref{eq_coll_limit_FS_review} and \eqref{eq_coll_limit_IS_review} that there is a peculiar  difference between initial- and final-state collinear limits, in that the latter give rise to weighted anomalous dimensions, while the former lead directly to generalized anomalous dimensions. This is due to the behavior of the damping factors under the action of the collinear operators.
    Indeed, we have~\cite{Devoto:2025kin}
    \begin{equation}
        C_{a \Fp} \Delta^{(\Fp)} = 1 \; , \qquad C_{i \Fp} \Delta^{(\Fp)} = E_i/(E_i + E_\Fp) \equiv z_{i,\Fp} \; ,
    \label{eq:NNLOdampinglimits}
    \end{equation}
    where $a(i)$ is the initial-state (resolved final-state) parton. The additional factor of  $z_{i,\Fp}$ leads to the weighted anomalous dimensions when one integrates over energies in the case of final-state collinear limits.   
    However, we showed in Ref.~\cite{Devoto:2025kin} that collinear splittings arising from \emph{different} potentially-unresolved partons can be combined to obtain generalized anomalous dimensions, and that it is advantageous to do so \emph{before} integrating over partonic energies. 
    In particular, to obtain the generalized anomalous dimension for a quark, we need to combine cases where $\Fp = q$ becomes collinear to a hard gluon with those where $\Fp = g$ becomes collinear to a hard quark. 
    In fact, in the combination 
    \begin{equation}
        \Gamma_{i,q}= \Gamma_{i, q \to qg}+ \Gamma_{i,q \to gq} \,,
    \label{eq_Gamma_q_FSR_def}
    \end{equation}  
    which appears \emph{naturally} in our set-up, the weight factors $z_{i, \Fp}$ disappear, leading to a standard quark collinear anomalous dimensions, related to the  integral of a splitting function. 
    Similarly, by accounting for $g \to gg$ and $g \to q\barq$ splitting, we obtain
    \begin{equation}
        \Gamma_{i, g} = \Gamma_{i, g \to gg} + 2\nf\, \Gamma_{i, g\to q \barq} \,,
    \label{eq_Gamma_g_FSR_def}
    \end{equation}
    which is directly related to the collinear anomalous dimension of a gluon. 
    
    At NNLO one has to consider the joint action of two soft-subtracted collinear operators $C_{i \Fp} C_{j \Sp}$ which, depending on the partition function, may either be applied  to different hard legs ($i \ne j$) or to the same leg ($i = j$). 
    The treatment of such combinations of collinear limits is complicated for two reasons. First, when the limits are applied to the same resolved parton, the phase spaces for partons $\xa$ and $\yb$ become intertwined, and care is needed in order to extract the relevant (generalized or weighted) anomalous dimensions~\cite{Devoto:2023rpv}.  
    Second, one needs to properly account for the many possible types of clustered partons that may appear in those cases; this is one of the problems that we discuss in detail in this paper. This issue is particularly important for the reconstruction of the generalized anomalous dimensions, discussed above. Indeed, the different splittings that lead to various weighted anomalous dimensions of the form $\Gamma_{i,f_1 \to f_2 \; f_3}$ can only be combined into generalized anomalous dimensions, as shown in Eqs~(\ref{eq_Gamma_q_FSR_def}, \ref{eq_Gamma_g_FSR_def}), if they multiply the same $\FLM$ function. Since one starts with $\FLM$ functions with different partonic configuration as arguments, showing that they become the same under the action of collinear operators is essential. 
    While it is relatively straightforward to demonstrate this at NLO, it becomes highly nontrivial to do so  when  dealing with arbitrary processes at NNLO. 
    Finally, once generalized anomalous dimensions are extracted, we combine them into the NLO collinear operator
    \begin{equation}
        \IColl(\ep) = \sum_{i} \frac{\Gamma_{i,\fl{i}}}{\ep} \, ,
    \label{eq_IColl_definition_review}
    \end{equation}
    or its NNLO counterparts $\IColl^2(\ep)$ and $\IColl(2\ep)$. 
    
    In addition to generalized collinear anomalous dimensions, hard-collinear configurations  also give rise to {\it boosted} contributions, see \eq\eqref{eq_coll_limit_IS_review}. This happens for the initial-state splittings, since in such cases the energy flowing into  the hard processes is rescaled by a factor $z = 1- E_{\Fp}/E_a$. When $z=1$, the integrated hard-collinear subtraction term corresponds to the generalized anomalous dimension functions, $\Gamma_{i,f_i}$, $i \in (a,b)$. 
    However, if $z < 1$, which implies  the emission of an unresolved parton with non-vanishing energy, a hard-collinear limit leads to a convolution denoted by the symbol $\conv$ in Eq.~\eqref{eq_coll_limit_IS_review}. Such structures do not occur in the case of final-state splittings, as the energy of the underlying hard process remains unchanged. 

    \item In addition to the real-emission contributions, one has to consider one- and two-loop virtual corrections, one-loop virtual corrections to single-parton emissions, and the renormalization of parton distribution functions. 
    The divergences of virtual corrections can be written as  operators in color space acting on LO  matrix elements squared. 
    For example, at NLO  we refer to such  operators as  $\IVirt(\ep)$, see the fourth pane in Fig.~\ref{fig_main_collinear_factorization_FSR_ISR}. 
    We note that $\IVirt$ is defined in Eq.~(A.36) of Ref.~\cite{Devoto:2025kin} and is closely related to Catani's operator $I_1(\ep)$ introduced in Ref.~\cite{Catani:1998bh}.     
    
    At NNLO, the singular contributions of the double-virtual corrections through $\order{\ep^{-2}}$     can be written in terms of $\IVirt(\ep)$ and $\IVirt(2\ep)$, and the commutator $[I_1(\ep), I_1^\dagger(\ep)]$ (see Section 4.3 in Ref.~\cite{Devoto:2023rpv}). 
    The collinear renormalization of parton distribution functions at NLO leads to the convolution of tree-level Altarelli-Parisi (AP) splitting functions, $\PAP_{ij}$, and Born-level matrix elements. 
    At NNLO, further contributions appear, for example the convolution of the one-loop AP splitting functions, $\PAPone_{ij}$, with LO matrix elements squared, and convolutions of $\PAP_{ij}$ with NLO partonic cross sections.

    \item At NNLO, we also need to account for triple-collinear singular limits. Due  to the iterative nature of the NSC subtraction scheme, we require such limits with all single-collinear and soft divergences removed. After integrating over the unresolved phase space, these terms contribute at $1/\ep$ and are given in Ref.~\cite{Delto:2019asp}. We note that the results of this reference need to be modified slightly for our purposes; we discuss this in Section~\ref{sec:TC}.

    \item For the processes that we considered in Refs~\cite{Devoto:2023rpv,Devoto:2025kin}, we were able to demonstrate the cancellation of $1/\ep$ poles analytically and to derive finite remainders. 
    As noted in Ref.~\cite{Devoto:2023rpv}, to achieve this  it is useful to separately consider contributions with different final-state kinematics (double-boosted, single-boosted, unboosted), as well as other distinguishing features   of the $\FLM$ functions (color-correlated pieces,  spin-correlated pieces, etc.) to identify subsets of integrated subtraction terms where the cancellation of divergences occurs independently.  
    In this  paper we will show that such a procedure is sufficiently flexible and can be used for an analysis of arbitrary processes.
    We note that 
    \begin{itemize}[label=--]
        \item at NLO, the  $1/\ep$ poles proportional to the boosted matrix elements do not involve color-correlated matrix elements,  and have to cancel among themselves. As we will see in \Sec\ref{NLO_any_process}, this is achieved upon combining the terms from the initial-state hard-collinear limits with those from the pdf renormalization,  using the relation $\CalPgen_{\alpha \beta} = - \PAP_{\alpha \beta} + \order{\ep}$ between the generalized splitting functions and the Altarelli-Parisi collinear splitting kernels. 
        At NNLO the combination of pdf renormalization and hard-collinear limits has to be supplemented by the real-virtual integrated subtraction terms.
        This leads to the appearance of divergent boosted terms that are also color-correlated. 
    
        \item at NLO, the  $1/\ep$ poles  proportional to the unboosted LO matrix elements feature both color-correlated and color-uncorrelated contributions. The former cancel in the combination $\IVirt(\ep) + \ISoft(\ep)$, which  however still contains color-uncorrelated $1/\ep$ divergences. These divergences cancel upon accounting for the collinear contribution  $\IColl(\ep)$, defined in \eq\eqref{eq_IColl_definition_review}.
        We then introduce  an operator 
        \begin{equation}
            \ITot(\ep) = \IVirt(\ep) + \ISoft(\ep) + \IColl(\ep)\, ,
        \label{ITot_defn_review}
        \end{equation}
        which has a finite $\ep \to 0$ limit. Its first non-vanishing contribution in the $\ep$-expansion, $\ITot^{(0)}$, contains color-correlated terms proportional to $\T_i \cdot \T_j$ (see Appendix~A in Ref.~\cite{Devoto:2025kin}).
        It turns out that many NNLO singularities are captured by the operators $\ISoft$, $\IColl$ and $\IVirt$ or their iterations, and that frequently they can be combined into iterations of the $\ITot$ operator.
        Identifying such structures early on in the calculation substantially streamlines both the cancellation of the $\ep$-poles, and the derivation of the finite remainders at NNLO.
    \end{itemize}
\end{enumerate}
\section{NLO QCD corrections to a general process at a hadron collider}
\label{NLO_any_process}

In this section, we discuss the computation of next-to-leading order QCD corrections to the process $pp \to \colsing + N~\mathrm{jets}$,  where $X$ is an arbitrary 
color-singlet state. At leading order, such a process is obtained from \eq\eqref{eq_FLM_LO_defn_gen}, where the partonic cross sections can be written as 
\begin{equation}
    2s_{ab} \,  {\rm d}\sigma_{ab}^{\rm LO}
    = \sum \limits_{n} 
    \lint \FLMlo{\inF\inS}{\setf{N}{n}{ }} \rint \,.
    \label{eq_LO_cross_section}
\end{equation}
Here, $\setf{N}{n}{}$ denotes a particular final state with $N$ QCD partons, each associated with an identified jet, that can be produced together with the color-singlet $X$ in the collision of partons $a$ and $b$.   
The index $n$ enumerates \emph{all} QCD final states which may contribute to the partonic process, including all combinations of flavors consistent with the initial state $(\inF,\inS)$ and the color-singlet $X$ in the final state.  
In what follows, we assume that such final states have been enumerated for arbitrary jet multiplicity $N$.\footnote{We demonstrate how this can be achieved in a toy model with one quark flavor in Appendix~\ref{appendix_A}.}
We also note that $\FLM$ contains all the symmetry factors associated with a particular final state $\Born_{N,n}$.

At NLO, several contributions are required. 
We will focus on the analysis of a real-emission process with initial state $\inF$ and $\inS$, which corresponds to the LO partonic process $ab \to (N+1)~{\rm partons} + X$. 
The cross section reads 
\begin{equation}
    2s_{\inF\inS} \, \dsigmahat_{\inF\inS}^{\R}
    = \sum_{n} \lint \FLMlo{\inF\inS}{\setf{N+1}{n}{ }} \rint \,.
\label{eq_dsigma_NLO_ab_genX}
\end{equation}
The sum appearing on the right-hand side of \eq\eqref{eq_dsigma_NLO_ab_genX} has the same meaning as in \eq\eqref{eq_LO_cross_section}.
Proceeding as in Refs~\cite{Devoto:2023rpv,Devoto:2025kin}, we insert a partition of unity for each term in the sum in \eq\eqref{eq_dsigma_NLO_ab_genX}
\begin{equation}
    \sum_{i \in \setf{N+1}{n}{}} \hspace{-3mm} \Delta^{(i)} = 1 \,,
\label{eq_damping_factors_genX}
\end{equation}
where the sum over the index $i$ runs over all final-state partons in the list $\setf{N+1}{n}{}$. As mentioned in \Sec\ref{sec:summary}, each damping factor $\Delta^{(i)}$ vanishes if any parton other than parton $i$ becomes unresolved. For convenience, we relabel the partons entering $\setf{N+1}{n}{}$ in such a way that the potentially-unresolved parton $i$ is always identified by $\Fp$.  Furthermore, we use the symmetry of $\FLM$ to write $\dsigmahat_{\inF\inS}^{\R}$ in terms of three contributions, distinguishing the cases where $\Fp$ is a gluon from those where $\Fp$ is a quark or an antiquark.
We find 
\begin{equation}
\begin{split} 
    2s_{\inF\inS} \, \dsigmahat_{\inF\inS}^{\R}
   &  = \sum_n \llint \Delta^{(\Fp)}  \FLMlo{\inF \inS}{\setf{N+1}{n}{}(\Fp_g)} \rrint
   + \sum_n \sum \limits_{\fgflin=1}^{\nf} 
   \llint \Delta^{(\xa)}\FLMlo{\inF \inS}{\setf{N+1}{n}{}(\Fp_{q_\fgflin})} \rrint 
   \\
   &  +\sum_n \sum \limits_{\fgflin=1}^{\nf} 
   \llint \Delta^{(\xa)} \FLMlo{\inF \inS}{\setf{N+1}{n}{}(\Fp_{\qb_\fgflin})} \rrint \,,
\end{split}
\label{eq_dsigma_qqb_gg_NLO_def_genX}
\end{equation}
where we have made explicit the sum over $\nf$ quark flavors.
The notation $\setf{N+1}{n}{}(\Fp)$, introduced in the above equation, indicates a list of $N+1$ partons where parton $\Fp$ has been \emph{identified} as potentially-unresolved. 
We use the convention that the symmetry factors in $\FLMlo{\inF \inS}{\setf{N+1}{n}{}(\Fp)}$ are determined by all final-state partons \emph{except} the marked one, $\Fp$. 
We also note that to identify a parton of a particular type as potentially-unresolved, ${\cal B}_{N+1,n}$ must contain at least one such parton. This trivial remark implies that the sum over $n$ in the first term on the \rhs\ of \eq\eqref{eq_dsigma_qqb_gg_NLO_def_genX} runs over all QCD final states that contain at least one gluon and can be produced in collisions of partons $a+b$ together with $X$. Analogously, the sum over $n$ in the second term runs over all possible final states with at least one quark of flavor $\fgflin$, and the same applies to the third term with respect to the antiquark $\qb_\fgflin$.  
Therefore, although we always sum over same index $n$ to lighten the notation,  the three sums in  \eq\eqref{eq_dsigma_qqb_gg_NLO_def_genX} run over different final states.  

We can now apply the subtraction procedure introduced in Ref.~\cite{Caola:2017dug} and outlined in Section~\ref{sec:summary} to $\dsigmahat_{\inF\inS}^{\R}$, by multiplying each contribution in \eq\eqref{eq_dsigma_qqb_gg_NLO_def_genX} with an identity operator written in the following way
\begin{equation}
    \iden = S_\Fp + \sum_{i\in\HP} \oS_\Fp C_{i \Fp} + \ONLO^{(\Fp)} \,,
\label{eq_iden_S_oS_oC_relation}
\end{equation}
where $\ONLO^{(\Fp)}$ is defined as
\begin{equation}
    \ONLO^{(\Fp)} = \sum_{i\in\HP} \oS_{\Fp} \oC_{i \Fp} \, \partFuncNLOfp{i} \,,
\label{eq_ONLO_defn}
\end{equation}
and depends on the partition functions $\partFuncNLOfp{i}$ introduced in Section \ref{sec:summary}. 
The sums are taken over sets $\HP$, which include both the initial- and final-state partons specified by the arguments of the $\FLM$ functions upon which the above operators act, excluding the parton $\Fp$.  
 
We need to understand what happens when the operators in Eq.~\eqref{eq_iden_S_oS_oC_relation} act on the function $\FLM$ and, in particular, how a list ${\cal B}_{N+1,n}(\xa)$ changes once the parton $\xa$ becomes soft or collinear to another parton. 
We begin by considering a potentially-unresolved gluon, which can be emitted by either of the initial-state partons, or by any of the resolved final-state partons,  without changing the identity of the emitter.
It follows that the action of the soft operator $S_\Fp$ is described by the formula (cf.~\eq\eqref{eq_soft_limit_review})
\begin{equation}
    \sum_n  \llint S_\Fp \, \Delta^{(\Fp)}  \FLMlo{\inF \inS}{\setf{N+1}{n}{}(\Fp_g)} \rrint
    = \sum_{\np} \asbr \lint \ISoft(\ep) \colorprod \FLMlo{\inF\inS}{\setf{N}{n'}{}} \rint
    \, ,
    \label{eq_mg_genX_2}
\end{equation}
where the sum on the right-hand side extends over \emph{all} sets with $N$ hard partons that can be produced in the process $\inF\inS \to N~{\rm jets} + X$.

We now move on to the hard-collinear limit $\oS_{\Fp_g} C_{i \Fp_g}$. 
As we already mentioned, the potentially-unresolved gluon can be emitted from, or clustered with, any hard parton without changing the hard parton's identity. Therefore, under the action of the hard-collinear operator, the list ${\setf{N+1}{n}{}}(\xa_g)$ becomes a list composed of resolved partons taken from the same list.
Accordingly, no changes in the $\FLM$ symmetry factors occur and, in analogy to Eqs.~(\ref{eq_coll_limit_FS_review}, \ref{eq_coll_limit_IS_review}), we obtain 
\begin{equation}
\begin{split}
    \sum_n  
    \llint 
    \oS_\Fp C_{a\Fp} \Delta^{(\Fp)}  \FLMlo{\inF \inS}{\setf{N+1}{n}{}(\Fp_g)} \rrint
    = \sum_{n'} \bigg[
    \frac{\asbr}{\ep} \,   \lint \Gamma_{a,\fl{\inF}} \FLMlo{\inF\inS}{\setf{N}{n'}{}} \rint 
    + \frac{\asbr}{\ep} \, 
    \lint \CalPgen_{\inF \inF} \conv \FLMlo{\inF \inS}{\setf{N}{n'}{}} \rint 
    \bigg],
\end{split}
\label{eq_nf_1_collinear_limits_general_rules_first_2}
\end{equation}
for the initial-state radiation and 
\begin{equation}
\begin{split}
    \sum_n  
    \! \sum_{\substack{i\in \setf{N+1}{n}{}(\Fp_g) }} \hspace{-2mm}
    \lint \oS_\Fp C_{i\Fp} \Delta^{(\Fp)}  \, \FLMlo{\inF \inS}{\setf{N+1}{n}{}(\Fp_g)} \rint
    = \sum_{n'} \!\! \sum_{i\in \setf{N}{n'}{}} \!\!\! \asbr \, \Lint \frac{\Gamma_{i,f_i\to f_i g}}{\ep} \, \FLMlo{\inF\inS}{\setf{N}{n'}{}} \Rint \,,
    \label{eq_nf_1_collinear_limits_general_rules_first}
\end{split}
\end{equation}
for the final-state radiation. 
The sum over $i$ on the right-hand side of  Eq.~(\ref{eq_nf_1_collinear_limits_general_rules_first}) runs over all  final-state partons in the configuration $\setf{N}{\np}{}$, while the sums over $n'$ on the right-hand sides of Eqs.~\eqref{eq_nf_1_collinear_limits_general_rules_first_2} and~\eqref{eq_nf_1_collinear_limits_general_rules_first}  indicate that all partonic channels consistent with the final-state color-singlet $X$ and the initial state given by the associated $\FLM$ function have to be included. 

We now turn to the case where the potentially-unresolved parton is a quark of flavor $\rho$, $\Fp_{q_\fgflin}$.
Then, infrared singularities arise if $\Fp_{q_\fgflin}$ becomes collinear to an initial-state gluon, or to an initial-state quark of the same flavor, $q_\fgflin$, or if it becomes collinear to a final-state gluon or an antiquark of the same flavor, $\qb_\fgflin$. 
Therefore, we need to understand what happens to the partonic final state ${\setf{N+1}{n}{}}(\Fp_{q_\fgflin})$ in these limits.

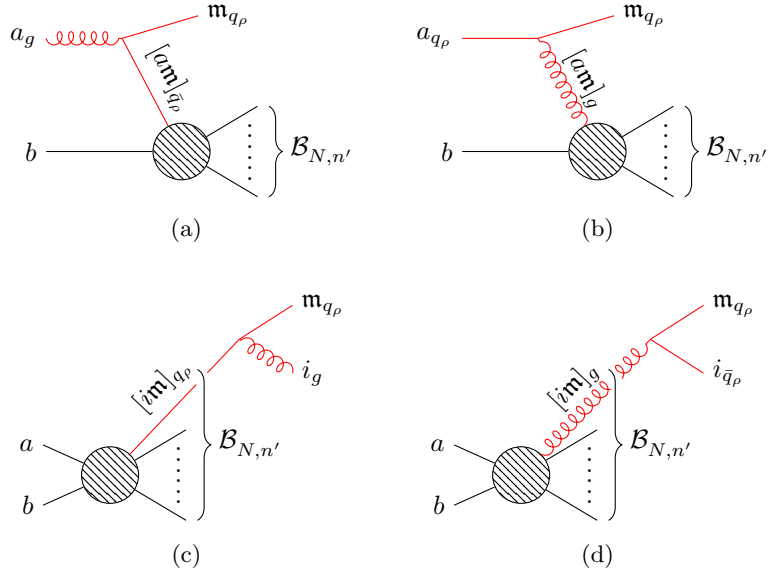
\figureNLOvsLO{ht}{Examples of  contributions to initial- and final-state collinear limits.}

Suppose $\Fp_{q_\fgflin}$ becomes collinear to an initial state gluon, which we identify with parton $\inF$ for concreteness. 
The singular contribution arises from diagrams where the initial-state gluon splits into a $q_\fgflin \qb_\fgflin$ pair, with the $\qb_\fgflin$ entering the scattering process and producing color-singlet $X$ as well as all partons in the list $\setf{N+1}{n}{}(\Fp_{q_\fgflin})$  other than $q_\fgflin$. This is shown in Fig.~\ref{figureNLOvsLO_a}. In this collinear limit, the resolved final-state partons are unchanged, while the initial state of the hard process changes to $\qb_\fgflin \inS$. 
Thus, in this limit, summing over the contributions of \emph{all} processes with $(a,b)$ in the initial state and final states denoted by $\setf{N+1}{n}{}(\Fp_{q_\fgflin})$ is equivalent to summing  over \emph{all} processes  with the initial state $\qb_\fgflin \inS$ and $N$ partons in the final state.  
We will refer to final states in such processes as ${\setf{N}{n'}{}}$. 
Thus, we can write 
\begin{equation}
    \sum \limits_{n}  \delta_{ag} \lint C_{\inF\Fp} \Delta^{(\Fp)} \FLMlo{\inF \inS}{\setf{N+1}{n}{}(\Fp_{q_\fgflin})}  \rrint 
    = \sum \limits_{\np}  \frac{\asbr}{\ep}  \lint \CalPgen_{\bar q g} \conv \FLMlo{\bar q_\fgflin\inS}{\setf{N}{\np}{}} \rint  \, .
\label{eq_limit_aqj_genX}
\end{equation}
We note that, since the symmetry factor of a function $\FLM$ is determined solely by the ``unmarked'' (i.e., resolved) final-state partons (thus excluding $\Fp$), and since a collinear limit with the initial state always leaves the list of hard final-state partons unchanged, the symmetry factors on the \lhs\ and the \rhs\ of \eq\eqref{eq_limit_aqj_genX} are identical.  
This statement holds in general, independently of the flavors of $\inF$ and $\Fp$.    
We note that the spin degrees of freedom as well as the color factors in the initial state do change; these changes are absorbed in the definition of $\CalPgen_{\bar q g}$.

The argument used to obtain \eq\eqref{eq_limit_aqj_genX} can be repeated verbatim if we consider the other singular initial-state collinear limit, which occurs when $\inF = q_\fgflin$. This contribution arises from diagrams where the initial state $q_\fgflin$ splits into a final-state quark $\Fp_{q_\fgflin}$ and a gluon that becomes an initial-state parton for the hard process, 
see Fig.~\ref{figureNLOvsLO_b}.  
Exploiting the (by now clear) connection between final states with $N+1$ and $N$ partons, we write the hard-collinear initial state term for an unresolved quark as  
\begin{equation}
  \sum_n  \lint C_{\inF\Fp} \Delta^{(\Fp)} \FLMlo{\inF \inS}{\setf{N+1}{n}{}(\Fp_{q_\fgflin})} \rrint 
    = 
    \sum \limits_{\np} \delta_{ag}  \frac{\asbr}{\ep}  \lint \CalPgen_{\bar q g} \conv \FLMlo{\bar q_\fgflin \inS}{\setf{N}{\np}{ }} \rint  
    + 
    \sum_{\np} \delta_{a q_\fgflin} \frac{\asbr}{\ep}  \lint \CalPgen_{g q} \conv \FLMlo{g \inS}{\setf{N}{\np}{}} \rint 
    \,,
\label{eq_NLO_IS_qj_genX}
\end{equation}
where, again, the summation goes over all final states with $N$ partons and a modified initial state.

This argument can be extended to final-state hard-collinear limits, described by operators $\oS_{\Fp} C_{i \Fp_{q_\fgflin}}$,  in a fairly simple way.
These limits are singular if the parton $i$ is either a gluon or an antiquark $\qb_\fgflin$.
In the former case, singularities reside in the Feynman diagrams where the quark line $q_\fgflin$ radiates the gluon $i$, with all other particles emerging separately (see Fig.~\ref{figureNLOvsLO_c}). We write 
\begin{equation}   
    \sum_n \! \sum_{i\in \setf{N+1}{n}{}(\Fp_{q_\fgflin})} \hspace{-2mm} \delta_{ig} \lint \oS_{\Fp} C_{i \Fp} \Delta^{(\Fp)} \FLMlo{\inF \inS}{\setf{N+1}{n}{}(\Fp_{q_\fgflin})} \rint
    = \sum_{\np} \sum_{i \in {\cal B}^{}_{N,\np}} \!\!\! \delta_{i q_\fgflin} \frac{\asbr}{\ep} \lint \Gamma_{i,q \to g q} \FLMlo{\inF \inS}{\setf{N}{\np}{}} \rint  \, , 
\label{eq_NLO_FS_qjg_genX}
\end{equation}
where the sum on the \rhs\ runs over all processes with $N$ final-state partons produced in the collisions of initial-state partons $a,b$.

An important aspect of the final-state collinear limits that needs to be understood to ensure the validity of \eq\eqref{eq_NLO_FS_qjg_genX} is the symmetry factors.   
To this end, consider a term on the left-hand side of \eq\eqref{eq_NLO_FS_qjg_genX}
that describes a process with the final state $\setf{N+1}{n}{}(\Fp_{q_\fgflin})$.  
We assume that it contains $\Ng$ gluons, $N_{q_\fgflin}$ quarks of flavor $\fgflin$ (not including 
$\Fp_{q_\fgflin}$ into the quark count), and other partons that are not important for our purposes.   
The symmetry factor included in $\FLMlo{\inF \inS}{\setf{N+1}{n}{}(\Fp_{q_\fgflin})}$ is $1/(\Ng! \times N_{q_\fgflin}! \times \mydots )$, where the ellipses stand for contributions to the symmetry factor from other final-state partons.  
In the limit $i_g \parallel \Fp_{q_\fgflin}$, these two partons are removed from the list $\setf{N+1}{n}{}(\Fp_{q_\fgflin})$ and replaced by the clustered parton $[i\Fp]_{q_\fgflin}$.  
This results in a final state with $N_g-1$ gluons, $N_{q_\fgflin}+1$ quarks of flavor $\rho$ and everything else unchanged.  
Denoting the corresponding final state as $\setf{N}{n'}{}$, the symmetry factor associated with the function $\FLMlo{\inF \inS}{\setf{N}{n'}{}}$ on the right-hand side of Eq.~(\ref{eq_NLO_FS_qjg_genX}) is $1/((\Ng-1)! \times (N_{q_\fgflin}+1)! \times \mydots)$. This apparent mismatch is easy to understand. 
Indeed, the sum over $i$ on the left-hand side of Eq.~(\ref{eq_NLO_FS_qjg_genX}) gives a factor $N_g$ in the numerator, owing to the symmetry of $\FLM$ under permutations of the gluons. For the same reason, the sum over quarks on the right-hand side of \eq\eqref{eq_NLO_FS_qjg_genX} compensates for the factor $(N_{q_\rho}+1)$ in the denominator.

The reasoning used to obtain \eq\eqref{eq_NLO_FS_qjg_genX} can be applied to the other final-state limit, where the singularities arise from a gluon splitting into a $q_\fgflin \qb_\fgflin$ pair, displayed in Fig.~\ref{figureNLOvsLO_d}. 
Combining these, we obtain
\begin{equation}
\begin{split}
    \sum_n \! \sum_{\substack{i\in \setf{N+1}{n}{}(\Fp_{q_\fgflin}) }} \hspace{-2mm} 
    \lint  \oS_{\Fp} C_{i \Fp} \Delta^{(\Fp)} \FLMlo{\inF \inS}{\setf{N+1}{n}{}(\Fp_{q_\fgflin})}  \rint
    = \sum_{n'} \sum_{i \in \setf{N}{\np}{}} \hspace{-2mm} \frac{\asbr}{\ep} \lint  \left (\delta_{ig} \Gamma_{i, g \to q \bar q} + \delta_{i  q_\fgflin} \Gamma_{i,q \to g q} \right ) \FLMlo{\inF \inS}{\setf{N}{\np}{}}\rint  \, .
\end{split}
\label{eq_NLO_FS_qj_genX}
\end{equation}
Thus we see that the summation over the relevant underlying Born processes with $N$ partons emerges quite naturally, allowing us to combine \eq\eqref{eq_NLO_FS_qj_genX} with the weighted anomalous dimensions present in \eq\eqref{eq_nf_1_collinear_limits_general_rules_first}. 
We can repeat this argument for the last term in \eq\eqref{eq_dsigma_qqb_gg_NLO_def_genX} where the potentially-unresolved parton is an antiquark $\Fp_{\qb_\fgflin}$.
Combining these contributions, we reconstruct the collinear anomalous dimensions for all hard partons and hence the $\IColl$ operator, introduced in Section~\ref{sec:summary}.
Putting everything together, we obtain\footnote{We emphasize one more time that, for the sake of compactness, we use one label $n$ to describe sums over all possible final states consistent with a particular partonic initial state. However, in reality these sums can be quite different, including their summation ranges.}
\begin{equation}
\begin{split}
    &\; 2s_{\inF\inS} \, \dsigmahat_{\inF\inS}^{\R}
    = \sum_n \Lint \ONLO^{(\Fp)} \Delta^{(\Fp)} \bigg[ 
    \FLMlo{\inF\inS}{\setf{N+1}{n}{}(\Fp_g)}
    + \sum_{\fgflin = 1}^{\nf} \Big[ \FLMlo{\inF\inS}{\setf{N+1}{n}{}(\Fp_{q_\fgflin})}  
    + \FLMlo{\inF\inS}{\setf{N+1}{n}{}(\Fp_{\qb_\fgflin})} \Big] \bigg] \Rint \\
    & + \sum_n \, \asbr \lint \big[\ISoft(\ep) + \IColl(\ep)\big] \colorprod
    \FLMlo{\inF\inS}{\setf{N}{n}{}} \rint
    + \frac{\asbr}{\ep} \sum_{x} \sum_n \llint
   \CalPgen_{x\inF} \conv \FLMlo{x\inS}{\setf{N}{n}{}}
   + \FLMlo{\inF x}{\setf{N}{n}{}}
   \conv \CalPgen_{x\inS} \rrint  \,.
\end{split}
\label{eq_dsigma_NLO_real_final}
\end{equation}

The treatment of the lists of LO and NLO partonic configurations that we have presented is necessarily quite abstract, as any enumeration of such lists is process-specific. Nevertheless, we can explicitly construct all lists of allowed partonic processes if we limit ourselves to the case of a single quark flavor $\nf=1$, and therefore chargeless color-singlet states $X$. Such a construction is described in Appendix~\ref{appendix_A}, which is useful to understand the details and subtleties of our approach.

For an NLO computation, the real-emission cross section in \eq\eqref{eq_dsigma_NLO_real_final} has to be supplemented with the virtual corrections to the LO cross section, and the contribution from the collinear renormalization of pdfs. 
We write them as 
\begin{align}
    & 2s_{\inF\inS} \, \dsigmahat_{\inF\inS}^\V
    =  \sum_n \Big[\asbr \lint \IVirt(\ep) \colorprod \FLMlo{\inF\inS}{\setf{N}{n}{}} \rint
    + \lint \FLVfinlo{\inF\inS}{\setf{N}{n}{}} \rint \Big] \,,
    \label{eq_dsigmahat_V_ab_nf_1} \\
    & 2s_{\inF\inS} \, \dsigmahat_{\inF\inS}^\pdf
    = \frac{\alpha_{\rm s}(\mu)}{2\pi} \frac{1}{\ep} \, \sum_x \sum_n \llint \PAP_{x\inF} \conv \FLMlo{x\inS}{\setf{N}{n}{}}
    + \FLMlo{\inF x}{\setf{N}{n}{}} \conv \PAP_{x\inS} \rrint \,,
\label{eq_pdf_renorm_nlo}
\end{align}
where $F_{\rm LV, fin}$ is the $\ep$-finite remainder of the one-loop amplitude.
Combining Eqs~(\ref{eq_dsigma_NLO_real_final}, \ref{eq_dsigmahat_V_ab_nf_1}, \ref{eq_pdf_renorm_nlo}), we find the \emph{finite} NLO partonic cross section  
\begin{equation}
\begin{split}
    &\; 2s_{\inF\inS} \, \dsigmahat_{\inF\inS}^\NLO
    = \sum_n \Lint \ONLO^{(\Fp)} \Delta^{(\Fp)} \bigg[ 
    \FLMlo{\inF\inS}{\setf{N+1}{n}{}(\Fp_g)}
    + \sum_{\fgflin = 1}^{\nf} \Big[ \FLMlo{\inF\inS}{\setf{N+1}{n}{}(\Fp_{q_\fgflin})}  
    + \FLMlo{\inF\inS}{\setf{N+1}{n}{}(\Fp_{\qb_\fgflin})} \Big] \bigg] \Rint \\
    & + \sum_n \Big[\asbr \lint \ITot^{(0)} \colorprod
    \FLMlo{\inF\inS}{\setf{N}{n}{}} \rint
    + \lint \FLVfinlo{\inF\inS}{\setf{N}{n}{}} \rint \Big]
    + \sum_{x} \sum_n \, \asbr \llint
   \PNLO_{x\inF} \conv \FLMlo{x\inS}{\setf{N}{n}{}}
   + \FLMlo{\inF x}{\setf{N}{n}{}}
   \conv \PNLO_{x\inS} \rrint \,,
\end{split}
\label{eq_dsigma_NLO_real_final_genX}
\end{equation}
where $\ITot^{(0)}$ is the $\order{\ep^0}$ term of $\ITot(\ep)$ defined in \eq\eqref{ITot_defn_review}; its explicit expression is given in \eq(A.45) of Ref.~\cite{Devoto:2025kin}. 
The functions $\PNLO_{\alpha \beta}$, which are defined in Table \ref{table_final_result}, arise from the combination of generalized splitting functions $\CalPgen_{\alpha \beta}$ and the Altarelli-Parisi splitting kernels, thanks to the following relation
\begin{equation}
    \CalPgen_{\alpha \beta}(z,E_i) \,+ \PAP_{\alpha \beta}(z) = \ep \, \PNLO_{\alpha \beta}(z,E_i) + \order{\ep^2} \,.
\label{eq_Pgen_PAP_PNLO_relation}
\end{equation} 
The energy arguments of the functions $\CalPgen_{\alpha \beta}$ and $\PNLO_{\alpha \beta}$ should be taken to be $E_a$ for left  convolutions (as in the second-last term of \eq\eqref{eq_dsigma_NLO_real_final_genX}) and $E_b$ for right convolutions (as in the final term of \eq\eqref{eq_dsigma_NLO_real_final_genX}).
This convention applies to all the splitting functions that are used in this paper. 
We conclude by noting that the hadronic cross section at NLO at scales $\muR = \muF = \mu$ is obtained by convoluting the partonic cross sections with parton distribution functions 
\begin{equation}
    \dsigma^\NLO = \sum_{\inF,\inS} \, (f_\inF \conv f_\inS) \conv \dsigmahat_{\inF\inS}^\NLO \,.
\label{eq_dsigma_NLO_real_final_genX_had}
\end{equation}
\section{NNLO QCD corrections to an arbitrary process at  colliders}
\label{sec_NNLO_general}

The goal of this section is to present formulas for the NNLO QCD corrections to $pp \to \colsing + N~\mathrm{jets}$ and $\ell^+\ell^- \to X+N~{\rm jets}$, where $X$ is an arbitrary color-singlet state.   
The underlying ideas behind these results closely follow Refs~\cite{Devoto:2023rpv,Devoto:2025kin}; the novelty is that here we deal with arbitrary initial and final states. In Section~\ref{NLO_any_process}, we have explained how this aspect of the problem is addressed, using NLO as an example. The NNLO case is obviously more complex and requires more attention.

We begin in Section~\ref{subsec_NNLO_nf_formulas} with a brief discussion of the general framework, followed by the presentation of the  final results  in Eqs.~(\ref{eq_final_result_nf} -- \ref{eq_Iuncfin_def}). 
Then, in Section~\ref{ssec_NNLO_important_aspect}, we discuss details of the calculation that we found challenging when extending the results of Refs~\cite{Devoto:2023rpv,Devoto:2025kin} to general processes.


\subsection{General setup and the final formula}
\label{subsec_NNLO_nf_formulas}

To compute the NNLO corrections to the production of $N$ jets and a color-singlet $X$ in hadron collisions, three contributions need to be considered -- the double-virtual, the real-virtual, and the double-real.  
We begin with the double-real contribution and write the partonic cross section as (cf.~\eq\eqref{eq_dsigma_NLO_ab_genX})
\begin{equation}
    2s_{\inF\inS} \, \dsigmahat^\RR_{\inF\inS}
    = \sum_{n} \lint \FLMlo{\inF\inS}{\setf{N+2}{n}{}} \rint \,.
\label{eq_dsigma_NNLO_partonic_nf}
\end{equation}
As in the NLO case, the index $n$ parametrizes a particular final state with $N+2$ QCD partons that can be produced in the collision of partons $(\inF,\inS)$ in association with $X$, and the sum over $n$  indicates that all such final states have to be included. We note that Eq.~(\ref{eq_dsigma_NNLO_partonic_nf}) 
can also be used to describe the production of $N+2$ jets in association with $X$, provided of course the measurement function is modified. 

Following Refs~\cite{Devoto:2023rpv,Devoto:2025kin} (see also the discussion in Section~\ref{sec:summary}), we insert the partition of unity
\begin{equation}
    \sum_{(ij) \in \setf{N+2}{n}{}} \!\!\! \hspace{-3.5mm} \Delta^{(ij)} = 1 \,,
\label{eq4.2}
\end{equation}
into \eq\eqref{eq_dsigma_NNLO_partonic_nf}. The sum over the indices $i,j$ in Eq.~(\ref{eq4.2}) runs over all final-state partons in the list $\setf{N+2}{n}{}$, and each damping factor $\Delta^{(ij)}$ vanishes if any parton other than partons $i,j$ becomes unresolved.
Then, similarly to the NLO case, we label the potentially-unresolved partons as $\Fp$ and $\Sp$, and use the symmetry of the gluon, quark, and antiquark lists within $\setf{N+2}{n}{}$ to write the double-real emission partonic cross section as
\begin{equation}
    2s_{\inF\inS}\, \dsigma^{\RR}_{\inF\inS}
    = \lint \Delta^{(\Fp\Sp)} \THmn  \bbFLMlo{\inF\inS, \DS}{\Fp, \Sp} \rint 
    + \lint \Delta^{(\Fp\Sp)} \bbFLMlo{\inF\inS, \noDS}{\Fp, \Sp} \rint \,.
\label{eq_dsigma_NNLO_partonic_2}
\end{equation}
The functions $\bbFLM$ in the above equation are defined as follows 
\begin{equation}
    \bbFLMlo{\inF\inS, \DS}{\Fp, \Sp}  
    = \sum_n\FLMlo{\inF\inS}{\setf{N+2}{n}{}(\Fp_g,\Sp_g)}
    + \sum_n \sum_{\fgflin=1}^{\nf} \FLMlo{\inF\inS}{\setf{N+2}{n}{}(\Fp_{(q_\fgflin},\Sp_{\qb_\fgflin)})}, 
\label{eq_FLM_DS_any_nf_def}
\end{equation}
and 
\begin{equation}
    \bbFLMlo{\inF\inS, \noDS}{\Fp, \Sp}
    = \bbFLMlonf{1}{\inF\inS, \noDS}{\Fp, \Sp} + \bbFLMlonf{2}{\inF\inS, \noDS}{\Fp, \Sp} \,,
\label{eq_FLM_noDS_any_nf_def}
\end{equation}
with 
\begin{align}
    & \begin{aligned}
        \bbFLMlonf{1}{\inF\inS, \noDS}{\Fp, \Sp} 
        = &\; \sum_{n} \sum_{\fgflin=1}^{\nf}
        \FLMlo{\inF\inS}{\setf{N+2}{n}{}(\Fp_{q_\fgflin},\Sp_g)}
        + \sum_{n} \sum_{\fgflin=1}^{\nf} \FLMlo{\inF\inS}{\setf{N+2}{n}{}(\Fp_{\qb_\fgflin},\Sp_g)} \\
        & + \sum_{n} \sum_{\fgflin=1}^{\nf} \frac12 \FLMlo{\inF\inS}{\setf{N+2}{n}{}(\Fp_{q_\fgflin},\Sp_{q_\fgflin})}
        + \sum_{n} \sum_{\fgflin=1}^{\nf} \frac12 \FLMlo{\inF\inS}{\setf{N+2}{n}{}(\Fp_{\qb_\fgflin},\Sp_{\qb_\fgflin})} \Big] \,,
    \end{aligned}
    \label{eq_FLM_noDS_1_any_nf_def} \\
    & \begin{aligned}
        \bbFLMlonf{2}{\inF\inS, \noDS}{\Fp, \Sp}  
        = &\; \sum_n \! \sum_{\substack{\fgflin,\sgflin= 1 \\ \sgflin > \fgflin}}^{\nf} 
        \!\! \FLMlo{\inF\inS}{\setf{N+2}{n}{}(\Fp_{q_\fgflin},\Sp_{q_\sgflin})}
        + \sum_n \! \sum_{\substack{\fgflin,\sgflin= 1 \\ \sgflin > \fgflin}}^{\nf} 
        \!\! \FLMlo{\inF\inS}{\setf{N+2}{n}{}(\Fp_{\qb_\fgflin},\Sp_{\qb_\sgflin})} \\
        & + \sum_n \! \sum_{\substack{\fgflin,\sgflin= 1 \\ \sgflin \neq \fgflin}}^{\nf} 
        \!\! \FLMlo{\inF\inS}{\setf{N+2}{n}{}(\Fp_{q_\fgflin},\Sp_{\qb_\sgflin})} \,.
    \end{aligned}
    \label{eq_FLM_noDS_2_any_nf_def}    
\end{align}
In writing \eq\eqref{eq_FLM_DS_any_nf_def}, we adopted the shorthand notation
\begin{equation}
    \FLMlo{\inF\inS}{\setf{N+2}{n}{}(\Fp_{(q_\fgflin},\Sp_{\qb_\fgflin)})}
    =
    \FLMlo{\inF\inS}{\setf{N+2}{n}{}(\Fp_{q_\fgflin},\Sp_{\qb_\fgflin})}
    + \FLMlo{\inF\inS}{\setf{N+2}{n}{}(\Fp_{\qb_\fgflin},\Sp_{q_\fgflin})} \,,
    \label{eq:symm_notation}
\end{equation}
which was already introduced in Ref.~\cite{Devoto:2025kin}.  

The above representation of the $\FLM$ functions has several important features that we would like to comment upon.  
First, the $\DS$ contribution in \eq\eqref{eq_FLM_DS_any_nf_def} collects combinations of unresolved partons  that possess a singular double-soft limit (i.e. $E_{\Fp,\Sp} \to 0$ with $E_\Fp/E_\Sp$ fixed), whereas the $\noDS$ terms in Eqs~(\ref{eq_FLM_noDS_any_nf_def} -- \ref{eq_FLM_noDS_2_any_nf_def}) contain those that do not.  
In both cases, the symmetry factors included in the definition of the $\FLM$ functions are determined \emph{entirely} by the final-state partons in $\setf{N+2}{n}{}(\Fp,\Sp)$ that do not carry labels $\Fp$ or $\Sp$.

The last two terms in \eq\eqref{eq_FLM_noDS_1_any_nf_def} contain factors $1/2$, which account for the symmetry of the $\FLM$ functions under the exchange $\Fp_{q_\fgflin} \leftrightarrow \Sp_{q_\fgflin}$ or $\Fp_{\qb_\fgflin} \leftrightarrow \Sp_{\qb_\fgflin}$.  
A similar factor for the $(\Fp_g, \Sp_g)$ unresolved final states is absent due to the energy ordering enforced by the function $\THmn = \Theta(E_\Fp - E_\Sp)$ in \eq\eqref{eq_dsigma_NNLO_partonic_2}.  
Finally, in Eqs~(\ref{eq_FLM_DS_any_nf_def} -- \ref{eq_FLM_noDS_2_any_nf_def}), each $\FLM$ function depends on its own list of final-state partons $\setf{N+2}{n}{}(\Fp,\Sp)$, and the sum over $n$ runs over all possible states consistent with the initial partonic state $\inF\inS$.

\tableFinalResult{t}{
    List of functions collected in the ancillary file \finalresult.  
    The first column shows the names of the functions that are used in the final result, the second column indicates the equation in which they appear, and the third provides their names in the file \finalresult.  
    For brevity, in the second block of the table, the splitting kernels $\calP_{gg}^{\,\dots}$, $\calP_{qg}^{\,\dots}$, etc., are collectively denoted by $\calP_{xy}^{\,\dots}$.  
    Further information can be found in the \readme file provided with the ancillary file. 
    We recall that the energy arguments of the initial-state splittings should be taken to be $E_a$ when the splitting appears on the left-hand side of the $\otimes$ or $\barotimes$ symbols, and $E_b$ when on the right (see the comment below \eq\eqref{eq_Pgen_PAP_PNLO_relation}). 
}

Using  Eqs~(\ref{eq_FLM_DS_any_nf_def} -- \ref{eq_FLM_noDS_2_any_nf_def}) as the starting point, we follow the steps described in Refs~\cite{Devoto:2023rpv,Devoto:2025kin} to extract the $1/\ep$ poles in the double-real contribution.
We then combine these with the divergences arising from the virtual corrections in order to cancel all $1/\ep$ singularities and extract the finite remainders which can be evaluated in four dimensions. We do so separately for  the fully-resolved (FR), single-unresolved (SU), and double-unresolved (DU) contributions, which we specify below.   In terms of these three contributions the result reads
\begin{equation}
    2s_{\inF\inS} \, \dsigmahat_{\inF\inS}^\NNLO
    = 2s_{\inF\inS} \big[\dsigmahat_{\inF\inS}^\fr + \dsigmahat_{\inF\inS}^\su + \dsigmahat_{\inF\inS}^\du \big] \,.
\label{eq_final_result_nf}
\end{equation} 

The first term in \eq\eqref{eq_final_result_nf} is the fully-resolved contribution, which contains $N+2$ resolved partons in the final state and subtraction terms that make it finite. This contribution reads
\begin{equation}
    \dsigmahat_{\inF\inS}^\fr 
    = 
    \lint \oS_{\Fp\Sp} \oS_\Sp \Omega_1 \Delta^{(\Fp\Sp)} \THmn \bbFLMlo{\inF\inS, \DS}{\Fp, \Sp} \rint 
    + \lint \oS_{\Fp\Sp} \oS_\Sp \Omega_1 \Delta^{(\Fp\Sp)} \bbFLMlo{\inF\inS, \noDS}{\Fp, \Sp} \rint \,.
\label{eq_fully_unresolved_final_nf}
\end{equation}
The $\Omega_1$ operator reads
\begin{equation}
\begin{split}
    \Omega_{1} = & ~ \sum_{\inotj} \oC_{i \Fp} \oC_{j \Sp} [\rmd p_\Fp] [\rmd p_\Sp] \, \omega^{\Fp i, \Sp j}  + \sum_{i} \Big[\oC_{i \Sp} \theta^{(a)} + \oC_{\Fp \Sp} \theta^{(b)} + \oC_{i \Fp} \theta^{(c)} + \oC_{\Fp \Sp} \theta^{(d)}\Big] [\rmd p_\Fp] [\rmd p_\Sp]  \, \oC_{\Fp \Sp, i}\; \omega^{\Fp i, \Sp i} \, .
\label{eq:Omega1}
\end{split}
\end{equation}
In \eq\eqref{eq:Omega1}, the sum over $i$ runs over all resolved partons, while the sum over $(ij)$ runs over all unordered pairs of resolved partons with $i \ne j$ (i.e.\ this sum would include both $i=1, j=2$ and $i=2,j=1$, and so forth).
The calculation of $\dsigmahat_{\inF\inS}^\fr $ is performed numerically, and the discussion of how this is done in practice is beyond the scope of this paper. We note, however, that the angular partition functions  $\omega^{\Fp i,\Sp j}$ and  $\omega^{\Fp i,\Sp i}$ and the sector  functions $\theta^{(a, \mydots, d)}$ identify the distinct  ways in which two partons can approach collinear singularities.\footnote{The sector functions are defined as $\theta^{(a)} = \Theta\left(\eta_{i\Sp} < \eta_{i \Fp}/2\right)$, $\theta^{(b)} =\Theta\left(\eta_{i \Fp}/2 < \eta_{i\Sp} < \eta_{i \Fp}\right)$, $\theta^{(c)} = \Theta\left(\eta_{i\Fp} < \eta_{i \Sp}/2\right)$, $\theta^{(d)} = \Theta\left(\eta_{i \Sp}/2 < \eta_{i\Fp} < \eta_{i \Sp}\right)$, where $\eta_{ij}=(1-\cos\theta_{ij})/2$.} 
Each sector is treated separately, using the parameterization of the unresolved angular phase space proposed in Refs.~\cite{Czakon:2010td,Czakon:2011ve}.

Upon fixing $\Fp$ and $\Sp$, only four distinct sectors are required to implement the fully-resolved contributions for each of the $N+2$ triple-collinear partitions $\omega^{\Fp i, \Sp i}$.
The total number of double-collinear partitions $\omega^{\Fp i,\Sp j}$ with $ i \ne j$ is $(N+2)(N+1)$, and each can be parameterized independently, allowing the integration variables to be optimally adapted to the relevant singular limits.
We note that in each sector and partition, one needs to consider at most seven different kinematic configurations, each resulting from applying one or more of the operators in \eq\eqref{eq:Omega1} to the double-real emission kinematics.
Furthermore, in \eq\eqref{eq:Omega1}, for consistency with the computation of the integrated triple-collinear subtraction terms (see the comment below \eq\eqref{eq:Omega2_def}), the $\oC_{\Fp \Sp, i}$ operator does not act on the unresolved phase space, while the operators $\oC_{ij}$ do. To emphasize this, we write the phase space measure to the right of the operators $\oC_{ij}$ but to the left of the operators $\oC_{\mathfrak{m}\mathfrak{n},i}$. 

The second term in \eq\eqref{eq_final_result_nf} is the single-unresolved contribution, which contains $N+1$ resolved final-state partons and subtraction terms that make it finite.  
It is written as
\begin{equation}
    \dsigmahat_{\inF\inS}^\su 
    = \dsigmahat_{\inF\inS}^{\su, \inFsb}
    + \dsigmahat_{\inF\inS}^{\su, \inSsb}
    + \dsigmahat_{\inF\inS}^{\su, \el} \,.   
\label{eq_single_unresolved_final_nf}
\end{equation}
The first two terms on the \rhs\ are the single-boosted contributions. They read 
\begin{equation}
\begin{aligned}
    \dsigmahat_{\inF\inS}^{\su, \inFsb}
    & = \asbr \sum_x \sum_\Fp \sum_n \bigg\{ \lint \ONLO^{(\inF,\Fp)} \partFuncACsp{\inF} \Delta^{(\Fp)} \log \left ( \frac{\eta_{\inF\Fp}}{2} \right ) \PAP_{x\inF} \conv \FLMlo{x\inS}{\setf{N+1}{n}{}(\Fp)} \rint \\
    & + \lint \ONLO^{(\Fp)} \, \Delta^{(\Fp)} \PNLO_{x\inF} \conv \FLMlo{x\inS}{\setf{N+1}{n}{}(\Fp)} \rint \bigg\},
\label{eq_dsigmahat_su_inFsb}
\end{aligned}
\end{equation}
and
\begin{equation}
\begin{aligned}
    \dsigmahat_{\inF\inS}^{\su, \inSsb}
    & = \asbr \sum_x \sum_\Fp \sum_n \bigg\{ \lint \ONLO^{(\inS,\Fp)} \partFuncACsp{\inS} \Delta^{(\Fp)} \log \left ( \frac{\eta_{\inS\Fp}}{2} \right ) \FLMlo{\inF x}{\setf{N+1}{n}{}(\Fp)} \conv \PAP_{x\inS} \rint \\
    & + \lint \ONLO^{(\Fp)} \, \Delta^{(\Fp)} \FLMlo{\inF x}{\setf{N+1}{n}{}(\Fp)} \conv \PNLO_{x\inS} \rint \bigg\} \,,
\label{eq_dsigmahat_su_inSsb}
\end{aligned}
\end{equation}
where we recall that the left and right convolutions are defined in \eq\eqref{eq_convolution}.
We note that the sum over $\xa$ is understood as the sum over different species of potentially-unresolved partons, i.e. $g$, $q_\rho$, $\bar q_\rho$, with $\rho$ running over distinct quark flavors, and that each species provides exactly one representative to the sum.
The sum over $n$ accommodates all final states with a given $\Fp$ that can be produced in a particular partonic collision, and therefore it can have a different meaning for each term in the above equations, in spite of the fact that we use just one sum to keep equations more compact.   
Finally, the sum over $x$ runs over the subset of partons that can be produced by the parton $\inF$ or the parton $\inS$ upon considering all possible collinear splittings.   

The $\ONLO$ operators are defined as
\begin{equation}
    \ONLO^{(i,\Fp)} = \oS_\Fp \oC_{i\Fp} \,,
    \qquad
    \ONLO^{(\Fp)} = \sum_{i\in\HP} \ONLO^{(i,\Fp)} \, \partFuncNLOfp{i} \,,
\label{eq_ONLO_def_copia}
\end{equation}
where, as before, $\HP$ denotes the list of initial- and final-state partons associated with a given $\FLM$, excluding the potentially-unresolved parton $\Fp$, and $\partFuncNLOfp{i}$ are the NLO partition functions discussed in item v) of \Sec\ref{sec:summary}.  
The functions $\partFuncACsp{i}$ are the NNLO partition functions (also referenced in item v) of \Sec\ref{sec:summary}) upon which the collinear operator $C_{i\Sp}$ has been applied. Note that after the action of $C_{i\Sp}$, $\partFuncACsp{i}$ does not depend on parton $\Sp$ anymore.
The variable $\eta_{i\Fp}$ is  defined as $\eta_{i \Fp} = (1-\cos \theta_{i \xa})/2$, where $\theta_{i\xa}$ is the relative angle between the directions of parton $i$ and $\xa$, computed in the preselected reference frame (e.g., the center-of-mass frame of the partonic collision).

The last term on the \rhs\ of \eq\eqref{eq_single_unresolved_final_nf} is the elastic contribution, which reads 
\begin{equation}
\begin{aligned}
    &\; \dsigmahat_{\inF\inS}^{\su, \el} 
    = \sum_\Fp \sum_n \bigg\{ \asbr \llint \ONLO^{(\Fp)} \, \Delta^{(\Fp)} \big[ \delta_{\Fp g} \ITot^{(0)}(E_\Fp) + \overline{\delta}_{\Fp g} \ITot^{(0)}(\Emax) \big] \colorprod \FLMlo{\inF\inS}{\setf{N+1}{n}{}(\Fp)} \rrint \\
    & - \sum_{i \in \HP} \, \asbr \llint \ONLO^{(i,\Fp)} \, \partFuncACsp{i} \, \log \left(\frac{\eta_{i \Fp}}{2}\right) \Delta^{(\Fp)} 
    \Big[ \delta_{\Fp g} \Big( \gamma_i + 2 \T_i^2 L_i(E_\Fp) \Big)  
    + \overline{\delta}_{\Fp g} \Big( \gamma_i + 2 \T_i^2 L_i \Big) \Big] \FLMlo{\inF\inS}{\setf{N+1}{n}{}(\Fp)} \rrint \\
    & - \sum_{i \in \HP} \, \asbr \llint \ONLO^{(i,\Fp)} \, \partFuncBD{i} \, \log \left ( \frac{\eta_{i \Fp}}{4 (1-\eta_{i \Fp})} \right ) \Delta^{(\Fp)} \Big[ \gamma_\Fp + 2 \T_q^2 L_\Fp \overline{\delta}_{\Fp g} \Big] \FLMlo{\inF\inS}{\setf{N+1}{n}{}(\Fp)} \rrint \\
    & + \lint \ONLO^{(\Fp)} \, \Delta^{(\Fp)} \FRVfinlo{\inF\inS}{\setf{N+1}{n}{}(\Fp)} \rint \bigg\} 
    + \sum_{n} \sum_{i \in \HP} \frac{\asbr}{2} \llint \ONLO^{(i,\Fp)} \, \partFuncBD{i} \Delta^{(\Fp)} \Big[ \gamma_g^{\perp,\rmr}  \FLMlo{\inF\inS}{\setf{N+1}{n}{}(\Fp_g)} \\
    & + \gamma_g^{\perp} (r_i^\mu r_i^\nu + g^{\mu\nu}) \FLMlomunu{\inF\inS}{\setf{N+1}{n}{}(\Fp_g)} \Big] \rrint \,.
\end{aligned}
\label{eq_dsigmahat_SU_el_final}
\end{equation}
In the above equation, $\overline{\delta}_{ij} = 1 - \delta_{ij}$, and the quantity $\ITot^{(0)}$ denotes the $\order{\ep^0}$ expansion coefficient of the infrared-finite operator $\ITot(\ep)$; one should replace $E_{\rm max}$ with $E_\xa$ in that equation to obtain $\ITot^{(0)}(E_\Fp)$.    
In the second line of Eq.~(\ref{eq_dsigmahat_SU_el_final}), $L_i = \log(\Emax/E_i)$,  $L_i(E_\Fp) = \log(E_\Fp/E_i)$, and $\gamma_i$ is the collinear anomalous dimension of parton $i$, with $\gamma_q = \gamma_\qb  = 3/2\,\Cf$ and  $\gamma_g = \beta_0 = 11/6\,\Ca - 2/3\,\TR\nf$. 
In the third line, the functions $\partFuncBD{i}$ are the NNLO partition functions computed in the collinear limit $\xa \parallel \yb$,\footnote{Analogously to $\partFuncACsp{i}$, the functions 
$\partFuncBD{i}$ are independent of 
parton $\Sp$.} and $L_\Fp = \log(\Emax/E_\Fp)$.  
The vector $r_i^\mu$ that appears in the last line of Eq.~(\ref{eq_dsigmahat_SU_el_final}) is defined in Appendix~E of Ref.~\cite{Devoto:2023rpv}.  
All remaining quantities can be found in an ancillary file, as summarized in Table~\ref{table_final_result}.
We note that the sums over the index $i \in \HP$ in the second, third, and fourth lines of Eq.~(\ref{eq_dsigmahat_SU_el_final}) run over all partons in the corresponding $\FLM$ functions, excluding the unresolved parton $\Fp$, and that sums over $n$ have the usual meaning. 

Next, the double-unresolved contribution in \eq\eqref{eq_final_result_nf} contains $N$ resolved partons in the final state, which equals the number of jets in the LO process.  We can write it as a sum of four terms, each having distinct kinematics
\begin{equation}
    \dsigmahat_{\inF\inS}^\du 
    = \dsigmahat_{\inF\inS}^{\du, \db}
    + \dsigmahat_{\inF\inS}^{\du, \inFsb}
    + \dsigmahat_{\inF\inS}^{\du, \inSsb}
    + \dsigmahat_{\inF\inS}^{\du, \el} \,.
\label{eq_double_unresolved_final_nf}
\end{equation}

The first term is the double-boosted contribution
\begin{equation}
    \dsigmahat_{\inF\inS}^{\du, \db}
    =
    \sum_{x,y} \sum_n \, \asbr^2  \lint \PNLO_{x\inF} \conv \FLMlo{xy}{\setf{N}{n}{}} \conv \PNLO_{y\inS} \rint \,,
\label{eq_final_formula_DU_DB}
\end{equation}
where $x,y$ include all partons that can be obtained from the collinear splittings of partons $a,b$.
The second and the third terms  are the single-boosted contributions which describe collinear splittings of partons $a$ and $b$, respectively. They read
\begin{equation}
\begin{aligned}
    \dsigmahat_{\inF\inS}^{\du, \inFsb}
    & = \sum_x \sum_n \Big[
    \asbr \lint \PNLO_{x\inF} \conv \calF^{x\inS}[\setf{N}{n}{}] \rint
    + \asbr^2 \lint \PNNLO_{x\inF} \conv \FLMlo{x\inS}{\setf{N}{n}{}} \rint \Big] \\
    & + \sum_{n} \, \asbr^2 \lint \calPW_{\inF\inF} \conv \big[\Wacfin{\inF} \colorprod \FLMlo{\inF\inS}{\setf{N}{n}{}} \big] \rint  \, ,
\label{eq_dsigmahat_ab_du_inFsb_nf}
\end{aligned}
\end{equation}
and
\begin{equation}
\begin{aligned}
    \dsigmahat_{\inF\inS}^{\du, \inSsb}
    & = \sum_x \sum_n \Big[
    \asbr \lint \calF^{\inF x}[\setf{N}{n}{}] \conv \PNLO_{x\inS} \rint
    + \asbr^2 \lint \FLMlo{\inF x}{\setf{N}{n}{}} \conv \PNNLO_{x\inS}  \rint \Big] \\
    & + \sum_{n} \, \asbr^2 \lint \big[\Wacfin{\inS} \colorprod \FLMlo{\inF\inS}{\setf{N}{n}{}} \big] \conv \calPW_{\inS\inS} \rint \,,
\label{eq_dsigmahat_ab_du_inSsb_nf}
\end{aligned}
\end{equation}
where
\begin{equation}
    \calF^{ij}[\setf{N}{n}{}] = \asbr \, \ITot^{(0)} \colorprod \FLMlo{ij}{\setf{N}{n}{}} + \FLVfinlo{ij}{\setf{N}{n}{}} \,.
\end{equation}
The functions $\PNNLO_{\alpha\beta}$\footnote{In Eqs (\ref{eq_dsigmahat_ab_du_inFsb_nf}, \ref{eq_dsigmahat_ab_du_inSsb_nf}), the $\alpha\beta$ pairs in $\PNLO_{\alpha\beta}$ that yield non-vanishing results are $qq$, $qg$, $gq$, $gg$, together with the same pairs where $q$ is replaced by $\qb$.
For the functions $\PNNLO_{\alpha\beta}$, in addition to the pairs listed above, the combinations $q\qb$, $q\qp$, and $q\qbp$ also contribute. Moreover, one finds that $\PNNLO_{q\qb} = \PNNLO_{\qb q}$ and $\PNNLO_{q\qp} = \PNNLO_{q\qbp}$.} 
and $\calPW_{\alpha\alpha}$ can be found in Table~\ref{table_final_result}, and we will comment on the operators $\Wacfin{a}$ and $\Wacfin{b}$ shortly. We note that Eqs (\ref{eq_final_formula_DU_DB} -- \ref{eq_dsigmahat_ab_du_inSsb_nf}) are a  natural extension of formulas reported in Refs.~\cite{Devoto:2023rpv,Devoto:2025kin}, except that in those references collinear splittings that change the type of the initial-state parton were omitted. 

Finally, the double-unresolved contribution corresponding to $N$-jet final states without additional boosts reads 
\begin{equation}
\begin{aligned}
    \dsigmahat_{\inF\inS}^{\du, \el} 
    & = \sum_n \bigg\{ \, \asbr^2 \lint\big[\Iccfin + I_{\rm ss}^{\rm fin} + I_{\rm tri}^{\rm fin} + \Iuncfin \big] \colorprod \FLMlo{\inF\inS}{\setf{N}{n}{}} \rint  \\
    & + \asbr^2 \sum_{i \in \HP} \llint \Big[\theta_\HPf \gamma_{z,f_i \to f_i g}^{\calW}(L_i) \, \Wacfin{i} + \delta^{(0)}\Wbdfin{i} + \delta^{\perp,(0)} \Wr{i}\Big] \colorprod \FLMlo{\inF\inS}{\setf{N}{n}{}} \rrint \\ 
    & + \asbr \lint \ITot^{(0)} \colorprod \FLVfinlo{\inF\inS}{\setf{N}{n}{}} \rint
    + \lint \FLVfinsqlo{\inF\inS}{\setf{N}{n}{}} \rint 
    + \lint \FVVfinlo{\inF\inS}{\setf{N}{n}{}} \rint \bigg\} \, ,
\end{aligned}
\label{eq_dsigmahat_DU_el_final}
\end{equation} 
where $\HP = \{\inF,\inS\} \cup \setf{N}{n}{}$.  
In the second line of \eq\eqref{eq_dsigmahat_DU_el_final}, we use the function $\theta_\HPf$ which evaluates to $\theta_\HPf = 1$ if $i \in \HPf$ (final-state parton) and $\theta_\HPf = 0$ otherwise.
We note that all limits that contribute to the elastic contribution $\dsigmahat_{\inF\inS}^{\du, \el}$ were considered in Refs.~\cite{Devoto:2023rpv,Devoto:2025kin}. Consequently, the structure of this result is identical to those discussed in these references, and the functions appearing in \eq\eqref{eq_dsigmahat_DU_el_final} have already been defined there. Nevertheless, for completeness, we briefly describe the various terms that appear in this equation. 

The operator $I_{\rm cc}^{\rm fin}$ in the first line of \eq\eqref{eq_dsigmahat_DU_el_final} is defined in \eq(7.13) of Ref.~\cite{Devoto:2025kin}. It contains terms with two and four color-charge operators $\T_i$ as well as various remnants of virtual $\IVirt$, soft $\ISoft$ and collinear $\IColl$ operators. 
The quantity $I_{\rm ss}^{\rm fin}$ denotes a particular finite remainder of the double-soft integrated subtraction term. It is defined as\footnote{These terms are given in the third line of Eq.~(7.12) of Ref.~\cite{Devoto:2025kin}.}
\begin{equation}
\begin{split}
    I_{\rm ss}^{\rm fin} 
    = \sum_{(ij)\in\HP} \!\! {\rm DS}^{\rm fin}_{ij} \, (\T_i \cdot \T_j) \,,
\end{split}
\label{eq_Issfin_def}
\end{equation}
and the coefficients ${\rm DS}^{\rm fin}_{ij}$ can be obtained from an ancillary file, see Table~\ref{table_final_result}.  
Here, the notation $(ij)\in\HP$ means that one sums over unordered pairs of initial- and final-state particles, i.e.  $\HP = \{\inF,\inS\} \cup \setf{N}{n}{}$, with $i \neq j$.
The operator $\Itrifin$ represents the component proportional to the product of three color-charge operators.
In Ref.~\cite{Devoto:2023rpv}, we have shown that such triple-color correlators originate from three distinct sources: 
i) double-virtual corrections~\cite{Catani:1998bh,Becher:2009qa,Gardi:2009qi}, 
ii) commutators of the soft $\ISoft$ and virtual $\IVirt$ operators, and 
iii) the soft limit of the real-virtual contributions \cite{Catani:1999ss}. 
The triple color-correlated  contribution was computed in Ref.~\cite{Devoto:2023rpv} in full generality, c.f. Eq.~(I.9) in that reference. For this reason, it can be used for calculating NNLO QCD corrections to arbitrary processes without further ado.

All remaining color-uncorrelated contributions are collected into the term $\Iuncfin$.  It is defined as 
\begin{equation}
\begin{split}
    I_{\rm unc}^{\rm fin} 
    = \sum_{i\in\HP} \Big[\T_i^2 \, D_{T^2} + D_{f_i}(E_i) \Big] \,,
\end{split}
\label{eq_Iuncfin_def}
\end{equation}
where, in the first function $ D_{T^2}$, we have collected  all the color-uncorrelated terms that  depend on the external legs only via the relevant Casimir factor. The remainder $D_{f_i}$ depends on the type of parton, i.e.\ $i$ being  a quark  or a gluon. These quantities can be extracted from the ancillary file as explained in Table~\ref{table_final_result} (note that in the Table~\ref{table_final_result} the $D_{f_i}$ functions are reported with superscripts ISR and FSR added to distinguish between initial- and final-state radiation contributions). 

In the second line of \eq\eqref{eq_dsigmahat_DU_el_final}, the partition-dependent operators $\Wacfin{i}$, $\Wbdfin{i}$, and $\Wr{i}$ appear. 
The former 
arise from residual partition-dependent correction from subsectors $(a)$ and $(c)$, while the latter two arise from spin-correlated singular contributions, which are discussed extensively in Refs \cite{Devoto:2023rpv, Devoto:2025kin}.

We emphasize that these functions are independent of parton $\Sp$: 
the notation $i \parallel \Sp$ and $\Fp \parallel \Sp$ is kept to identify which collinear limit has been performed.
The quantities $\delta^{(0)}$, $\delta^{\perp,(0)}$, and $\gamma_{z,f_i \to f_i g}^{\calW}$ can be extracted from the ancillary file as explained in Table \ref{table_final_result}. Finally, the quantities $\FLVfinsq$ and $\FVVfin$ that appear in the final line of \eq\eqref{eq_dsigmahat_DU_el_final} are the process-dependent finite remainders of the Catani-subtracted one-loop squared and two-loop virtual amplitudes, respectively. \\

Eqs~(\ref{eq_final_result_nf} -- \ref{eq_Iuncfin_def}) provide finite remainders of the integrated NNLO QCD subtraction terms for hadron collider processes, $pp \to \colsing + N ~\mathrm{jets}$. 
However, since these formulas are written as expressions that are applied to each of the resolved partons, they can accommodate processes at lepton-hadron or lepton-lepton colliders with minimal adjustments.
Taking the latter case as an example, we explain how to modify  Eqs~(\ref{eq_final_result_nf} -- \ref{eq_Iuncfin_def})
to arrive at the finite remainders for $\ell^+\ell^- \to \colsing + N~\mathrm{jets}$. 
\begin{enumerate}[label=\roman*)]
    \item The fully-regulated contribution in \eq\eqref{eq_fully_unresolved_final_nf} remains unchanged. However, the sums over the indices $i$ and $j$ in \eq\eqref{eq:Omega1} should include initial-state particles only if they have color charge; hence, in the case of a lepton-lepton collider, these sums should run over final-state partons only.

    \item In the single-unresolved contribution in  Eq.~\eqref{eq_single_unresolved_final_nf}, the boosted terms, defined in Eqs~(\ref{eq_dsigmahat_su_inFsb}, \ref{eq_dsigmahat_su_inSsb}), must be set to zero. 

    \item In the elastic single-unresolved contribution given in Eq.~\eqref{eq_dsigmahat_SU_el_final}, all terms involving initial states must be discarded.  
    Hence, in the operators $\ITot^{(0)}$, $\ONLO^{(\Fp)}$ and $\ONLO^{(i,\Fp)}$, as well as in the sums over $i \in \HP$, only final-state partons should be considered. 
    
    \item In the double-unresolved contribution of Eq.~\eqref{eq_double_unresolved_final_nf}, the first three initial-state-boosted terms, defined in Eqs~(\ref{eq_final_formula_DU_DB} -- \ref{eq_dsigmahat_ab_du_inSsb_nf}), must be set to zero. 

    \item For the elastic double-unresolved contribution in Eq.~\eqref{eq_dsigmahat_DU_el_final}, the comments from point~iii) apply, i.e.\ all terms involving a initial state must be dropped.
    Furthermore, the operator $I_{\rm tri}^{\rm fin}$ simplifies and 
     the corresponding expression can be found in Eq.~(H.16) of Ref.~\cite{Devoto:2023rpv}.    
\end{enumerate}


\subsection{Important aspects of the calculation}
\label{ssec_NNLO_important_aspect}

Having presented the final result in Section~\ref{subsec_NNLO_nf_formulas}, we would like to discuss those aspects of the derivation that are new with respect to Refs~\cite{Devoto:2023rpv,Devoto:2025kin}.  
In particular, we describe the emergence of complete sums over intermediate (clustered) partons in the collinear limits involving initial states, the appearance of full collinear anomalous dimensions, and the triple-collinear integrated subtraction terms.   
Additional technical details pertinent to these problems are provided in Appendices~\ref{app_details_DC} and~\ref{app_details_TC}.


\subsubsection{Double-collinear contribution}
\label{ssec_double_collinear_ep_minus_2}

We begin with the discussion of the double-collinear soft-subtracted contributions.  
We find it convenient to treat terms arising from the double-collinear and triple-collinear partitions separately.
Thus, we define 
\begin{align}
    & \SigmaDCdc = \sum_{n} \!\! \sum_{\inotj \in \HP} \!\! \Lint \oS_\Fp \oS_\Sp C_{i \Fp} C_{j \Sp} \Delta^{(\Fp\Sp)} \bigg[\frac12 \bbFLMlo{\inF\inS, \DS}{\Fp, \Sp}
    + \bbFLMlo{\inF\inS, \noDS}{\Fp, \Sp} \bigg] \Rint,
    \label{eq_SigmaDC_dc_nf_def}
    \\
    & \SigmaDCtc  = \frac{1}{2} \sum_{n} \sum_{i \in \HP} \Big[\lint \oS_\Fp \oS_\Sp C_{i \Fp} C_{i \Sp} \Delta^{(\Fp\Sp)} \bbFLMlo{\inF\inS, \DS}{\Fp, \Sp} \rint
    + \lint \oS_\Sp \big(C_{i\Sp} C_{i\Fp} + C_{i\Fp} C_{i\Sp}\big) \Delta^{(\Fp \Sp)} \bbFLMlo{\inF\inS, \noDS}{\Fp, \Sp} \rint \Big] \,,
\label{eq_SigmaDC_tc_nf_def}
\end{align}
where the functions $\bbFLMlo{}{\Fp, \Sp}$  are given in Eqs~(\ref{eq_FLM_DS_any_nf_def}, \ref{eq_FLM_noDS_any_nf_def}).
The sets $\calH$ in Eqs~(\ref{eq_SigmaDC_dc_nf_def},~\ref{eq_SigmaDC_tc_nf_def}) include partons $\{\inF,\inS\}$ and the resolved partons in the lists $\setf{N+2}{n}{}(\Fp,\Sp)$ present in the $\bbFLMlo{}{\Fp, \Sp}$ functions, whereas in \eq\eqref{eq_SigmaDC_dc_nf_def}, the second sum runs over all unordered pairs of $i,j \in \HP$ with $i \ne j$.  

We need to rewrite $\SigmaDCdc$ and $\SigmaDCtc$ in such a way that their collinear singularities are made explicit, and their finite remainders are clearly defined. 
The calculation of the $\DS$ terms on the \rhs\ of Eqs~(\ref{eq_SigmaDC_dc_nf_def}, \ref{eq_SigmaDC_tc_nf_def}) was discussed in detail in Section~5.4 of Ref.~\cite{Devoto:2023rpv},\footnote{Specifically, the discussion in this reference  focuses on the case of an $\FLM$ function whose unresolved partons $\Fp,\Sp$ are a pair of gluons. As noted in Ref.~\cite{Devoto:2025kin}, the same procedure can be straightforwardly extended for an unresolved $q\qb$ pair of the same flavor.} while the $\noDS$ terms were partially addressed in Section~5.2 of Ref.~\cite{Devoto:2025kin}, and we complete the analysis here. Hence, in this section we briefly summarize the key steps of the calculation without repeating all technical details.  
Additional technical aspects relevant to the simplification of \eq\eqref{eq_SigmaDC_tc_nf_def} can be found in Appendix \ref{app_details_DC}.


\subsubsection*{The double-collinear sector}

We begin with the discussion of the double-collinear partition, \eq\eqref{eq_SigmaDC_dc_nf_def}. Since $i \ne j$, these terms, essentially, are the product or convolution of two NLO-like contributions that appear because of the action of the soft-regulated collinear operators $\oS_\Fp C_{i\Fp}$ and $\oS_\Sp C_{j\Sp}$.
Their analysis results in a simple, natural formula that reads 
\begin{equation}
\begin{split}
    \SigmaDCdc 
    & = \sum_{n} \! \sum_{\inotj \in \HP} \asbr^2 \Lint \frac{\Gamma_{i,f_i} \, \Gamma_{j,f_j}}{2\ep^2} \, \FLMlo{\inF\inS}{\setf{N}{n}{}} \Rint 
    + \sum_{x,y} \sum_n \frac{\asbr^2}{\ep^2} \lint \CalPgen_{x\inF} \conv \FLMlo{xy}{\setf{N}{n}{}} \conv \CalPgen_{y\inS} \rint \\
    & + \sum_x \sum_n  \frac{\asbr^2}{\ep} \bigg[\llint \CalPgen_{x\inF} \conv \Big[\Big(\IColl(\ep) - \frac{\Gamma_{\inF,x}}{\ep} \Big) \colorprod \FLMlo{x\inS}{\setf{N}{n}{}} \Big] \rrint
    + \llint \Big[\Big(\IColl(\ep) - \frac{\Gamma_{\inS,x}}{\ep} \Big) \colorprod \FLMlo{\inF x}{\setf{N}{n}{}} \Big] \conv \CalPgen_{x\inS} \rrint \bigg] \,.
\end{split}
\label{eq_SigmaDCdc_qqbgg_final_nf}
\end{equation}
The equation above shares several important features with the results for finite remainders presented in the previous subsection. In particular, it includes the double-boosted contributions, the convolutions of collinear operators with splitting functions $\CalPgen_{xy}$, and pair-wise products of collinear anomalous dimensions of external partons. Furthermore, we also observe the summation over relevant Born processes. 

The derivation of \eq\eqref{eq_SigmaDCdc_qqbgg_final_nf} is conceptually straightforward but somewhat tedious.  Hence, we will outline how its features arise from \eq\eqref{eq_SigmaDC_dc_nf_def}, where $\Sigma_{\rm DC}^{\rm dc}$
is written in terms of soft and collinear operators. 
There are three cases to consider: i) partons $i$ and $j$ are in the initial state, ii) parton $i$ is in the initial and parton $j$ in the final state (or vice versa), and iii) partons $i$ and $j$ are in the final state. 

We begin with the first case, where the unresolved partons become collinear to \emph{different} incoming partons. Then, hard final-state partons are not affected, but the initial-state partons undergo clustering and may change their types.
Since this clustering occurs independently on each initial-state leg, the NLO analysis can be repeated by considering the simultaneous action of the operators $C_{\inF \Fp} C_{\inS \Sp}$ and $C_{\inF \Sp} C_{\inS \Fp}$ on an $\FLM$ function (cf.\ Eqs~(\ref{eq_limit_aqj_genX}, \ref{eq_NLO_IS_qj_genX})).
Among other things, this leads to a double-boosted contribution (the second term on the \rhs\ of \eq\eqref{eq_SigmaDCdc_qqbgg_final_nf}) where the sums over all intermediate parton types ($x$ and $y$) emerge. 
This happens because the potentially-unresolved partons $\xa$ and $\yb$ can be of all possible flavors, so that the sum over all types of $\xa,\yb$ naturally becomes a sum over all possible types of clustered partons. 

Initial-state collinear limits also produce collinear anomalous dimensions if the unresolved parton is a gluon (cf.~\eq\eqref{eq_nf_1_collinear_limits_general_rules_first_2}).
Thus, the case where the unresolved partons $\Fp$ and $\Sp$ both become collinear to distinct initial-state partons leads, in addition to the double-boosted contribution, to terms that contain $\Gamma_{a,\fl{a}}\Gamma_{b,\fl{b}}$, $\CalPgen_{x\inF} \conv \Gamma_{\inS,\fl{\inS}}$, and $\Gamma_{\inF,\fl{\inF}} \conv \CalPgen_{x\inS}$.  
These contributions appear in the first term of \eq\eqref{eq_SigmaDCdc_qqbgg_final_nf} and in the $\IColl$ operators in the second line of that equation, respectively.  
We note that the initial-state anomalous dimensions $\Gamma_{\inF,x}$ and $\Gamma_{\inS,x}$ are subtracted from these $\IColl$ operators. 
This happens because terms such as $\CalPgen_{x\inF} \conv \Gamma_{\inF,\fl{\inF}}$ and $\Gamma_{\inS,\fl{\inS}} \conv \CalPgen_{x\inS}$ cannot appear in the double-collinear sectors, because the two collinear operators are applied to \emph{different} external partons. 

We continue with the case where one of the partons $\Fp$ and $\Sp$ becomes collinear to an initial-state parton, while the other becomes collinear to a final-state parton. 
This case can be analyzed following the NLO calculation, leading to the remaining products of an initial- and a final-state anomalous dimension in the first term of \eq\eqref{eq_SigmaDCdc_qqbgg_final_nf}, as well as the remaining terms in the $\IColl$ operators in the second line of that equation.

Finally, we consider the case where the two partons $\xa$ and $\yb$ become \emph{collinear to different final-state partons}. Here, the calculation is more involved, because hard final-state partons are affected by clustering with the unresolved partons $\xa$ and $\yb$, which can change their types. Such a clustering removes two partons from the final state; the list $\setf{N+2}{n}{}(\xa,\yb)$ then becomes one of the LO-like lists. 
The challenge is to make sure that a proper LO-like list appears at the right place. 
Furthermore, one also needs to reconstruct all generalized anomalous dimensions by combining incomplete contributions, since the latter are standard outcomes of individual collinear limits (cf.\ Eqs~(\ref{eq_Gamma_q_FSR_def}, \ref{eq_Gamma_g_FSR_def})).

Although a detailed analysis of final-state collinear limits requires significant care, at its core, it is again an iterative extension of what has already been discussed in \Sec\ref{NLO_any_process}. 
Nevertheless, we find it instructive to discuss an explicit example of how complete anomalous dimensions arise. To this end, we consider the term $\Gamma_{i,g} \Gamma_{j,q_\fgflin}/(2\ep^2)$ in \eq\eqref{eq_SigmaDCdc_qqbgg_final_nf} and, for simplicity, ignore the $2\nf\,\Gamma_{i,g \to q \qb}$ contribution to $\Gamma_{i,g}$.  
To obtain this term, we focus first on $\bbFLMlo{\inF\inS, \DS}{\Fp, \Sp}$ (cf.~\eq\eqref{eq_FLM_DS_any_nf_def}) with the $(\Fp_g, \Sp_g)$ pair, and consider the collinear limits $\Fp_g \parallel i_g$ and $\Sp_g \parallel j_{q_\fgflin}$.  
Applying \eq\eqref{eq_nf_1_collinear_limits_general_rules_first} separately to the two legs $i$ and $j$, we obtain  
\begin{equation}
    \sum_{(ij) \in \setf{N}{n}{}} \hspace{-3mm} \delta_{ig} \delta_{jq_\fgflin} \, \Gamma_{i,g\to gg} \, \Gamma_{j,q\to qg} \, \FLMlo{\inF\inS}{\setf{N}{n}{}} \,.
\label{eq_double_collinear_sector_term_1}
\end{equation}
As anticipated, the quark anomalous dimension $\Gamma_{j,q\to qg}$  in this expression is \emph{incomplete}, as it lacks the contribution $\Gamma_{j,q\to gq}$.  
To complete it, we focus on the term with the unresolved partonic state $(\Fp_{q_\fgflin},\Sp_g)$ appearing in $\bbFLMlo{\inF\inS, \noDS}{\Fp, \Sp}$ (cf.~\eq\eqref{eq_FLM_DS_any_nf_def}), and we consider final states with at least two \emph{distinct} hard gluons, so that both unresolved partons can independently become collinear to a hard gluon. To identify all singular contributions of this type, the NLO construction that we described in \Sec\ref{NLO_any_process} is particularly helpful. 
Indeed, a sequential application of \eq\eqref{eq_nf_1_collinear_limits_general_rules_first} and \eq\eqref{eq_NLO_FS_qjg_genX}\footnote{These hard-collinear operators may be applied in either order, since the two collinear limits commute when acting on different legs.} automatically yields the following result
\begin{equation}
    \sum_{(ij) \in \setf{N}{n}{}} \hspace{-3mm} \delta_{ig} \delta_{jq_\fgflin} \, \Gamma_{i,g\to gg} \, \Gamma_{j,q\to gq} \, \FLMlo{\inF\inS}{\setf{N}{n}{}} \,.
\label{eq_double_collinear_sector_term_2}
\end{equation} 
This complements the collinear anomalous dimension in \eq(\ref{eq_double_collinear_sector_term_1}), since the sum of the two equations gives $\Gamma_{j,q_\fgflin}$, as desired.  
Proceeding in the same way while considering all possible pairs of collinear limits acting on different final-state legs of the $\FLM$ functions in $\bbFLMlo{\inF\inS, \DS}{\Fp, \Sp}$ and $\bbFLMlo{\inF\inS, \noDS}{\Fp, \Sp}$, one obtains the first term on the \rhs\ of \eq\eqref{eq_SigmaDCdc_qqbgg_final_nf} when $i$ and $j$ are two final-state partons.


\subsubsection*{The triple-collinear sector}

The triple-collinear partition contribution, defined in \eq\eqref{eq_SigmaDC_tc_nf_def}, involves sequential soft-regulated collinear limits where partons $\Fp$ and $\Sp$ become collinear to a \emph{single} hard parton $i$.
This calculation is more complicated than the one for the double-collinear partitions, and we describe it in detail in Appendix \ref{app_ssec_DC_ISR}.
Combining the results of this appendix, we find\footnote{We note that the $\barotimes$ convolution that we use here is defined differently (cf.~\eq\eqref{eq_barotimes_conv_def}) compared to what we have used earlier in Refs \cite{Devoto:2023rpv, Devoto:2025kin}.}
\begin{align}
    \SigmaDCtc
    & = \frac{\asbr^2}{2\ep^2} \sum_{x} \sum_{n} \bigg\{ \Lint \bigg[\sum_{y} \PgenoxPgen{xy}{ya} + G_{x\inF}\bigg] \conv \FLMlo{x\inS}{\setf{N}{n}{}} \Rint
    + \Lint \FLMlo{\inF x}{\setf{N}{n}{}} \conv \bigg[\sum_{y} \PgenoxPgen{xy}{y\inS} + G_{x\inS}\bigg]\Rint \bigg\} \notag \\
    & + \frac{\asbr^2}{\ep} \sum_{x} \sum_{n} \bigg\{\Lint \CalPgen_{x\inF} \conv \bigg[\frac{\Gamma_{\inF,x}}{\ep} \, \FLMlo{x\inS}{\setf{N}{n}{}}\bigg] \Rint
    + \Lint \bigg[\frac{\Gamma_{\inS,x}}{\ep} \, \FLMlo{\inF x}{\setf{N}{n}{}}\bigg] \conv \CalPgen_{x\inS}  \Rint \bigg\} 
    \label{eq_SigmaDCtc_qqbgg_final_nf} \\
    & 
    + \frac{\asbr^2}{2\ep^2} \sum_{n} \Lint \bigg[ \sum_{i\in \HP} \Gamma_{i,f_i}^2 + 
    \sum_{i\in \setf{N}{n}{}} \bigg( \delta_{ig} \Big[2 \nf \,\Gamma_{i,g\to q\qb}\big(\Gamma_{i,q} - \Gamma_{i,g}\big)
    + \GFSR{i}{g \hspace{8mm}}{z,g\to gg} + 2 \nf \,\GFSR{i}{q \hspace{8mm}}{z,g\to q\qb} \Big] \notag \\
    & + \sum_{\fgflin=1}^{\nf} (\delta_{iq_\fgflin} + \delta_{i\qb_\fgflin}) \Big[\Gamma_{i,q\to gq}\big(\Gamma_{i,g} - \Gamma_{i,q}\big)
    + \GFSR{i}{q \hspace{8mm}}{z,q\to qg} + \GFSR{i}{g \hspace{8mm}}{z,q\to gq} \Big] \bigg) \bigg] \FLMlo{\inF\inS}{ \setf{N}{n}{} } \Rint \,. \notag
\end{align}
Even without an in-depth discussion of the derivation of the above equation, we can still explain how some of its features follow from \eq\eqref{eq_SigmaDC_tc_nf_def}.  
In the first line of \eq\eqref{eq_SigmaDCtc_qqbgg_final_nf}, the convolutions of two generalized splitting functions
and $\FLMlo{x\inS}{\setf{N}{n}{}}$ are present. 
Such terms arise from the initial-state collinear limits, which lead to two sequential clusterings of partons. In the first step, we cluster partons $\inF$ and $\Sp$ to form $[\inF \bar{\Sp}]$, and in the second step, we cluster it with parton $\Fp$ to produce $[\inF \bar{\Fp} \bar{\Sp}]$. Schematically, the following equation holds \begin{equation}
    \lint \oS_\Fp \oS_\Sp C_{a \xa} C_{a \yb} \FLMlo{ab}{... | \xa, \yb}\rint 
    \sim \calP_{[a \bar{\xa} \bar{\yb}] [a \bar{\yb}]} \, \barotimes \, 
    \calP_{[a \bar{\yb}] a} \otimes \FLM^{[a\bar{\xa} \bar{\yb}] b}.
\end{equation}
One can check that summing over all types of partons $\xa $ and $\yb$, encoded in the function  $\bbFLMlo{}{\Fp, \Sp}$, is equivalent to summing  over all types of clustered partons $x$ and $y$.
Furthermore, as explained in detail in Refs~\cite{Devoto:2023rpv, Devoto:2025kin}, the collinear limits $i \parallel \xa$ and $i \parallel \yb$ are not fully independent because of phase-space constraints. 
This gives rise to the functions $G_{xa}$ and $G_{xb}$ that appear in \eq\eqref{eq_SigmaDCtc_qqbgg_final_nf}. 

In the second line of \eq\eqref{eq_SigmaDCtc_qqbgg_final_nf} terms appear that  allow the  completion of the $\IColl$ operators present in the second line of \eq\eqref{eq_SigmaDCdc_qqbgg_final_nf}. We recall that such contributions are due to initial-state collinear limits. In particular, if $\Fp$ or $\Sp$ is a gluon, these limits lead to diagonal transitions which introduce anomalous dimensions, as shown in \eq\eqref{eq_coll_limit_IS_review}.  Convolutions between these terms and the splitting functions lead to terms on the second line of \eq\eqref{eq_SigmaDCtc_qqbgg_final_nf}, whereas the product of initial-state anomalous dimensions forms part of the first term on the third line of \eq\eqref{eq_SigmaDCtc_qqbgg_final_nf}. 

The remaining terms in the last two lines in \eq(\ref{eq_SigmaDCtc_qqbgg_final_nf}) describe final-state collinear limits. By analogy with the initial-state contribution, we expect two sequential final-state collinear limits to give rise to products of weighted anomalous dimensions, i.e.  
\begin{equation}
    \lint \oS_\Fp \oS_\Sp C_{i \xa } C_{i \yb} \FLMlo{ab}{\mydots\,, i, \mydots |  \xa , \yb  } \rint
    \sim \Gamma_{[i \xa \yb] \to [i \yb] \xa } \, \Gamma_{[i \yb] \to i \yb} \,  \FLMlo{ab}{\mydots \,,  [i\xa \yb] \,, \mydots} \,,
\label{FS_TC_Limit}
\end{equation}
as well as some functions $G_i$ which, as mentioned previously, take into account the phase-space intertwinement of partons $\Fp$ and $\Sp$. Finally, moving from \eq(\ref{FS_TC_Limit}) to the last two lines in \eq(\ref{eq_SigmaDCtc_qqbgg_final_nf}) requires an explicit enumeration of all possibilities for the clustered partons, an analysis of the transformation of the final-state hard partons under the action of the collinear limits, and the reconstruction of the squares of the complete anomalous dimensions, which will be needed for the $\IColl^2$ operator. 


\subsubsection{Triple-collinear subtraction contribution}
\label{sec:TC}

We now turn to the triple-collinear limits, which correspond to the situation in which the two unresolved partons $\Fp,\Sp$ become simultaneously collinear to a resolved parton.  In the context of the nested soft-collinear subtraction scheme, the relevant terms are obtained as integrals of the triple-collinear limits of the double-real matrix elements squared, followed by the subtraction of the double-soft, single-soft, as well as single-collinear singularities.
Hence, such contributions are written as  
\begin{equation}
    \sum_n \sum_{i\in\HP} \llint \oS_{\Fp \Sp} \oS_{\Sp} \Omega_2^{(i)} \Delta^{(\Fp \Sp)} \Big[\THmn \bbFLMlo{\inF\inS, \DS}{\Fp, \Sp} + \bbFLMlo{\inF\inS, \noDS}{\Fp, \Sp} \Big] \rrint \,,
\label{eq_triple_collinear_lim_nf}
\end{equation}  
where the two $\bbFLMlo{}{\Fp, \Sp}$ functions can be found in Eqs \eqref{eq_FLM_DS_any_nf_def} and \eqref{eq_FLM_noDS_any_nf_def},
while the operator $\Omega_2^{(i)}$ reads  
\begin{equation}
    \Omega_{2}^{(i)} = \Big[\oC_{i \Sp} \theta^{(a)} + \oC_{\Fp \Sp} \theta^{(b)} + \oC_{i \Fp} \theta^{(c)} + \oC_{\Fp \Sp} \theta^{(d)}\Big] [\rmd p_\Fp] [\rmd p_\Sp]  \, C_{\Fp \Sp, i} \; \omega^{\Fp i, \Sp i} \,.
\label{eq:Omega2_def}
\end{equation}  
The functions $\theta^{(a,b,c,d)}$ define the phase-space sector, restricting the number of possible singular collinear limits that one needs to consider. We note that the triple-collinear operator $C_{\Fp \Sp, i}$ does not act on the phase-space measure of the unresolved gluons, while operators $\oC_{ij}$ do. For further details, see the discussion below \eq\eqref{eq:Omega1} as well as Refs~\cite{Caola:2019nzf, Devoto:2023rpv, Devoto:2025kin}. 

All triple-collinear splitting functions were computed in Ref.~\cite{Catani:1999ss}.
They were integrated over the appropriate phase spaces in Ref.~\cite{Delto:2019asp}, for both initial- and final-state splittings.
However, as we already pointed in Ref.~\cite{Devoto:2025kin}, the integrated triple-collinear subtraction terms in \eq\eqref{eq_triple_collinear_lim_nf} 
differ slightly from the quantities computed in Ref.~\cite{Delto:2019asp} for some final-state splittings.
This happens because some integrals were calculated in this reference without the damping factors $\Delta^{(\Fp \Sp)}$, which produce additional energy-dependent weights in the triple-collinear limits.
Furthermore, in Ref.~\cite{Delto:2019asp}, all unresolved partons were energy-ordered when considering final-state splittings, whereas here we do not employ energy ordering for the combinations of unresolved partons that cannot produce double-soft singular limits.
These differences lead to very minor modifications to the soft and strongly-ordered collinear subtraction terms, while the unsubtracted integrals of the triple-collinear splitting functions over the phase space of the unresolved partons remain unchanged. 

We discuss the required changes in Appendix~\ref{app_details_TC}. Here, we sketch them briefly by considering the $g^* \to gq\qb$ splitting. Using the definitions of functions $\bbFLM$, we find that the following quantity is required 
\begin{equation}
\begin{aligned}
    &\; \sum_{n} \sum_{i\in\HP_{\rm f}} \llint \oS_{\Fp \Sp} \oS_{\Sp} \Omega_2^{(i)} \Delta^{(\Fp \Sp)} \Big[
    \delta_{ig} \THmn \FLMlo{\inF\inS}{\setf{N+2}{n}{}(\Fp_{(q_\sgflin},\Sp_{\qb_\sgflin)})} \\
    & + \sum_{\fgflin=1}^{\nf} \big( \delta_{i\qb_\fgflin} \FLMlo{\inF\inS}{\setf{N+2}{n}{}(\Fp_{q_\fgflin},\Sp_g)}
    + \delta_{iq_\fgflin} \FLMlo{\inF\inS}{\setf{N+2}{n}{}(\Fp_{\qb_\fgflin},\Sp_g)} \big) \Big] \rrint \,.
\end{aligned}
\label{eq_triple_coll_nf_g_gqqb}
\end{equation}
A similar but not identical quantity was computed in Ref.~\cite{Delto:2019asp}, where the energy-ordering $\THmn$ function was applied to all three $\FLM$ functions in \eq\eqref{eq_triple_coll_nf_g_gqqb}. 
One can show that the difference between the two calculations is given by  
\begin{equation}
    - \sum_{n} \sum_{i\in\HP_{\rm f}}\llint \THnm S_\Sp \Omega_2^{(i)} \Delta^{(\Fp \Sp)} \sum_{\fgflin=1}^{\nf} \Big[ \delta_{i\qb_\fgflin} \FLMlo{\inF\inS}{\setf{N+2}{n}{}(\Fp_{q_\fgflin},\Sp_g)}
    + \delta_{iq_\fgflin} \FLMlo{\inF\inS}{\setf{N+2}{n}{}(\Fp_{\qb_\fgflin},\Sp_g)} \Big] \rrint \,,
\label{eq_triple_coll_nf_g_gqqb_new}
\end{equation} 
where $\THnm = \Theta(E_\Sp-E_\Fp)$.
The important point is that the difference involves the soft operator $S_\yb$, which implies that a \emph{simplified} version of the triple-collinear splitting function is needed to compute the difference. 
We note that similar simplifications occur for all other splitting functions where modifications are required.

We find it convenient to define triple-collinear terms in the following way. 
For the final-state gluon splitting, we combine the $g^* \to ggg$ and $g^* \to gq\qb$ processes,  and write  
\begin{equation}
    \sum_{n} \sum_{i\in\HP_{\rm f}} \delta_{{[i\Fp\Sp]g}}\llint \oS_{\Fp \Sp} \oS_{\Sp} \Omega_2^{(i)} \Delta^{(\Fp \Sp)} \Big[\THmn \bbFLMlo{\inF\inS, \DS}{\Fp, \Sp} + \bbFLMlo{\inF\inS, \noDS}{\Fp, \Sp} \Big] \rrint
    = \sum_{n} \sum_{i \in \setf{N}{n}{}} \hspace{-2mm} \delta_{ig} \frac{\asbr^2}{\ep} \lint \GammaTC_{i,g} \, \FLMlo{\inF\inS}{\setf{N}{n}{}} \rint\, .
\label{eq_triple_coll_nf_GammaTC_g}
\end{equation} 
For the final-state quark splitting, $q^* \to qgg$ and $q^* \to qq'\qb'$ (with $q'$ running over all flavors including $q'=q$), we define
\begin{equation}
    \sum_{n} \sum_{i\in\HP_{\rm f}} \delta_{{[i\Fp\Sp]q}}\llint \oS_{\Fp \Sp} \oS_{\Sp} \Omega_2^{(i)} \Delta^{(\Fp \Sp)} \Big[\THmn \bbFLMlo{\inF\inS, \DS}{\Fp, \Sp} + \bbFLMlo{\inF\inS, \noDS}{\Fp, \Sp} \Big] \rrint
    = \sum_{n} \sum_{i \in \setf{N}{n}{}} \hspace{-2mm} \delta_{iq} \frac{\asbr^2}{\ep} \lint \GammaTC_{i,q} \, \FLMlo{\inF\inS}{\setf{N}{n}{}} \rint\,,
\label{eq_triple_coll_nf_GammaTC_q}
\end{equation}
where we have suppressed the flavor index of $q$ to simplify the notation. 
We note that the case $[i\Fp\Sp]=\qb$ is identical to the one just described, so that $\GammaTC_{i,\qb} \equiv \GammaTC_{i,q}$. 
The explicit expressions for all relevant integrated triple-collinear terms $\GammaTC_{i,f_i}$ are reported in the ancillary file \triplecollinearsplittings provided with this paper (see Table \ref{table_triplecollinearsplittings}).

Next, we consider the initial-state triple-collinear splittings, i.e.\ $i \in \{\inF,\inS\}$.
For definiteness, we focus on the case $i = \inF$; then, the splitting process we are interested in is $\inF \to [\inF \barFp \barSp]^* + \Fp +\Sp$. 
We have to sum over all possible types of partons $\Fp$ and $\Sp$, keeping $[\inF \barFp \barSp]$ fixed,  
and collect all contributions to the integrated triple-collinear splitting function $\CalPTC_{[\inF \barFp \barSp]\inF}$.
Accounting for all the relevant terms with appropriate parton permutations and averaging factors, we arrive at the following formula for the integrated initial-state triple-collinear subtraction terms 
\begin{equation}
    \sum_n \llint \oS_{\Fp \Sp} \oS_{\Sp} \Omega_2^{(\inF)} \Delta^{(\Fp \Sp)} \Big[\THmn \bbFLMlo{\inF\inS, \DS}{\Fp, \Sp} + \bbFLMlo{\inF\inS, \noDS}{\Fp, \Sp} \Big] \rrint 
    = \sum_x \sum_n \frac{\asbr^2}{\ep} \lint \CalPTC_{x\inF} \conv \FLMlo{x\inS}{\setf{N}{n}{}} \rint \,.
\label{eq_triple_collinear_lim_nf_i_equal_a}
\end{equation}
Similar to cases discussed earlier, the sum over $x$ runs over all parton types, and the triple-collinear splitting functions $\CalPTC_{x\inF}$ accommodate the allowed $a \to x$ splittings. Explicit expressions for these splitting functions are reported in the ancillary file \triplecollinearsplittings, see Table \ref{table_triplecollinearsplittings}.
We conclude this section by writing the final expression for \eq\eqref{eq_triple_collinear_lim_nf} that takes into account all initial- and final-state splittings 
\begin{equation}
\begin{aligned}
   & \sum_n \sum_{i\in\HP} \llint \oS_{\Fp \Sp} \oS_{\Sp} \Omega_2^{(i)} \Delta^{(\Fp \Sp)} \Big[\THmn \bbFLMlo{\inF\inS, \DS}{\Fp, \Sp} + \bbFLMlo{\inF\inS, \noDS}{\Fp, \Sp} \Big] \rrint \\
    & = \sum_x \sum_n \frac{\asbr^2}{\ep} \Big[\lint \CalPTC_{x\inF} \conv \FLMlo{x\inS}{\setf{N}{n}{}} \rint + \lint \FLMlo{\inF x}{\setf{N}{n}{}} \conv \CalPTC_{x\inS} \rint \Big]
    + \sum_n \hspace{-1mm} \sum_{i\in\setf{N}{n}{}} \hspace{-2mm} \frac{\asbr^2}{\ep} \lint \GammaTC_{i,f_i} \, \FLMlo{\inF\inS}{\setf{N}{n}{}} \rint \,.
\end{aligned}
\label{eq_triple_collinear_lim_nf_final}
\end{equation}
\section{Conclusions}
\label{sec:concl}

In this paper, we have presented a fully general derivation of the finite remainders of the integrated NNLO subtraction terms within the nested soft-collinear framework. Our results are applicable to \emph{arbitrary} processes with massless QCD partons at lepton and hadron colliders. 
In the process, we verify the analytic cancellation of all infrared divergences for infrared-safe observables in a process-independent manner, confirming the consistency of the subtraction scheme at NNLO. 

Our analysis focused on the process $pp \to X + N\text{ jets}$, where $X$ represents a generic color-singlet system, and the number of jets $N$ is a free parameter.  
The finite remainders for this process at NNLO QCD accuracy are given in Section~\ref{subsec_NNLO_nf_formulas}. We also discuss there the required (minor) modifications to make our results applicable to processes at lepton colliders.
 
The calculation of finite remainders at NNLO QCD accuracy for arbitrary processes required us to overcome two central challenges. 
The first one involved developing a systematic understanding of the singular limits of radiative scattering amplitudes at NNLO, including their interplay, and finding a suitable way to combine the corresponding subtraction terms with divergent contributions from virtual corrections. In Ref.~\cite{Devoto:2023rpv}, this problem was addressed by adopting, as the guiding principle, the idea of expressing the integrated subtraction terms in a form closely resembling that of Catani’s operator~\cite{Catani:1998bh}, which describes the $1/\ep$ poles from virtual corrections. 
This approach enabled the cancellation of such poles through the combination of process-independent soft and collinear operators. 
The analysis in Ref.~\cite{Devoto:2023rpv} showed that, in this way, the apparent mismatch between the ``simple" form of the (double-) virtual singularities and the increasing complexity of the real-virtual and double-real contributions can be resolved through a careful combination of contributions that share certain functional properties, such as color correlations. The method introduced in Ref.~\cite{Devoto:2023rpv} was proven to be valid in the specific case of quark-antiquark annihilation into an arbitrary number of gluons.

The second challenge concerned the combinatorial complexity of bookkeeping. This can be understood as the need for a systematic enumeration of all relevant partonic channels contributing to a given process at fixed perturbative order, as well as their modifications induced by soft and collinear limits. This step is essential to the subtraction procedure, as the cancellation of collinear singularities requires precise control over all initial- and final-state divergent components. To investigate how these cancellations occur, the method introduced in Ref.~\cite{Devoto:2023rpv} was extended in Ref.~\cite{Devoto:2025kin} to a more complex process that includes a quark in the Born-level final state. This study revealed important structural patterns, and highlighted the main combinatorial challenges. The main conclusion of Ref.~\cite{Devoto:2025kin} was that the nested soft-collinear subtraction scheme, in its revised formulation introduced in Ref.~\cite{Devoto:2023rpv}, is sufficiently robust to be applied to complex final states involving both quarks and gluons. It was also shown that physically relevant quantities, such as collinear anomalous dimensions, emerge naturally when singular configurations are combined prior to integration over the unresolved phase space. These findings indicated that the method not only facilitated the computation of NNLO corrections but also improved the physical transparency of the results.

Before concluding, we provide an overview of the main results of this paper.   
The master formulas for partonic and hadronic NLO cross sections within the nested soft-collinear subtraction scheme are given in Eqs.~\eqref{eq_dsigma_NLO_real_final_genX} and~\eqref{eq_dsigma_NLO_real_final_genX_had}, respectively. 
The NNLO master formula, the central result of this work, is distributed across Section~\ref{subsec_NNLO_nf_formulas}. While it mirrors the structure of the NLO expression, the additional complexity of the singular limits at this order leads to a more intricate decomposition. Thus, we divide the result into three parts: fully-resolved, single-unresolved, and double-unresolved, whose highest-multiplicity contributions involve $(N+2)$, $(N+1)$, and $N$ jets, respectively. Each part is made finite by virtue of dedicated subtraction terms.  
The fully-resolved term appears in Eq.~\eqref{eq_fully_unresolved_final_nf}. A thorough discussion of its numerical implementation goes beyond the scope of this paper. We simply comment that it requires enumerating all possible partonic channels and singular configurations for a given process, together with an appropriate sector-by-sector phase-space parametrization, similar in spirit to the  FKS construction at NLO \cite{Frixione:1995ms}. 
In the triple-collinear limit, this becomes subtle since new overlapping singularities appear, making further partitioning necessary. A suitable phase-space parametrization for this partitioning was presented in Ref.~\cite{Czakon:2010td}. Particular attention must be paid to spin correlations in triple-collinear splittings.
The single- and double-unresolved contributions arise from integrating the subtraction terms over the unresolved phase space. The single-unresolved contributions, residing in the $(N+1)$-parton phase space, are given in Eqs.~\eqref{eq_dsigmahat_su_inFsb},~\eqref{eq_dsigmahat_su_inSsb}, and~\eqref{eq_dsigmahat_SU_el_final}. The first two involve boosted kinematics, arising due to collinear radiation by the initial state partons, while the latter exhibits kinematics identical to that of the NLO real-emission contributions.

The double-unresolved term lives in the $N$-parton phase space and is decomposed into four parts, as summarized in \eq\eqref{eq_double_unresolved_final_nf}. The first contribution, \eq\eqref{eq_final_formula_DU_DB}, involves a double convolution and hence double-boosted kinematics. The single-boosted terms are given in Eqs.~\eqref{eq_dsigmahat_ab_du_inFsb_nf} and~\eqref{eq_dsigmahat_ab_du_inSsb_nf}, resembling the NLO boosted contributions. The elastic contribution appears in Eq.~\eqref{eq_dsigmahat_DU_el_final}. Despite containing the most intricate functions from the double-soft limits, it is the simplest to implement in a numerical code due to its LO-like kinematics.

In summary, building on the findings of Refs.~\cite{Devoto:2023rpv,Devoto:2025kin}, we have extended the approach developed in these references to arbitrary final states, completing the construction of a general NNLO subtraction scheme.
In this paper, we have addressed and resolved the two challenges described above, organizing all unresolved limits and their integrated counterparts into a compact expression. Its modular structure enables its application to any process with an $X + N\text{ jets}$ final state without requiring process-specific modifications. Furthermore, the formula separates final-state and initial-state contributions, making it directly applicable to non-hadronic collisions, such as $\ell^+\ell^- \to X + N\text{ jets}$, thereby extending its applicability beyond the current LHC program.
Future developments include the treatment of massive final-state quarks, embedding this scheme in parton-level event generators, and exploring its extension to N$^3$LO QCD accuracy.


\acknowledgments
K.M. and C.S-S. wish to thank the CERN Theoretical Physics Department for hospitality during work on this paper. 
The research of M.T. is supported by ``Karlsruhe School of Elementary and Astroparticle Physics'' (KSETA) through a grant from ``Deutscher Akademischer Austauschdienst'' (DAAD).
The research of K.M. and D.M.T. is supported by the Deutsche Forschungsgemeinschaft (DFG, German Research Foundation) under grant no.\ 396021762 - TRR 257. 
F.D.\ is supported by the United States Department of Energy, Contract DE-AC02-76SF00515.
R.R.\ is partially supported by the Italian Ministry of Universities and Research (MUR) through grant PRIN2022BCXSW9. 
This research was supported in part by grant NSF PHY-2309135 to the Kavli Institute for Theoretical Physics (KITP).


\newpage
\appendix
\section{Operations with lists}
\label{appendix_A}

In \Sec\ref{NLO_any_process} and \Sec\ref{sec_NNLO_general} we discussed how to derive the finite remainders of subtraction terms for NLO and NNLO QCD corrections, respectively, to  a generic process of the type  $pp \to X+ N \, {\rm jets}$.
Focusing on a given partonic channel $(\inF,\inS)$, we introduced abstract representations of the final-state $N$-parton configurations  $\setf{N}{n}{}$ where index $n$ parametrizes different final states at fixed $N$ and $(a,b)$.
We then examined how soft and collinear operators act on these lists, and showed that applying such operators to a complete set of  $\FLM$ functions -- each depending on a given higher-multiplicity final-state configuration -- leads to a \emph{complete} sum of $\FLM$ functions involving lists with lower final-state multiplicity. 

In this Appendix, we will construct such lists explicitly  for a toy example --  a process $pp \to  N ~ {\rm jets}$ in QCD with gluons and a single quark flavor. While this is only a toy example (and is not designed to accommodate processes including W bosons, for instance), we believe that an explicit construction of these lists is useful for checking the general statements made in the main body of the paper.

Given an initial state $(a,b)$, we can describe the corresponding final states with $N$ jets in terms of the lists
\begin{equation}
    \Born = \big(\setg{\Ng}, \setq{\Nq}, \setqb{\Nqb}\big) \,.
\label{eq_Bornf_def}
\end{equation}
Here, $\Ng$, $\Nq$, and $\Nqb$ denote the numbers of final-state gluons, quarks, and antiquarks, respectively.
These multiplicities must be such that the process $\inF\inS \to \calB$ is allowed in QCD. This implies that the possible combinations of gluons, quarks, and antiquarks for a given jet multiplicity $N$ depend on the initial-state baryon charge $\Qab$.  
In the toy model with a single quark flavor, the baryon charge of the initial partonic state can take values $\Qab = 0, \pm 1, \pm 2$, and we will now examine the possible final-state configurations corresponding to each of these cases.  

We begin by considering the initial states with the vanishing baryon charge, $\Qab = 0$. 
They are $(\inF,\inS) \in \{ (q,\qb), (\qb,q), (g,g)\}$.
Such initial-state configurations are compatible with $N$-gluon final states, as well as with all the partonic channels that can be obtained by replacing $2n$ of these gluons with $n$ $q\barq$ pairs.
Upon doing this, we find  the following possible final-state configurations 
\begin{align}
N~{\rm even}: \quad
\begin{pmatrix}
            \setg{N}   & \setq{0}             & \setqb{0}             \\
            \setg{N-2} & \setq{1}             & \setqb{1}             \\
            \setg{N-4} & \setq{2}             & \setqb{2}             \\
            \vdots     & \vdots               & \vdots                \\
            \setg{0}   & \setq{\frac{N}{2}}   & \setqb{\frac{N}{2}}   \\
        \end{pmatrix} 
        \qquad
        N~{\rm odd}: \quad
        \begin{pmatrix}
            \setg{N}   & \setq{0}             & \setqb{0}             \\
            \setg{N-2} & \setq{1}             & \setqb{1}             \\
            \setg{N-4} & \setq{2}             & \setqb{2}             \\
            \vdots     & \vdots               & \vdots                \\
            \setg{1}   & \setq{\frac{N-1}{2}}   & \setqb{\frac{N-1}{2}}   \\
        \end{pmatrix} ,
\label{eq_part_config_LO_symm}
\end{align}
which differ for even and odd $N$. 
We can unify the two cases by using the so-called floor function $\myfloor{x}$ which is defined as the largest integer $n$ such that $n \leq x$.
Then, it is easy to see that all $N$-jet partonic final states that can be produced from a $\Qab = 0$ initial state  are described by the following list
\begin{equation}
  \setf{N}{n}{0} = \big(\setg{N-2n}, \setq{n}, \setqb{n}\big) \,,
  \qquad
  n \in \left [0,\myfloor{\frac{N}{2}}
  \right ] \,, 
\label{eq_qqbggLO}
\end{equation}
where the superscript in $\setf{N}{n}{0}$ refers to the baryon charge. 
To give a concrete example of this notation, we note that in Ref.~\cite{Devoto:2023rpv} we considered the case $\setf{N}{0}{0}$, i.e.\ the process $q\qb \to \colsing + N\,g$.

We continue with the initial states that carry baryon charge $\Qab = \pm 1$, namely $(\inF,\inS) \in \{(g,q), (q, g), (g, \qb), \allowbreak (\qb, g) \}$. 
Considering for definiteness the $(g,q)$ initial state which has $\Qab = +1$, the allowed partonic final states are given by the following lists
\begin{align}
    N~{\rm even}: \quad
    \begin{pmatrix}
        \setg{N-1} & \setq{1}               & \setqb{0}             \\
        \setg{N-3} & \setq{2}               & \setqb{1}             \\
        \setg{N-5} & \setq{3}               & \setqb{2}             \\
        \vdots     & \vdots                 & \vdots                \\
        \setg{1}   & \setq{\frac{N}{2}}     & \setqb{\frac{N-2}{2}} \\
    \end{pmatrix} ,
    \qquad
    N~{\rm odd}: \quad
           \begin{pmatrix}
        \setg{N-1} & \setq{1}               & \setqb{0}             \\
        \setg{N-3} & \setq{2}               & \setqb{1}             \\
        \setg{N-5} & \setq{3}               & \setqb{2}             \\
        \vdots     & \vdots                 & \vdots                \\
        \setg{0}   & \setq{\frac{N+1}{2}}     & \setqb{\frac{N-1}{2}} \\
    \end{pmatrix} ,
\label{eq_nf_1_LO_list_QB_pm_1}
\end{align}
which can be summarized as 
\begin{equation}
    \setf{N}{n}{+1}  = \big(\setg{N-1-2n}, \setq{n+1}, \setqb{n}\big) \,,
    \qquad
    n \in \left [0,\myfloor{\frac{N-1}{2}} \right ] \,.
\label{eq_gqLO}
\end{equation}
We note that in Ref.~\cite{Devoto:2025kin} we analyzed the case $\setf{N}{0}{+1}$, i.e.\ the process $gq \to \colsing + (N-1)g + q$. 
The final states compatible with $\Qab = -1$ are obtained from \eq\eqref{eq_nf_1_LO_list_QB_pm_1} and \eq\eqref{eq_gqLO} by replacing quarks with antiquarks and vice versa.
  
Finally, initial states with baryon charge $\Qab = \pm 2$ include $(\inF,\inS) = \{(q,q), (\qb,\qb) \}$. 
Focusing for definiteness on the $\Qab = +2$ case, we find
\begin{align}
    N~{\rm even}: \quad
    \begin{pmatrix}
        \setg{N-2} & \setq{2}               & \setqb{0}             \\
        \setg{N-4} & \setq{3}               & \setqb{1}             \\
        \setg{N-6} & \setq{4}               & \setqb{2}             \\
        \vdots     & \vdots                 & \vdots                \\
        \setg{0}   & \setq{\frac{N+2}{2}}   & \setqb{\frac{N-2}{2}}   \\
    \end{pmatrix} ,
    \qquad
    N~{\rm odd}: \quad
\begin{pmatrix}
        \setg{N-2} & \setq{2}               & \setqb{0}             \\
        \setg{N-4} & \setq{3}               & \setqb{1}             \\
        \setg{N-6} & \setq{4}               & \setqb{2}             \\
        \vdots     & \vdots                 & \vdots                \\
        \setg{1}   & \setq{\frac{N+1}{2}}   & \setqb{\frac{N-3}{2}}   \\
    \end{pmatrix} .
\label{eq_nf_1_LO_list_QB_pm_2}
\end{align}
We write the above lists as 
\begin{equation}
    \setf{N}{n}{+2} =\big(\setg{N-2-2n}, \setq{n+2}, \setqb{n}\big) \,, n \in \left [0,\myfloor{\frac{N-2}{2}} \right ].
\label{eq_qqLO}
\end{equation}
The final state  with $\Qab=-2$ are easily obtained from the above formula by exchanging $\qarrowqb$. 

It is clear that all the final-state lists in Eqs.~\eqref{eq_qqbggLO}, \eqref{eq_gqLO}, and \eqref{eq_qqLO} can be unified by writing 
\begin{equation}
    \begin{aligned}
       \Qab \ge 0: & \quad \setf{N}{n}{\Qab} = \big(\setg{N-\Qab-2n}, \setq{n+\Qab}, \setqb{n} \big) \,, \\
        \Qab < 0: & \quad \setf{N}{n}{\Qab} = \big(\setg{N-\abs{\Qab}-2n}, \setq{n}, \setqb{n+\abs{\Qab}} \big) \,,
    \end{aligned}
    \qquad
    n \in [0, \xi(N,\Qab)] \,,
    ~
    \xi(N,\Qab) = \myfloor{\frac{N-|\Qab|}{2}} \,.
\label{eq_paramet_channels_nf_1}
\end{equation}
The upper bound in \eq\eqref{eq_paramet_channels_nf_1} satisfies $\xi(N,\Qab) \ge 0$, which implies $N \ge \abs{\Qab}$ for any $\setf{N}{n}{\Qab}$.

Through the parametrization in \eq\eqref{eq_paramet_channels_nf_1}, the sum over $n$ that appears in \eq\eqref{eq_LO_cross_section} becomes explicit. Indeed, fixing the initial state $(a,b)$  determines the baryon charge $\Qab$, which is used in the parametrization of \eq\eqref{eq_paramet_channels_nf_1} to write the lists $\setf{N}{n}{\Qab}$ enumerating  the possible final states. 
In this way, the sum over all possible configurations in \eq\eqref{eq_LO_cross_section} becomes a sum over the parameter $n \in [0, \xi(N,\Qab)]$.  
One can use the explicit parametrizations constructed in this appendix to check all the statements made in the main text of the paper concerning  the behavior of various lists in different limits and, especially, how they transform into each other. 
We emphasize that the example discussed in this appendix is limited, while the construction and formulas presented in the main body of the paper are valid for any number of quark flavors and are  also independent of the nature of the color singlet $X$.
\section{Details on the double-collinear contribution in the triple-collinear sector}
\label{app_details_DC}

In \eq\eqref{eq_SigmaDC_tc_nf_def} we introduced the double-collinear contribution which occurs when partons $\Fp$ and $\Sp$ simultaneously become collinear to one of the hard partons $i$. 
For convenience, we repeat the expression here
\begin{equation}
\begin{split}
    \SigmaDCtc & = \frac{1}{2} \sum_n \sum_{i \in \HP} \Big[\lint \oS_\Fp \oS_\Sp C_{i \Fp} C_{i \Sp} \Delta^{(\Fp\Sp)} \bbFLMlo{\inF\inS, \DS}{\Fp, \Sp} \rint
    + \lint \oS_\Sp \big(C_{i\Sp} C_{i\Fp} + C_{i\Fp} C_{i\Sp}\big) \Delta^{(\Fp \Sp)} \bbFLMlo{\inF\inS, \noDS}{\Fp, \Sp} \rint \Big] \,.
\end{split}
\label{eq_SigmaDC_tc_def_appendix}
\end{equation}
The functions $\bbFLM$ are defined in Eqs~(\ref{eq_FLM_DS_any_nf_def} -- \ref{eq_FLM_noDS_2_any_nf_def}).
We presented the result for $\Sigma^{\rm tc}_{\rm DC}$ in \eq\eqref{eq_SigmaDCtc_qqbgg_final_nf}
without providing a derivation. 
Here we explain in detail how to obtain it, considering the initial state first and then the final state.

Before proceeding with the calculation, it is useful to write \eq\eqref{eq_SigmaDC_tc_def_appendix} in a more suitable way.  
Specifically, we would like to write  the $\noDS$ contribution in the same way as the $\DS$ part, using a single pair of operators $C_{i \Fp} C_{i \Sp}$.  
This is achieved by first symmetrizing $\bbFLMlonf{1}{\inF\inS, \noDS}{\Fp,\Sp}$ with respect to $\xa$ and $\yb$.
We write  
\begin{equation}
\begin{aligned}
    &\;  \oS_\Sp \big( C_{i\Sp} C_{i\Fp} + C_{i\Fp} C_{i\Sp} \big) \Delta^{(\Fp \Sp)} \FLMlo{\inF\inS}{\setf{N+2}{n}{}(\Fp_q,\Sp_g)} \\
    & = \oS_\Fp \oS_\Sp \big( C_{i\Sp} C_{i\Fp} + C_{i\Fp} C_{i\Sp} \big) \Delta^{(\Fp \Sp)} \frac12 \Big[\FLMun{\inF\inS}{\setf{N+2}{n}{}(\Fp_q,\Sp_g)} + \FLMlo{\inF\inS}{\setf{N+2}{n}{}(\Fp_g,\Sp_q)}\Big] \,,
\end{aligned}
\label{eq_SigmaDC_tc_def_appendix_appo_1}
\end{equation}
and do the same for the $(\Fp_\qb, \Sp_g)$ term.  
We note that on the \rhs\ of \eq\eqref{eq_SigmaDC_tc_def_appendix_appo_1}, we have added the operator $\oS_\Fp$, which coincides with the identity operator for $\Fp \neq g$.  
Therefore, it can also be inserted into \eq\eqref{eq_SigmaDC_tc_def_appendix} for the remaining terms of $\bbFLMlonf{1}{\inF\inS, \noDS}{\Fp,\Sp}$, since in all such cases $\Fp$ is not a gluon.  
Finally, we symmetrize $\bbFLMlonf{2}{\inF\inS, \noDS}{\Fp, \Sp}$ by writing 
\begin{equation}
\begin{aligned}
        \bbFLMlonf{2}{\inF\inS, \noDS}{\Fp, \Sp}  
        & = \sum_n \! \sum_{\substack{\fgflin,\sgflin= 1 \\ \sgflin \neq \fgflin}}^{\nf} 
        \! \frac12 \FLMlo{\inF\inS}{\setf{N+2}{n}{}(\Fp_{q_\fgflin},\Sp_{q_\sgflin})}
        + \sum_n \! \sum_{\substack{\fgflin,\sgflin= 1 \\ \sgflin \neq \fgflin}}^{\nf} 
        \! \frac12 \FLMlo{\inF\inS}{\setf{N+2}{n}{}(\Fp_{\qb_\fgflin},\Sp_{\qb_\sgflin})} \\
        & + \sum_n \! \sum_{\substack{\fgflin,\sgflin= 1 \\ \sgflin \neq \fgflin}}^{\nf} 
        \! \frac12 \Big( \FLMlo{\inF\inS}{\setf{N+2}{n}{}(\Fp_{q_\fgflin},\Sp_{\qb_\sgflin})} + \FLMlo{\inF\inS}{\setf{N+2}{n}{}(\Fp_{\qb_\sgflin},\Sp_{q_\fgflin})} \Big)\,.
    \end{aligned}
    \label{eq_FLM_noDS_2_any_nf_appendix}    
\end{equation}
Again, in this case we can insert the operator $\oS_\Fp$ in front of $\bbFLMlonf{2}{\inF\inS, \noDS}{\Fp, \Sp}$, since it acts as the identity operator.  

After these changes, the function $\bbFLMlo{\inF\inS, \noDS}{\Fp, \Sp}$ becomes symmetric under the exchange $\Fp \leftrightarrow \Sp$, allowing us to replace the sum of operators $\oS_\Fp \oS_\Sp \big(C_{i\Sp} C_{i\Fp} + C_{i\Fp} C_{i\Sp}\big)$ with $2\,\oS_\Fp \oS_\Sp C_{i \Fp} C_{i \Sp}$.  
Combining this result with the $\DS$ contribution, we can write \eq\eqref{eq_SigmaDC_tc_def_appendix} as  
\begin{equation}
    \SigmaDCtc
    = \frac{1}{2} \sum_n \sum_{\Fp,\Sp} \sum_{i \in \HP} \lint \oS_\Fp \oS_\Sp C_{i \Fp} C_{i \Sp} \Delta^{(\Fp\Sp)} \FLMlo{\inF\inS}{\setf{N+2}{n}{}(\Fp, \Sp)} \rint \,,
\label{eq_SigmaDC_tc_nf_1_def_new_form}
\end{equation}
where $\Fp,\Sp \in \{g,q_\fgflin,\qb_\fgflin \}$ with $\fgflin = 1, \mydots\,, \nf$.  
\eq\eqref{eq_SigmaDC_tc_nf_1_def_new_form} is our starting point for our discussion.  
We note that definitions of several quantities that appear below are given in Appendix~A of Ref.~\cite{Devoto:2025kin}.  


\subsection*{Initial-state }
\label{app_ssec_DC_ISR}

We begin with the initial-state case, and  assume  $i=a$ for definiteness. 
 The action of the operators $\oS_\Sp \oS_\Fp C_{\inF \Fp} C_{\inF \Sp}$ on a generic function $\FLM$   gives 
\begin{equation}
\begin{aligned}
    &\; \oS_\Fp \oS_\Sp C_{\inF \Fp} C_{\inF \Sp} \Delta^{(\Fp\Sp)} \FLMlo{\inF\inS}{\Fp,\Sp}
    = \frac{\big(\gsb^2 \mu_0^{2\ep}\big)^2}{E_\Fp^2 E_\Sp^2 \, \rho_{\inF\Fp} \rho_{\inF\Sp}} \bigg[
    \frac{P_{[\inF\barSp]\inF,\rmi}(z_\Sp)}{(1-z_\Sp)^{-1}} \frac{P_{[\inF\barFp\barSp][\inF\barSp],\rmi}(z_\Fp)}{(1-z_\Fp)^{-1}} \, \frac{\FLM^{(z_\Fp z_\Sp \cdot [\inF\barFp\barSp]) b}}{z_\Fp z_\Sp} \\
    & - \delta_{[\inF\barSp]\inF} 2\T_{[\inF\barSp]}^2 \frac{P_{[\inF\barFp\barSp][\inF\barSp],\rmi}(w_\Fp)}{(1-w_\Fp)^{-1}} \, \frac{\FLM^{(w_\Fp\cdot [\inF\barFp\barSp]) b}}{w_\Fp}
    - \delta_{[\inF\barFp\barSp][\inF\barSp]} 2\T_{[\inF\barFp\barSp]}^2 \frac{P_{[\inF\barSp]\inF,\rmi}(z_\Sp)}{(1-z_\Sp)^{-1}} \, \frac{\FLM^{(z_\Sp\cdot [\inF\barFp\barSp]) b}}{z_\Sp} \\
    & + \delta_{[\inF\barSp]\inF} \delta_{[\inF\barFp\barSp][\inF\barSp]} 4\T_{[\inF\barSp]}^2 \T_{[\inF\barFp\barSp]}^2 \FLM^{[\inF\barFp\barSp]\inS} \bigg] \,,
    \end{aligned}
\label{eq_app_ISR_DC_first_step}
\end{equation}
where the $P_{\alpha\beta,\rmi}$ splitting functions are reported in \eq(A.12) of Ref.~\cite{Devoto:2025kin}, and
\begin{equation}
    \rho_{ij} = 1 - \cos\theta_{ij} \,,
    \qquad
    E_{\Fp,\Sp} \in [0,\Emax] \,,
    \qquad
    z_\Sp = 1 - \frac{E_\Sp}{E_\inF} \,,
    \qquad
    z_\Fp = 1 - \frac{E_\Fp}{z_\Sp E_\inF} \,,
    \qquad
    w_\Fp = 1 - \frac{E_\Fp}{E_\inF} \,.
\end{equation}
The arguments of the $\FLM$ functions on the right-hand side -- which we do not display for simplicity -- are lists that are obtained from those on the left-hand side by removing partons $ \xa$ and $\yb$. 

Since the integration over the angular phase space in \eq\eqref{eq_app_ISR_DC_first_step} is straightforward, we focus on the energy integrals. We will discuss the first term on the right-hand side, which is the most complicated.   
First, we change the integration variables from $(E_\Fp,E_\Sp)$ to $(z_\Sp, \xi = z_\Fp z_\Sp)$, where $z_\Sp \in [1-\Emax/E_\inF, 1]$ and $\xi \in [z_\Sp - \Emax/E_\inF, z_\Sp]$.  
Since $E_\inF < \Emax$, and the physical integration region of a splitting function requires $z_\Sp \ge 0$, we can restrict the values of this variable to $z_\Sp \in [0,1]$.  
In the case of the variable $\xi$, we have $z_\Sp \le 1$ and $\Emax/E_\inF > 1$, which implies $z_\Sp - \Emax/E_\inF < 0$.  
However, since the function $\FLM^{(\xi \cdot [\inF\barFp\barSp]) b}$ has no support for $\xi < 0$, we can assume $\xi \in [0,z_\Sp]$.  
The integral over the energies therefore reads 
\begin{equation}
\begin{split}
    & \int_{0}^{\Emax}  \dE_\Sp \, E_\Sp^{1-2\ep} \int_{0}^{\Emax} \dE_\Fp \, E_\Fp^{1-2\ep} \, 
    \frac{1}{E_\Fp^2 E_\Sp^2} \frac{P_{[\inF\barSp]\inF,\rmi}(z_\Sp)}{(1-z_\Sp)^{-1}} 
    \frac{P_{[\inF\barFp\barSp][\inF\barSp],\rmi}(z_\Fp)}{(1-z_\Fp)^{-1}} \, 
    \frac{\FLM^{(z_\Fp z_\Sp \cdot [\inF\barFp\barSp]) b}}{z_\Fp z_\Sp} \\
    & = E_\inF^{-4\ep} \int_{0}^{1} \frac{\dz_\Sp}{z_\Sp} \, z_\Sp^{-2\ep} \frac{P_{[\inF\barSp]\inF,\rmi}(z_\Sp)}{(1-z_\Sp)^{2\ep}}
    \int_{0}^{z_\Sp} \dxi \, \frac{P_{[\inF\barFp\barSp][\inF\barSp],\rmi}(\xi/z_\Sp)}{(1-\xi/z_\Sp)^{2\ep}} \,
    \frac{\FLM^{(\xi \cdot [\inF\barFp\barSp]) b}}{\xi} \,.
    \label{eq_b7}
\end{split}
\end{equation}
We change the order of integration and rename the integration variables as $\xi \mapsto z$ and $z_\Sp \mapsto t$ to obtain a convolution, and then  use the  definition of the splitting function $\calP_{\alpha\beta}^{(2)}$ given in Eq.~(A.15) of Ref.~\cite{Devoto:2025kin} to rewrite \eq\eqref{eq_b7} as
\begin{equation}
\begin{split}
    &  E_\inF^{-4\ep} \int_{0}^{1} \dz \Bigg[\bigg[\int_{z}^{1} \frac{\dt}{t} \, t^{-2\ep} \calP^{(2)}_{[\inF\barSp]\inF,\rmi}(t,E_\inF) \, \calP^{(2)}_{[\inF\barFp\barSp][\inF\barSp],\rmi}(z/t,E_\inF) \bigg]
    - \delta_{[\inF\barSp]\inF} \frac{\T_{[\inF\barSp]}^2}{\ep} e^{-2\ep L_\inF} \calP^{(2)}_{[\inF\barFp\barSp][\inF\barSp]}(z,E_\inF) \\
    & - \delta_{[\inF\barFp\barSp][\inF\barSp]} \frac{\T_{[\inF\barFp\barSp]}^2}{\ep} e^{-2\ep L_\inF} z^{-2\ep} \calP^{(2)}_{[\inF\barSp]\inF}(z,E_\inF)
    + \delta_{[\inF\barSp]\inF} \delta_{[\inF\barFp\barSp][\inF\barSp]} \frac{\T_{[\inF\barSp]}^2 \T_{[\inF\barFp\barSp]}^2}{\ep^2} e^{-4\ep L_\inF} \delta(1-z)\Bigg] \frac{\FLM^{(z \cdot [\inF\barFp\barSp]) b}}{z} \,,
\end{split}
\label{eq_app_DC_I}
\end{equation}
where $L_a = \log(\Emax/E_a)$.
Treating the remaining terms in \eq\eqref{eq_app_ISR_DC_first_step} in the same manner, and including the integral over the angular phase space leads to the following expression 
\begin{equation}
\begin{split}
    &\; \lint \oS_\Fp \oS_\Sp C_{\inF \Fp} C_{\inF \Sp} \Delta^{(\Fp\Sp)} \FLMlo{\inF\inS}{\Fp,\Sp} \rint \\
    & = \frac{\asbr^2}{\ep^2} \bigg[\left(\frac{2E_\inF}{\mu}\right)^{\!\!-2\ep} \frac{\Gamma^2(1-\ep)}{\Gamma(1-2\ep)}\bigg]^2 \int_{0}^{1} \dz \, \Lint \bigg[\bigg(\int_{z}^{1} \frac{\dt}{t} \, t^{-2\ep} \calP^{(2)}_{[\inF\barSp]\inF,\rmi}(t,E_\inF) \, \calP^{(2)}_{[\inF\barFp\barSp][\inF\barSp],\rmi}(z/t,E_\inF) \bigg) \\
    & + \delta_{[\inF\barFp\barSp][\inF\barSp]} \frac{\T_{[\inF\barFp\barSp]}^2}{\ep} e^{-2\ep L_\inF}
    (1-z^{-2\ep}) \calP^{(2)}_{[\inF\barSp]\inF}(z,E_\inF) \bigg] \frac{\FLM^{(z \cdot [\inF\barFp\barSp]) b}}{z} \Rint \,.
\end{split}
\label{eq_app_DC_I_II_III_IV}
\end{equation}

The next step consists of rewriting the splitting functions $\calP^{(2)}_{\alpha\beta}$ in \eq\eqref{eq_app_DC_I_II_III_IV} using  \eq(A.16) of Ref.~\cite{Devoto:2023rpv} and expressing $\calP^{(2)}_{\alpha\beta}$ in terms of the quantities $\CalPgen_{\alpha\beta}$ and $\Gamma_{\inF,\alpha}$.
From this, we obtain
\begin{align}
    &\; \lint \oS_\Fp \oS_\Sp C_{\inF \Fp} C_{\inF \Sp} \Delta^{(\Fp\Sp)} \FLMlo{\inF\inS}{\Fp,\Sp} \rint
    = \frac{\asbr^2}{\ep^2} \int_{0}^{1} \dz \, \Lint \bigg\{\PgenoxPgen{[\inF\barFp\barSp][\inF\barSp]}{[\inF\barSp]\inF}(z,E_\inF)
    + \delta_{[\inF\barSp]\inF} \Gamma_{\inF,[\inF\barSp]} \CalPgen_{[\inF\barFp\barSp][\inF\barSp]}(z,E_\inF) \notag \\
    & + \delta_{[\inF\barFp\barSp][\inF\barSp]} z^{-2\ep} \Gamma_{\inF,[\inF\barFp\barSp]} \CalPgen_{[\inF\barSp]\inF}(z,E_\inF)
    + \delta_{[\inF\barSp]\inF} \delta_{[\inF\barFp\barSp][\inF\barSp]} \Gamma_{\inF,[\inF\barSp]} \Gamma_{\inF,[\inF\barFp\barSp]} \delta(1-z) 
    \label{eq_app_DC_I_II_III_IV_appo} \\
    & - \delta_{[\inF\barFp\barSp][\inF\barSp]} \frac{\T_{[\inF\barFp\barSp]}^2}{\ep} e^{-2\ep L_\inF} (1-z^{-2\ep}) 
    \Big[\Gamma_{\inF,[\inF\barSp]} \delta(1-z) + \CalPgen_{[\inF\barSp]\inF}(z,E_\inF) \Big] \bigg(\frac{2E_\inF}{\mu}\bigg)^{\!\! -2\ep} \frac{\Gamma^2(1-\ep)}{\Gamma(1-2\ep)} \bigg\} \frac{\FLM^{(z \cdot [\inF\barFp\barSp]) b}}{z} \Rint \,. \notag
\end{align}
In Eq.~(\ref{eq_app_DC_I_II_III_IV_appo}) we have defined the  $\barotimes$ convolution as
\begin{equation}
    \big[f_{xy} \,\barotimes\, g_{y\inF}](z,E_\inF)
    \eqdef
    \int_{z}^{1} \frac{\dt}{t} \, f_{xy}(z/t,E_\inF) \times t^{-2\ep} g_{y\inF}(t,E_\inF) \,,
\label{eq_barotimes_conv_def}
\end{equation}
which differs from the one used in Ref.~\cite{Devoto:2025kin} by the order in which the convoluted functions are written. This ensures that the indices appear in  the same order as in the terms  arising from the renormalization of the pdfs.  
The long expression in the curly brackets in \eq\eqref{eq_app_DC_I_II_III_IV_appo} can be simplified by noting that 
\begin{equation}
    z^{-2\ep} \Gamma_{\inF,[\inF\barFp\barSp]} - \frac{\T_{[\inF\barFp\barSp]}^2}{\ep} e^{-2\ep L_\inF} (1-z^{-2\ep}) \bigg(\frac{2E_\inF}{\mu}\bigg)^{\!\! -2\ep} \frac{\Gamma^2(1-\ep)}{\Gamma(1-2\ep)}
    = \Gamma_{z \cdot \inF, [\inF\barFp\barSp]} \,,
    \;\;\;\; \text{where} \;\;\;\;
    \Gamma_{z \cdot \inF, \alpha} \eqdef \Gamma_{\inF, \alpha} \big|_{E_\inF \mapsto z E_\inF} \,,
\end{equation}
and also that $(1-z^{-2\ep}) \delta(1-z) F(z) = 0$. Employing these relations, we obtain
\begin{equation}
\begin{split}
    &\; \lint \oS_\Fp \oS_\Sp C_{\inF \Fp} C_{\inF \Sp} \Delta^{(\Fp\Sp)} \FLMlo{\inF\inS}{\Fp,\Sp} \rint
    = \frac{\asbr^2}{\ep^2} \! \int_{0}^{1} \! \dz \, \Lint \bigg\{\PgenoxPgen{[\inF\barFp\barSp][\inF\barSp]}{[\inF\barSp]\inF}(z,E_\inF)
    + \delta_{[\inF\barSp]\inF} \Gamma_{\inF,[\inF\barSp]} \CalPgen_{[\inF\barFp\barSp][\inF\barSp]}(z,E_\inF) \\
    & + \delta_{[\inF\barFp\barSp][\inF\barSp]} \Gamma_{z\cdot\inF,[\inF\barFp\barSp]} \CalPgen_{[\inF\barSp]\inF}(z,E_\inF)
    + \delta_{[\inF\barSp]\inF} \delta_{[\inF\barFp\barSp][\inF\barSp]} \Gamma_{\inF,[\inF\barSp]} \Gamma_{\inF,[\inF\barFp\barSp]} \delta(1-z) \bigg\} \frac{\FLM^{(z \cdot [\inF\barFp\barSp]) b}}{z} \Rint \,.
\end{split}
\label{eq_app_DC_Cam_Can}
\end{equation}
Finally, we include the sum over the unresolved partons $\Fp,\Sp$, which allows us to write the $i=\inF$ term of \eq\eqref{eq_SigmaDC_tc_nf_1_def_new_form} as 
\begin{equation}
\begin{split}
    & \frac{1}{2} \sum_n \sum_{\Fp\Sp} \lint \oS_\Fp \oS_\Sp C_{\inF \Fp} C_{\inF \Sp} \Delta^{(\Fp\Sp)} \FLMlo{\inF\inS}{\setf{N+2}{n}{}(\Fp, \Sp)} \rint 
    = \asbr^2 \sum_n \Lint \frac{\Gamma_{\inF,f_\inF}^2}{2\ep^2} \, \FLMlo{\inF\inS}{\setf{N}{n}{}} \Rint \\
    & + \frac{\asbr^2}{2\ep^2} \sum_{x} \sum_n \Lint \bigg[\sum_{y} \PgenoxPgen{xy}{y\inF} + G_{x\inF}\bigg] \conv \FLMlo{x\inS}{\setf{N}{n}{}} \Rint 
    + \frac{\asbr^2}{\ep} \sum_{x} \sum_n \Lint \CalPgen_{x\inF} \conv \bigg[\frac{\Gamma_{\inF,x}}{\ep} \, \FLMlo{x\inS}{\setf{N}{n}{}}\bigg] \Rint \,,
\end{split}
\label{eq_DC_ISR_leg_a_final}
\end{equation}
where
\begin{equation}
    G_{x\inF}(z,E_\inF)
    \equiv \sum_{y} \Big[\delta_{\inF y} \Gamma_{\inF,y} \CalPgen_{xy}(z,E_\inF) - \delta_{xy} \Gamma_{z\cdot \inF,x} \CalPgen_{y\inF}(z,E_\inF) \Big] 
    = \big[\Gamma_{\inF,f_\inF} - \Gamma_{z\cdot\inF,x}] \, \CalPgen_{x\inF} (z,E_\inF) \,.
\label{eq_G_ISR_def}
\end{equation}
Note that we already encountered the function $G$ in Eq.~(5.15) of Ref.~\cite{Devoto:2025kin} in the case of identical flavors, i.e.\ $x=\inF$.  
\eq\eqref{eq_G_ISR_def} generalizes the earlier definition to the case of different flavors, $x \neq \inF$.
\eq\eqref{eq_DC_ISR_leg_a_final} corresponds  to the initial-state contribution in \eq\eqref{eq_SigmaDCtc_qqbgg_final_nf}.


\subsection*{Final-state}
\label{app_ssec_DC_FSR}

We proceed with the calculation of the expression in \eq\eqref{eq_SigmaDC_tc_nf_1_def_new_form} for the final-state case, i.e.\ $i\in\setf{N+2}{n}{}(\Fp,\Sp)$.  
We first consider the action of the operator $C_{i\Sp}$.  
It yields  
\begin{equation}
\begin{split}
    & \lint \oS_\Sp \oS_\Fp C_{i\Fp} C_{i\Sp} \Delta^{(\Fp\Sp)} \FLMun{\inF\inS}{...\,, i, ...}{\Fp,\Sp} \rint \\
    & = \frac{\asbr}{\ep} \frac{\Gamma^2(1-\ep)}{\Gamma(1-2\ep)} \Lint \oS_\Fp C_{[i\Sp]\Fp} \Delta^{(\Fp)} (2E_{[i\Sp]}/\mu)^{-2\ep} \gamma^{22}_{z,[i\Sp] \to i\Sp}(L_{[i\Sp]}) \, \FLMun{\inF\inS}{...\,, [i\Sp], ...}{\Fp} \Rint \,.
\end{split}
\label{eq_app_FSR_DC_starting_first_step}
\end{equation}  
\eq\eqref{eq_app_FSR_DC_starting_first_step} has the form of an  NLO-like expression with the second collinear operator yet to be applied.   
The quantities $\gamma^{22}$ are defined in Eq.~(A.20) of Ref.~\cite{Devoto:2025kin}.  
They depend on the logarithm $L_{[i\Sp]} = \log(\Emax/E_{[i\Sp]})$, where $E_{[i\Sp]} = E_i + E_\Sp$.   
Furthermore, in \eq\eqref{eq_app_FSR_DC_starting_first_step} we have left the lists of resolved partons in the arguments of the two $\FLM$ functions implicit.

Before applying the second collinear operator, it is convenient to rewrite  $\gamma^{22}$ in a way that separates the terms that depend on the energy $E_{[i\Sp]}$ from those that do not. Substituting  
\begin{equation}
    \gamma^{22}_{z,[i\Sp] \to i\Sp}(L_{[i\Sp]})
    = \bigg(\gamma^{22}_{z,[i\Sp] \to i\Sp}(0) + \delta_{[i\Sp]i} \frac{\T_{[i\Sp]}^2}{\ep} \bigg)
    - \delta_{[i\Sp]i} \frac{\T_{[i\Sp]}^2}{\ep} \bigg(\frac{\Emax}{E_{[i\Sp]}}\bigg)^{\!\! -2\ep} \,,
\label{eq_app_FSR_DC_starting_first_step_appo}
\end{equation}  
into \eq\eqref{eq_app_FSR_DC_starting_first_step}, we obtain  
\begin{equation}
\begin{split}
    \lint \oS_\Sp \oS_\Fp C_{i\Fp} C_{i\Sp} \Delta^{(\Fp\Sp)} \FLMun{\inF\inS}{...\,, i, ...}{\Fp,\Sp} \rint     %
    & = \frac{\asbr}{\ep} \frac{\Gamma^2(1-\ep)}{\Gamma(1-2\ep)} \Lint \oS_\Fp C_{[i\Sp]\Fp} \Delta^{(\Fp)} \bigg[(2E_{[i\Sp]}/\mu)^{-2\ep} \bigg(\gamma^{22}_{z,[i\Sp] \to i\Sp}(0) \\ & + \delta_{[i\Sp]i} \frac{\T_{[i\Sp]}^2}{\ep} \bigg) 
    - (2\Emax/\mu)^{-2\ep} \delta_{[i\Sp]i} \frac{\T_{[i\Sp]}^2}{\ep} \bigg] \FLMun{\inF\inS}{...\,, [i\Sp], ...}{\Fp} \Rint \,.
\end{split}
\label{eq_app_FSR_DC_starting_second_step}
\end{equation}  
At this point, we can apply the collinear operator $C_{[i\Sp]\Fp}$, which identifies the clustered parton $[i\Fp\Sp]$ with energy $E_{[i\Fp\Sp]} = E_\Fp/(1-z) =  E_{[i\Sp]}/z$, where $E_{[i\Fp\Sp]} = E_i + E_\Fp + E_\Sp$.  
Under its action, the term in \eq\eqref{eq_app_FSR_DC_starting_second_step} multiplying $(2E_{[i\Sp]}/\mu)^{-2\ep}$ gives rise to an anomalous dimension of the type $\gamma^{42}$, rather than the NLO-like $\gamma^{22}$, due to the additional factor $z^{-2\ep}$ contained in $E_{[i\Sp]}^{-2\ep} = z^{-2\ep} E_{[i\Fp\Sp]}^{-2\ep}$.  
The term in \eq\eqref{eq_app_FSR_DC_starting_second_step} that does not contain $E_{[i\Sp]}$, on the other hand, leads to the usual anomalous dimension $\gamma^{22}$.  
Therefore, we find  
\begin{equation}
\begin{split}
    &\; \lint \oS_\Sp \oS_\Fp C_{i\Fp} C_{i\Sp} \Delta^{(\Fp\Sp)} \FLMun{\inF\inS}{...\,, i, ...}{\Fp,\Sp} \rint
    = \frac{\asbr^2}{\ep^2} \Lint \bigg[\bigg(\frac{2E_{[i\Fp\Sp]}}{\mu}\bigg)^{\!\! -2\ep} \frac{\Gamma^2(1-\ep)}{\Gamma(1-2\ep)} \bigg]^2
    \bigg[\bigg(\gamma^{22}_{z,[i\Sp] \to i\Sp}(0) + \delta_{[i\Sp]i} \frac{\T_{[i\Sp]}^2}{\ep}\bigg) \\ 
    & \times \gamma^{42}_{z,[i\Fp\Sp] \to [i\Sp]\Fp}(L_{[i\Fp\Sp]})
    - \delta_{[i\Sp]i} \frac{\T_{[i\Sp]}^2}{\ep} e^{-2\ep L_{[i\Fp\Sp]}} \gamma^{22}_{z,[i\Fp\Sp] \to [i\Sp]\Fp}(L_{[i\Fp\Sp]}) \bigg] \FLMlo{\inF\inS}{...\,, [i\Fp\Sp], ...} \Rint \,.
\end{split}
\label{eq_app_FSR_DC_starting_third_step}
\end{equation} 

Finally, to simplify the expression in \eq\eqref{eq_app_FSR_DC_starting_third_step}, we introduce a quantity which  is the final-state counterpart of the $G$-function defined in \eq\eqref{eq_G_ISR_def}.  
It reads\footnote{The final-state $G$-function has already appeared in Eq.~(5.21) of Ref.~\cite{Devoto:2025kin}.}  
\begin{equation}
\begin{split}
    \GFSR{[i\Fp\Sp]}{f(z), [i\Sp] \to i\Sp \hspace{5mm}}{\tildef(z), [i\Fp\Sp] \to [i\Sp]\Fp} 
    & = \!\bigg[\bigg(\frac{2E_{[i\Fp\Sp]}}{\mu}\bigg)^{\!\! - 2\ep} \frac{\Gamma^2(1-\ep)}{\Gamma(1-2\ep)} \bigg]^2 
    \bigg[\gamma^{22}_{f(z), [i\Sp] \to i \Sp}(L_{[i\Fp\Sp]}) 
    + \delta_{[i\Sp]i} \frac{\T_{[i\Sp]}^2}{\ep} e^{-2\ep L_{[i\Fp\Sp]}} \bigg] \\ 
    & \times \bigg [\gamma^{42}_{\tildef(z),[i\Fp\Sp] \to [i\Sp]\Fp}(L_{[i\Fp\Sp]}) -  \gamma^{22}_{\tildef(z),[i\Fp\Sp] \to [i\Sp]\Fp}(L_{[i\Fp\Sp]})\bigg ] \,.
\end{split}
\label{eq_GFSR_general_def}
\end{equation}
Using this quantity, we can finally rewrite \eq\eqref{eq_app_FSR_DC_starting_first_step} as
\begin{equation}
\begin{split}
    \lint \oS_\Sp \oS_\Fp C_{i\Fp} C_{i\Sp} \Delta^{(\Fp\Sp)} \FLMun{\inF\inS}{...\,, i, ...}{\Fp,\Sp} \rint
    & = \frac{\asbr^2}{\ep^2} \llint \Big[\Gamma_{[i\Fp\Sp],[i\Sp] \to i\Sp} \, \Gamma_{[i\Fp\Sp],[i\Fp\Sp] \to [i\Sp]\Fp} \\
    & + \GFSR{[i\Fp\Sp]}{z, [i\Sp] \to i\Sp \hspace{6mm}}{z, [i\Fp\Sp] \to [i\Sp]\Fp} \Big] \FLMlo{\inF\inS}{...\,, [i\Fp\Sp], ...} \rrint \,.
\label{eq_app_FSR_DC_starting_fourth_step}
\end{split}
\end{equation}

The expression in \eq\eqref{eq_app_FSR_DC_starting_fourth_step} can now be inserted into \eq\eqref{eq_SigmaDC_tc_nf_1_def_new_form}.  
Summing over the indices $i$, $\Fp$, and $\Sp$, and relabeling the clustered parton $[i\Fp\Sp]$ as $i$, we obtain the final result
\begin{equation}
\begin{split}
    &\; \frac{1}{2} \sum_n \sum_{\Fp\Sp} \sum_{i \in \HPf} \lint \oS_\Fp \oS_\Sp C_{i \Fp} C_{i \Sp} \Delta^{(\Fp\Sp)} \FLMlo{\inF\inS}{\setf{N+2}{n}{}(\Fp, \Sp)} \rint \\
    & = \frac{\asbr^2}{2\ep^2} \sum_n \! \sum_{i\in\setf{N}{n}{}} \!\!\! \Lint \Bigg[
    \delta_{ig} \Big[\Gamma_{i,g}^2 + 2 \nf \,\Gamma_{i,g\to q\qb}\big(\Gamma_{i,q} - \Gamma_{i,g}\big)
    + \GFSR{i}{g \hspace{8mm}}{z,g\to gg} + 2 \nf \,\GFSR{i}{q \hspace{8mm}}{z,g\to q\qb} \Big] \\
    & + \sum_{\fgflin=1}^{\nf} \big(\delta_{iq_\fgflin} + \delta_{i\qb_\fgflin}\big) 
    \Big[\Gamma_{i,q}^2 + \Gamma_{i,q\to gq}\big(\Gamma_{i,g} - \Gamma_{i,q}\big)
    + \GFSR{i}{q \hspace{8mm}}{z,q\to qg} + \GFSR{i}{g \hspace{8mm}}{z,q\to gq} \Big] \Bigg] \FLMlo{\inF\inS}{\setf{N}{n}{}} \Rint \,.
\end{split}
\label{eq_app_FSR_DC_final}
\end{equation}
We note that, in order to obtain \eq\eqref{eq_app_FSR_DC_final}, we used the two properties of the $G$-functions described in Eqs (5.22, 5.23) of Ref.~\cite{Devoto:2025kin}. 
Combining  \eq\eqref{eq_app_FSR_DC_final}  with \eq\eqref{eq_DC_ISR_leg_a_final} and the analogous result for the initial-state parton $\inS$ gives the final formula for the triple-collinear partitions, shown in  \eq\eqref{eq_SigmaDCtc_qqbgg_final_nf}.
\section{ Integrated triple-collinear counterterms}
\label{app_details_TC}

In Section~\ref{sec:TC} we explained why the integrated triple-collinear terms computed in Ref.~\cite{Delto:2019asp} need to be modified to accommodate our  current setup. The goal of this Appendix is to make the relationship between the two results explicit. 

The required triple-collinear contributions read (c.f. Eq.~\eqref{eq_triple_collinear_lim_nf}) 
\begin{equation}
    \sum_n \sum_{i\in\HP} \llint \oS_{\Fp \Sp} \oS_{\Sp} \Omega_2^{(i)} \Delta^{(\Fp \Sp)} \Big[\THmn \bbFLMlo{\inF\inS, \DS}{\Fp, \Sp} + \bbFLMlo{\inF\inS, \noDS}{\Fp, \Sp} \Big] \rrint \,,
\label{eq_triple_collinear_lim_nf_app}
\end{equation}  
where the operator $\Omega_2^{(i)}$ is defined as  
\begin{equation}
    \Omega_{2}^{(i)} = \Big[\oC_{i \Sp} \theta^{(a)} + \oC_{\Fp \Sp} \theta^{(b)} + \oC_{i \Fp} \theta^{(c)} + \oC_{\Fp \Sp} \theta^{(d)}\Big] [\rmd p_\Fp] [\rmd p_\Sp]  \, C_{\Fp \Sp, i} \; \omega^{\Fp i, \Sp i} \, .
\end{equation}  
The functions $\bbFLMlo{}{\Fp, \Sp}$ in \eq\eqref{eq_triple_collinear_lim_nf_app} are given in Eqs~(\ref{eq_FLM_DS_any_nf_def} -- \ref{eq_FLM_noDS_2_any_nf_def}).  

We first consider the splitting $g^* \to ggg$.  
In this case, only the function $\FLMlo{\inF\inS}{\setf{N+2}{n}{}(\Fp_g,\Sp_g)}$ in \eq\eqref{eq_FLM_DS_any_nf_def} contributes, and we further require that the hard parton $i$ is a final-state gluon. 
In this case, the unresolved partons $\Fp$ and $\Sp$ are energy-ordered, and the damping factor $\Delta^{(\Fp\Sp)}$ is included. 
This is identical to what has been done in Ref.~\cite{Delto:2019asp}, so that no modifications are needed. Hence, we write 
\begin{equation}
    \sum_n \hspace{-1mm} \sum_{i \in \HP_{\rm f}} \delta_{ig} \llint \oS_{\Fp \Sp} \oS_{\Sp} \Omega_2^{(i)} \Delta^{(\Fp \Sp)} \THmn \FLMlo{\inF\inS}{\setf{N+2}{n}{}(\Fp_g, \Sp_g)} \rrint
    = \sum_n \sum_{i \in \setf{N}{n}{}} \hspace{-2mm} \delta_{ig} \frac{\asbr^2}{\ep} \lint \GammaTC_{i,g\to ggg} \, \FLMlo{\inF\inS}{\setf{N}{n}{}} \rint \,. 
\label{eq_triple_coll_nf_g_ggg}
\end{equation}  
Here, $\HP_{\rm f}$ is the list of the hard final-state particles in $\setf{N+2}{n}{}(\Fp_g, \Sp_g)$,   
\begin{equation}
    \GammaTC_{i,g\to ggg} = (E_i/\mu)^{-4\ep} \ep \, \FSR{5}(E_i) \,,
\label{eq_GammaTC_gTOggg}
\end{equation}  
and $\FSR{5}$ is defined in Table 2 of Ref.~\cite{Delto:2019asp}.

Next, we consider the splitting $g^* \to gq_\sgflin\qb_\sgflin$. As already discussed in \Sec\ref{sec:TC}, this triple-collinear final-state contribution reads
\begin{equation}
\begin{aligned}
    & \sum_n \sum_{i\in\HP_{\rm f}} \llint \oS_{\Fp \Sp} \oS_{\Sp} \Omega_2^{(i)} \Delta^{(\Fp \Sp)} \sum_{\sgflin=1}^{\nf} \Big[
    \delta_{ig} \THmn \FLMlo{\inF\inS}{\setf{N+2}{n}{}(\Fp_{(q_\sgflin},\Sp_{\qb_\sgflin)})}
    + \delta_{i\qb_\sgflin} \FLMlo{\inF\inS}{\setf{N+2}{n}{}(\Fp_{q_\sgflin},\Sp_g)} \\
    & + \delta_{iq_\sgflin} \FLMlo{\inF\inS}{\setf{N+2}{n}{}(\Fp_{\qb_\sgflin},\Sp_g)} \Big] \rrint
    = \sum_n \sum_{i \in \setf{N}{n}{}} \hspace{-2mm} \delta_{ig} \frac{\asbr^2}{\ep} \lint 2 \nf \GammaTC_{i,g\to gq\qb} \,  \FLMlo{\inF\inS}{\setf{N}{n}{}} \rint \,.
\end{aligned}
\label{eq_triple_coll_nf_g_gqqb_app}
\end{equation}  
The integrated triple-collinear function $\GammaTC_{i,g\to gq\qb}$ is written as 
\begin{equation}
    \GammaTC_{i,g\to gq\qb} = (E_i/\mu)^{-4\ep} \ep \, \FSRtilde{4}(E_i) \,, 
\label{eq_GammaTC_gTOgqqb_app}
\end{equation}
where the function $\FSRtilde{4}$ in \eq\eqref{eq_GammaTC_gTOgqqb_app} is related, but is not identical to the function $\FSR{4}$  in Ref.~\cite{Delto:2019asp}.
The difference is due to the fact that in the current framework we do not employ the energy-ordering in all $\FLM$ functions in \eq\eqref{eq_triple_coll_nf_g_gqqb_app}. To match the two calculations, we multiply the second and third terms on the \rhs\ of the first line of \eq\eqref{eq_triple_coll_nf_g_gqqb_app} with $1 = \THmn+\THnm$. 
This allows us to identify the quantity computed in Ref.~\cite{Delto:2019asp} and an additional term that reads
\begin{equation}
    - \sum_{n} \sum_{i\in\HP_{\rm f}}\llint \THnm S_\Sp \Omega_2^{(i)} \Delta^{(\Fp \Sp)} \sum_{\sgflin=1}^{\nf} \Big[ \delta_{i\qb_\sgflin} \FLMlo{\inF\inS}{\setf{N+2}{n}{}(\Fp_{q_\sgflin},\Sp_g)}
    + \delta_{iq_\sgflin} \FLMlo{\inF\inS}{\setf{N+2}{n}{}(\Fp_{\qb_\sgflin},\Sp_g)} \Big] \rrint \, .
\label{eq_triple_coll_nf_g_gqqb_app_2}
\end{equation} 
Compared to the integral of the triple-collinear splitting function, this term is simpler to compute because it contains an operator $S_\Sp$. 
Combining $\FSR{4}$ from Ref.~\cite{Delto:2019asp} with the above term, we obtain the function $\FSRtilde{4}$ that appeared in \eq\eqref{eq_GammaTC_gTOgqqb_app}.

The two contributions in Eqs (\ref{eq_GammaTC_gTOggg}, \ref{eq_GammaTC_gTOgqqb_app}) can be combined in a single term (cf.\ \eq\eqref{eq_triple_coll_nf_GammaTC_g})
\begin{equation}
    \GammaTC_{i,g} = \GammaTC_{i,g\to ggg} + 2\nf\GammaTC_{i,g\to g q\qb} \,,
\end{equation}
which defines the triple-collinear subtraction term for a final-state gluon leg. 

Cases where the mother parton is a quark or an antiquark can be analyzed in a similar manner. For definiteness, we consider the quark splittings $q_\fgflin^* \to q_\fgflin gg$ and $q_\fgflin^* \to q_\fgflin q_\sgflin \qb_\sgflin$, where $q_\sgflin$ can have any flavor, including that of $q_\fgflin$. We can write the term due to the first splitting as
\begin{equation}
\begin{aligned}
    & \sum_n \sum_{i\in\HP_{\rm f}} \llint \oS_{\Fp \Sp} \oS_{\Sp} \Omega_2^{(i)} \Delta^{(\Fp \Sp)} \sum_{\fgflin=1}^{\nf} \Big[
    \delta_{iq_\fgflin} \THmn \FLMlo{\inF\inS}{\setf{N+2}{n}{}(\Fp_g,\Sp_g)}
    + \delta_{ig} \FLMlo{\inF\inS}{\setf{N+2}{n}{}(\Fp_{q_\fgflin},\Sp_g)} \Big] \rrint \\
    & = \sum_n \sum_{i \in \setf{N}{n}{}} \sum_{\fgflin=1}^{\nf} \delta_{iq_\fgflin} \frac{\asbr^2}{\ep} \lint \GammaTC_{i,q\to qgg} \, \FLMlo{\inF\inS}{\setf{N}{n}{}} \rint \,,
\end{aligned}
\label{eq_triple_coll_nf_q_qgg_app}
\end{equation}
where
\begin{equation}
    \GammaTC_{i,q\to qgg} = (E_i/\mu)^{-4\ep} \ep \, \FSRtilde{1}(E_i) \,.
\label{eq_GammaTC_qTOqgg}
\end{equation}
This time $\FSRtilde{1}$ differs from $\FSR{1}$ of Ref.~\cite{Delto:2019asp} both due to the use of energy-ordering, and because the latter was calculated without a damping factor, unlike the term in \eq\eqref{eq_triple_coll_nf_q_qgg_app}.

The last final-state splitting that we need to consider is $q_\fgflin^* \to q_\fgflin q_\sgflin \qb_\sgflin$. In this case, many contributions come from the terms in Eqs (\ref{eq_FLM_noDS_any_nf_def} -- \ref{eq_FLM_noDS_2_any_nf_def}) that, for simplicity, we do not write. We collect these terms and find that
\begin{equation}
    \GammaTC_{i,q\to qq'\qb'} = (E_i/\mu)^{-4\ep} \frac{\ep}{2} \, \Big[\FSR{3}(E_i) + 2\nf\, \FSRtilde{2}(E_i) \bigg]  \,,
\label{eq_GammaTC_qTOqqqb}
\end{equation}  
where the function $\FSRtilde{2}$ can be obtained similarly to $\FSRtilde{1}$, while $\FSR{3}$ can be taken directly from Ref.~\cite{Delto:2019asp}. We note that both functions $\FSR{2}$ and $\FSR{3}$ were previously defined with no damping factors but with energy-ordering for all terms.  However, for the latter quantity, one can prove that the two definitions are in fact equivalent.
We also note that \eq\eqref{eq_GammaTC_qTOqqqb} contains an extra factor $1/2$ with respect to \eq\eqref{eq_GammaTC_qTOqgg}, which arises because this factor has been incorporated into the definition of $\FSR{1}$, but not into that of $\FSR{2,3}$, as can be seen in Table 2 of Ref.~\cite{Delto:2019asp}.  

Similarly to the gluon case, the two contributions in Eqs (\ref{eq_GammaTC_qTOqgg}, \ref{eq_GammaTC_qTOqqqb}) can be combined in a single term (cf.\ \eq\eqref{eq_triple_coll_nf_GammaTC_q})
\begin{equation}
    \GammaTC_{i,q} = \GammaTC_{i,q\to qgg} + \GammaTC_{i,q\to q q'\qb'} \,,
\end{equation}
which defines the triple-collinear subtraction term for a final-state quark leg. 
We note that the contribution for the antiquark final state leg is identical to the one just described, so that $\GammaTC_{i,\qb} = \GammaTC_{i,q}$.

\tableTripleCollinear{t!}{
    Functions collected in the ancillary file \triplecollinearsplittings.  
    The first column identifies the functions; the second indicates splittings that contribute to each function; the third provides the names of the functions in the ancillary file. We note that, in the listed splittings, the sum over the flavor index $\sgflin$ runs over the interval $\sgflin \in [1,\nf]$ and accounts for the appearance of $\nf$ prefactors. For initial-state radiation, such an $\nf$ factor can only arise if the incoming parton and the parton entering the hard scattering are the same. Therefore, in the case of the function $\CalPTC_{q\qp}$, since $q \ne \qp$, no contribution proportional to $\nf$ arises.
}

Finally, we consider the initial-state triple-collinear limits. These cases are somewhat simpler because a) damping factors $\Delta^{(\Fp \Sp)}$ always reduce to $1$ and b) as reported in Table 1 of Ref.~\cite{Delto:2019asp}, energy ordering was only used there for cases with double-soft singularities. This is identical to what we do in this paper, so that all result of Ref.~\cite{Delto:2019asp} pertinent to initial-state limits can be used. 
We reorganize these terms and write the result for leg $i=a$ as
\begin{equation}
    \sum_n \llint \oS_{\Fp \Sp} \oS_{\Sp} \Omega_2^{(\inF)} \Delta^{(\Fp \Sp)} \Big[\THmn \bbFLMlo{\inF\inS, \DS}{\Fp, \Sp} + \bbFLMlo{\inF\inS, \noDS}{\Fp, \Sp} \Big] \rrint 
    = \sum_x \sum_n \frac{\asbr^2}{\ep} \lint \CalPTC_{x\inF} \conv \FLMlo{x\inS}{\setf{N}{n}{}} \rint \,.
\label{eq_triple_collinear_lim_nf_1_i_equal_a_app}
\end{equation} 
The $\CalPTC_{xy}$ functions have the following properties
\begin{align}
    & \CalPTC_{g\qb_\fgflin}(z,E_i) \equiv \CalPTC_{gq_\fgflin}(z,E_i) \equiv \CalPTC_{gq}(z,E_i) \,, \notag \\
    & \CalPTC_{\qb_\fgflin g}(z,E_i) \equiv \CalPTC_{q_\fgflin g}(z,E_i) \equiv \CalPTC_{qg}(z,E_i) \,, \notag \\
    & \CalPTC_{\qb_\fgflin \qb_\fgflin}(z,E_i) \equiv \CalPTC_{q_\fgflin q_\fgflin}(z,E_i) \equiv \CalPTC_{qq}(z,E_i) \,, 
    \label{eq_CalPTC_properties} \\
    & \CalPTC_{\qb_\fgflin q_\fgflin}(z,E_i) \equiv \CalPTC_{q_\fgflin \qb_\fgflin}(z,E_i) \equiv \CalPTC_{q\qb}(z,E_i) \,, \notag \\
    & \CalPTC_{\qb_\fgflin q_\tgflin}(z,E_i) \equiv \CalPTC_{q_\fgflin \qb_\tgflin}(z,E_i) \equiv \CalPTC_{\qb_\fgflin \qb_\tgflin}(z,E_i) \equiv \CalPTC_{q_\fgflin q_\tgflin}(z,E_i) \equiv \CalPTC_{qq'}(z,E_i) \,, \quad \text{with $q \neq \qp$}  \,, \notag
\end{align}
where quark flavors $\rho,v$ are assumed to be different. 
In terms of quantities computed  in Ref.~\cite{Delto:2019asp}, the functions on the right-hand side of the above equation read 
\begin{equation}
\begin{aligned}
    & \CalPTC_{gg}(z,E_i) = (E_i/\mu)^{-4\ep} \ep \, \bigg[\ISR{2}(z,E_i) + \frac{2\nf}{2}\,\ISR{7}(z,E_i)\bigg] \,, \\
    & \CalPTC_{gq}(z,E_i) = - \frac{(1-\ep)\,\Cf}{\TR} \, (E_i/\mu)^{-4\ep} \ep \, \ISR{8}(z,E_i) \,, \\
    & \CalPTC_{qg}(z,E_i) =  - \frac{\TR}{(1-\ep) \,\Cf} \, (E_i/\mu)^{-4\ep} \ep \, \ISR{9}(z,E_i) \,, \\
    & \CalPTC_{qq}(z,E_i) = (E_i/\mu)^{-4\ep} \ep \, \bigg[\ISR{1}(z,E_i) + \frac{2\nf}{2}\,\ISR{3}(z,E_i) + \ISR{5}(z,E_i) + \ISR{4}(z,E_i)\bigg] \,, \\
    & \CalPTC_{q\qb}(z,E_i) = (E_i/\mu)^{-4\ep} \ep \, \bigg[\ISR{6}(z,E_i) + \ISR{4}(z,E_i)\bigg] \,, \\
    & \CalPTC_{qq'}(z,E_i) = (E_i/\mu)^{-4\ep} \ep \, \ISR{4}(z,E_i) \,.
\end{aligned}
\label{eq_CalPTC_defs}
\end{equation} 
The functions $\ISR{1 \mydots 9}$  are  listed in Table 1 of Ref.~\cite{Delto:2019asp}.


\newpage
\bibliographystyle{JHEP}
\bibliography{biblio.bib}

\end{document}